\documentclass[10pt]{article}
\usepackage{amssymb}
\usepackage{amscd}
\usepackage{latexsym}
\usepackage{boldtensors}
\usepackage{amsmath}
\usepackage{amsthm}
\usepackage{graphicx}
\newcommand{\newsec}{\setcounter{equation}{0}\section}%
\newcommand{\Zz}{{\mathbb Z}}
\newcommand{\Nn}{{\mathbb N}}
\newcommand{\Rr}{{\mathbb R}}

\newcommand{\Ee}{{\mathbb E}}
\newcommand{\Pp}{{\mathbb P}}
\def\be{\begin{equation}}
\def\ee{\end{equation}}
\def\bea{\begin{eqnarray}}
\def\eea{\end{eqnarray}}
\def\Tr{{\rm \,Tr\,}}

\def\d{{\,\rm d}}
\def\i{{\,\rm i}}

\def\0{{\bf 0}}

\def\p{{\bf p}}

\def\P{{\bf P}}

\def\h2m{\frac{\hbar^2}{2m}}
\def\p0{{P_{\beta H^0_N}}}

\def\calZ{{\cal Z}}

\def\supp{{\rm supp\,}}

\newtheorem{theorem}{Theorem}[section]
\newtheorem{lemma}{Lemma}[section]
\newtheorem{definition}{Definition}[section]
\newtheorem{proposition}{Proposition}[section]

\theoremstyle{remark}  \newtheorem{remark}{Remark}

\begin{document}
\title{
\large\bf Bose-Einstein condensation of interacting bosons: A two-step proof} 
\author{Andr\'as S\"ut\H o\\Wigner Research Centre for Physics\\P. O. B. 49, H-1525 Budapest, Hungary\\
E-mail: suto.andras@wigner.hu\\}
\date{}
\maketitle
\thispagestyle{empty}
\begin{abstract}
\noindent
We prove two equilibrium properties of a system of interacting atoms in three or higher dimensional continuous space. (i) If the particles interact via pair potentials of a nonnegative Fourier transform, their self-organization into infinite permutation cycles is simultaneous with off-diagonal long-range order. If the cycle lengths tend to infinity not slower than the square of the linear extension of the system, there is also Bose-Einstein condensation.
(ii) If the pair potential is also nonnegative, cycles composed of a nonzero fraction of the total number of particles do appear if the density exceeds a temperature-dependent threshold value. The two together constitute the proof that in such a system Bose-Einstein condensation takes place at high enough densities.

\vspace{2mm}
\noindent\textbf{Contents}

\renewcommand\theenumii{\arabic{enumi}.\arabic{enumii}}
\renewcommand\labelenumii{\theenumii}

\begin{enumerate}
\item
Introduction
\item
Two-step proof for the ideal Bose gas
\item
Remarks to Lemma~\ref{lemma-G}
\item
Proof of Lemma~\ref{lemma-G}
\begin{enumerate}
\item
Discrete-time approximation
\item
Fourier expansions
\item
The case $x\neq 0$
\item
The limit of continuous time
\item
Merger graphs: Analysis of the constraint
\end{enumerate}
\item
Proof of Theorem~\ref{first-step}
\begin{enumerate}
\item
Condition for ODLRO
\item
Condition for BEC
\end{enumerate}
\item
Proof of Lemma~\ref{bounds-to-free-energy}
\item
Proof of Theorem~\ref{second-step}
\begin{enumerate}
\item
Cycle-decoupling model
\item
Coupling of the cycles
\end{enumerate}
\item
Physical meaning of the permutation cycles
\item
Historical notes
\end{enumerate}
\end{abstract}

\newpage
\newsec{Introduction}
The aim of this paper is to prove Bose-Einstein condensation (BEC) in a system of particles interacting via pair potentials that are both positive and of the positive type.
The main example is a Gaussian function whose Fourier transform is also a Gaussian.
More generally, pair potentials that are both positive and positive-type result as the autocorrelation function of some integrable nonnegative function $v$,
\be\label{posdef}
u(x)=\int_{\Rr^d} v(x+y)v(y)\d y,
\ee
in which case $\hat{u}(z)=|\hat{v}(z)|^2$.
Our guiding idea is to separate the proof into two steps based on the seminal works of Feynman [F2] and of Oliver Penrose and Onsager [Pen].

In the early days of the research on BEC for interacting bosons a main concern was how to infer from the formulas $\langle N_0\rangle$, the expected number of particles in the zero-momentum plane-wave state, if one does not use the second-quantized formalism. The answer was given by Oliver Penrose and Onsager in 1956 who established a connection between $\langle N_0\rangle$ and the largest eigenvalue of the one-particle reduced density matrix. We present their conclusion for particles in the canonical ensemble on the cube
$\Lambda=(-L/2,L/2]^d$ with periodic boundary conditions, when the connection appears in its clearest form.
For $N$ particles of mass $m_0$ on the torus $\Lambda$ the Hamiltonian is
\be
H_{N,L}=-\frac{\hbar^2}{2m_0}\sum_{i=1}^N\Delta_i+\sum_{1\leq i<j\leq N}u_L(x_j-x_i).
\ee
Working on the torus we must periodize the pair potential $u$ which gives
\be\label{u_L-def}
u_L(x)=\sum_{z\in\Zz^d} u(x+Lz).
\ee
The partition function for this system is
\be
Q_{N,L}=\Tr P_+ e^{-\beta H_{N,L}},\qquad P_+=\frac{1}{N!}\sum_{\pi\in S_N}P_\pi.
\ee
$P_+$ is the orthogonal projection to the symmetric subspace, $S_N$ is the symmetric group and $P_\pi$ are permutation matrices. The one-body reduced density matrix is
\be
\sigma^{N,L}_1=\frac{N}{Q_{N,L}}\Tr_{2,\dots,N}P_+ e^{-\beta H_{N,L}}
\ee
where $\Tr_{2,\dots,N}$ is the partial trace over $N-1$ particles. In the first-quantized formalism the number operator for particles in the plane-wave state $L^{-d/2}e^{\i k\cdot x}$ is
\be
N_k=\sum_{j=1}^N N_{k,j},\quad N_{k,j}=I_{j-1}\otimes| k\rangle\langle k|\otimes I_{N-j}.
\ee
Here $| k\rangle\langle k|$ is the orthogonal projection to the one-dimensional subspace spanned by $L^{-d/2}e^{\i k\cdot x}$ and $I_{j-1}$ and $I_{N-j}$ are identity operators in the $j-1$- and $N-j$-particle Hilbert spaces, respectively. It is not difficult to show [Su6] that
\be
\sigma^{N,L}_1=\sum_{k\in (2\pi/L)\Zz^d}\langle N_{k}\rangle |k\rangle\langle k|
\ee
where
\be
\langle N_k\rangle=Q_{N,L}^{-1}\Tr P_+ e^{-\beta H_{N,L}}N_k.
\ee
Let $\langle x|\sigma^{N,L}_1|y\rangle$ denote the integral kernel of $\sigma^{N,L}_1$, it then follows that
\be
\langle N_{0}\rangle=\int_\Lambda \langle x|\sigma^{N,L}_1|0\rangle \d x.
\ee

Feynman in 1953  applied his path-integral method (the Feynman-Kac formula) to analyze the superfluid transition. His conclusion was that the superfluid transition must be accompanied by the appearance of "long" permutation cycles. To combine the result of Penrose and Onsager with the idea of Feynman we need the path-integral form of $Q_{N,L}$ and $\langle x|\sigma^{N,L}_1|y\rangle$.

Let $P^t_{xy}(\d\omega)$ be the Wiener measure in $\Rr^d$ for trajectories $\omega$ that start in $x$ at time 0 and end in $y$ at time $t$.
By definition, the analogous measure on the torus of side $L$ is
\be
W^t_{xy}(\d\omega)=\sum_{z\in \Zz^d}P^t_{x,y+Lz}(\d\omega).
\ee
The Feynman-Kac form of the partition function reads
\be\label{QNL-1}
Q_{N,L}=\sum_{p=1}^{N}\frac{1}{p!}\sum_{n_1,\dots,n_p\geq 1:\sum n_l=N}\prod_{l=1}^p \frac{1}{n_l}\int_\Lambda \d x_l \int W_{x_l x_l}^{n_l\beta}(\d\omega_l)
e^{-\beta U(\omega_l)}\left(\prod_{1\leq l'<l\leq p}e^{-\beta  U(\omega_{l'},\omega_l)}\right)\,.
\ee
Because
\be
\sum_{p=1}^{N}\frac{1}{p!}\sum_{n_1,\dots,n_p\geq 1:\sum n_l=N}\prod_{l=1}^p \frac{1}{n_l}=1,
\ee
this is an average over the conjugation classes of $S_N$ that replaces the average over $S_N$.
The functionals $U(\omega_l)$ and $U(\omega_{l'},\omega_l)$ represent the potential energy:
\be
U(\omega_l)=\beta^{-1}\int_0^\beta \sum_{0\leq j<k\leq n_l-1} u_L(\omega_l(k\beta+t)-\omega_l(j\beta+t))\d t
\ee
is the potential energy of the $n_l$ particles forming the trajectory $\omega_l$, and
\be
U(\omega_{l'},\omega_l)=\beta^{-1}\int_0^\beta \sum_{j=0}^{n_{l'}-1}\sum_{k=0}^{n_l-1} u_L(\omega_l(k\beta+t)-\omega_{l'}(j\beta+t))\d t
\ee
is the energy of interaction between the trajectories $\omega_l$ and $\omega_{l'}$.
Distinguishing the trajectory that contains particle no.1, $Q_{N,L}$ can be rewritten in a form which is almost a recurrence relation (it {\em is} for the noninteracting gas),
\be\label{QNL-2}
Q_{N,L}=\frac{1}{N}\sum_{n=1}^N \int_\Lambda\d x \int W^{n\beta}_{xx}(\d\omega_0)e^{-\beta U(\omega_0)} Q_{N-n,L}(\omega_0).
\ee
Here $Q_{0,L}(\omega_0)=1$, and for $n<N$
\begin{eqnarray}
Q_{N-n,L}(\omega_0)
=
\sum_{p=1}^{N-n}\frac{1}{p!}\sum_{n_1,\dots,n_p\geq 1:\sum n_l=N-n}\,\prod_{l=1}^p\frac{1}{n_l}
\int_\Lambda \d x_l \int W_{x_l x_l}^{n_l\beta}(\d\omega_l) e^{-\beta U(\omega_l)}
\nonumber\\
\times
\left(\prod_{1\leq l'<l\leq p}e^{-\beta  U(\omega_{l'},\omega_l)}\right)\,\,
e^{-\beta \sum_{l=1}^p U(\omega_0,\omega_l)}.
\end{eqnarray}
Without $e^{-\beta \sum_{l=1}^p U(\omega_0,\omega_l)}$ this is just $Q_{N-n,L}$.

The path-integral form of $\langle x|\sigma^{N,L}_1|0\rangle$ is
\be
\langle x|\sigma^{N,L}_1|0\rangle=Q_{N,L}^{-1}\sum_{n=1}^N\int W^{n\beta}_{0x}(\d\omega)e^{-\beta U(\omega)}Q_{N-n,L}(\omega).
\ee
It shows that not only $\sigma^{N,L}_1$ is a positive operator but $\langle x|\sigma^{N,L}_1|y\rangle$ is a "matrix" of positive elements. The Banach-space analogue of the Perron-Frobenius theorem applies, $\langle N_{0}\rangle$ is the largest eigenvalue with eigenvector identically equalling $L^{-d/2}$.
The condensate density is
\be
\rho_0^{N,L}=\frac{\langle N_{0}\rangle}{L^d}
=\frac{1}{L^d}\int_\Lambda \langle x|\sigma^{N,L}_1|0\rangle \d x
=\frac{\rho}{N Q{_{N,L}}}\sum_{n=1}^N\int_{\Lambda}\d x\int W^{n\beta}_{0x}(\d\omega)e^{-\beta U(\omega)}Q_{N-n,L}(\omega)
\ee
where $\rho=N/L^d$, the total density. Note that $\omega$ in $\langle x|\sigma^{N,L}_1|0\rangle$ is an {\em open} trajectory associated with a permutation {\em cycle}, just as the closed trajectories $\omega_1,\dots,\omega_p$ do; all are effective single-particle trajectories. Further on we use the words trajectory and cycle as synonyms.

We need the density (number per unit volume) $\rho^{N,L}_n$ of particles in permutation cycles of length $n$, where the length is the number of particles in the cycle. Because the particles are indistinguishable, $\rho_n^{N,L}$ equals $\rho$ times the probability according to the canonical Gibbs distribution that particle no.1 is in a cycle of length $n$. This latter is
\[
\frac{1}{N Q_{N,L}}\int_\Lambda\d x \int W^{n\beta}_{xx}(\d\omega_0)e^{-\beta U(\omega_0)} Q_{N-n,L}(\omega_0),
\]
so using translation invariance
\be
\rho^{N,L}_n=\frac{\rho}{N Q_{N,L}}\int_\Lambda\d x\int W^{n\beta}_{xx}(\d\omega_0)e^{-\beta U(\omega_0)}Q_{N-n,L}(\omega_0)
=\frac{1}{Q_{N,L}}\int W^{n\beta}_{00}(\d\omega_0)e^{-\beta U(\omega_0)}Q_{N-n,L}(\omega_0).
\ee
Inserting this expression into the formulas for $\langle x|\sigma^{N,L}_1|0\rangle$ and $\rho^{N,L}_0$,
\be\label{sigma(x)-with-rho_n}
\langle x|\sigma^{N,L}_1|0\rangle=\sum_{n=1}^N \rho^{N,L}_n\,\frac{\int W^{n\beta}_{0x}(\d\omega)e^{-\beta U(\omega)}Q_{N-n,L}(\omega)}{\int W^{n\beta}_{00}(\d\omega_0)e^{-\beta U(\omega_0)}Q_{N-n,L}(\omega_0)},
\ee
\be\label{rho_0-with-rho_n}
\rho^{N,L}_0=\sum_{n=1}^N \rho^{N,L}_n\,\frac{\int_\Lambda\d x\int W^{n\beta}_{0x}(\d\omega)e^{-\beta U(\omega)}Q_{N-n,L}(\omega)}{L^d\int W^{n\beta}_{00}(\d\omega_0)e^{-\beta U(\omega_0)}Q_{N-n,L}(\omega_0)}.
\ee
Because $\rho=\sum_{n=1}^N \rho^{N,L}_n$, with $\rho_n=\lim_{N,L\to\infty, N/L^d=\rho}\rho^{N,L}_n$
\begin{eqnarray}
\rho
&=& \lim_{M\to\infty}\lim_{N,L\to\infty, N/L^d=\rho}\left[\sum_{n=1}^M \rho^{N,L}_n
+ \sum_{M+1}^N  \rho^{N,L}_n\right]
\\
&=& \sum_{n=1}^\infty\rho_n + \lim_{M\to\infty}\lim_{N,L\to\infty, N/L^d=\rho}\sum_{M+1}^N  \rho^{N,L}_n\,
\geq \sum_{n=1}^\infty\rho_n.
\end{eqnarray}
The infinite sum is the density of particles in finite cycles in the infinite system. If the inequality is strict, there are also infinite cycles in the infinite system.

Now we are able to state the two steps mentioned in the title.\\
(i) Prove that if (and only if) $\sum_{n=1}^\infty\rho_n<\rho$ {\em and} the diverging cycle lengths tend to infinity sufficiently fast, there is BEC.\\
(ii) Prove that for a class of pair interactions at sufficiently high densities $\sum_{n=1}^\infty\rho_n<\rho$ does occur and the diverging cycle lengths tend to infinity fast enough.

Infinite cycles, off-diagonal long-range order (ODLRO) and BEC are three distinct notions. By definition, BEC implies ODLRO which implies infinite cycles, but implication in the opposite direction is subject to conditions.
Both ODLRO and BEC are related to the infinite-volume limit of $\langle x|\sigma^{N,L}_1| 0\rangle$. There is ODLRO if
\[
\lim_{x\to\infty}\,\lim_{N,L\to\infty, N/L^d=\rho}\langle x|\sigma^{N,L}_1|0\rangle \neq 0,
\]
including the possibility that the limit does not exist, and there is BEC if
\[
\rho_0=\lim_{N,L\to\infty,N/L^d=\rho}\,\rho^{N,L}_0=\lim_{N,L\to\infty, N/L^d=\rho}\frac{1}{L^d}\int_\Lambda \langle x|\sigma^{N,L}_1| 0\rangle \d x> 0.
\]
Thus, in principle, ODLRO can exist without BEC, but the opposite is obviously false. Also, infinite cycles do not necessarily mean ODLRO. It will be seen that for positive-type pair potentials the existence of infinite cycles suffices for ODLRO, but for BEC there must be cycles of length diverging at least as fast as $N^{2/d}$.
The significance of $n\propto N^{2/d}\propto L^2/\lambda_\beta^2$ is clear: this is the order of magnitude of the necessary number of steps for a random walk of step length $\lambda_\beta$ that starts from zero to attain any point of a cube of side $L$.

Before going further we introduce three more notations.
\begin{itemize}
\item[--]
$\lambda_\beta=\sqrt{2\pi\hbar^2\beta/m_0}$ the thermal wave length,
\item[--]
$\sigma_1(x)=\lim_{N,L\to\infty,N/L^d=\rho}\,\langle x|\sigma^{N,L}_1|0\rangle$,
\item[--]
$\hat{u}$ the Fourier transform of $u$.
\end{itemize}
The first step of the proof is

\begin{theorem}\label{first-step}
Consider $N$ identical bosons on a $d\geq 3$-torus of side $L$ at inverse temperature $\beta$ that interact via a pair potential $u:\Rr^d\to\Rr$ such that
\begin{itemize}
\item[(i)]
$u$ is positive-type ( $\hat{u}\geq 0$) and $\int \hat{u}(x) x^2 \d x<\infty$,
\item[(ii)]\label{decay}
$u(x)=O\left(|x|^{-\eta}\right)$ with some $\eta>d$ as $x\to\infty$ \hspace{5pt} (condition for periodization).
\end{itemize}

\noindent
For this system the following hold true:

\noindent
\begin{enumerate}
\item
\be\label{cond1}
\rho-\sum_{n=1}^\infty \rho_n
\ \leq\
\sigma_1(x)
\ \leq
\sum_{n=1}^\infty \rho_n \exp\left\{-\frac{\pi x^2}{n\lambda_\beta^2}\right\}
+\rho-\sum_{n=1}^\infty \rho_n
\ee
implying
\be\label{lim-cond1}
\lim_{x\to\infty}\sigma_1(x)= \rho-\sum_{n=1}^\infty \rho_n.
\ee
\item
For any $c>0$
\be\label{cond2}
\rho^\infty_c
\ \leq\ \rho_0
\ \leq\ \lim_{c'\downarrow 0} \rho^\infty_{c'}.
\ee
Here
\be\label{rho-infty-c}
\rho^\infty_c= \lim_{N,L\to\infty, N/L^d=\rho}
\sum_{n= \lfloor cN^{2/d}\rfloor}^N\rho^{N,L}_n\int_{\Rr^d} \frac{\nu^{N,L}_n(y)\,\d y}{\sum_{z\in\Zz^d}\exp\left\{-\frac{\pi n \lambda_\beta^2}{L^2}z\cdot\left(z+2Ly\right)\right\}}.
\ee
$\int\nu^{N,L}_n\d y=1$, $\nu^{N,L}_n$ is concentrated to an $O(1/\sqrt{n})$-neighborhood of the origin, making sure that if
\be\label{cond-BEC}
\lim_{N,L\to\infty, N/L^d=\rho}\sum_{n=\lfloor cN^{2/d}\rfloor}^N\rho^{N,L}_n>0
\ee
then $\rho_0>0$. If the infinite cycles are exclusively macroscopic then
\be\label{rho0-macr-cycles}
\rho_0=\lim_{\varepsilon\downarrow 0}\,\lim_{N,L\to\infty, N/L^d=\rho}\sum_{n= \lfloor \varepsilon N\rfloor}^N\rho^{N,L}_n.
\ee
\end{enumerate}
\end{theorem}

Concerning ODLRO, the theorem is about the analysis of the fraction within the sum (\ref{sigma(x)-with-rho_n}) when $x$ is fixed, and $N$ and $n$ go to infinity. It will be seen that for any sequence $g_N\to\infty$ and $n\geq g_N$ the fraction tends to one so it does not prevent cycles of diverging lengths contributing to the sum. Note the (most probably monotonic) decay of $\sigma_1(x)$ from $\sigma_1(0)=\rho$ to $\sigma_1(\infty)= \rho-\sum_{n=1}^\infty \rho_n$.

The bounds in (\ref{cond2}) show that a sufficient and necessary condition for BEC is the existence of cycles whose length diverges at least as fast as $N^{2/d}$.
If $n\geq cN^{2/d}$ then $1/\sqrt{n}\leq 1/(\sqrt{c}\rho^{1/d}L)$. Therefore,
$L|y|=O(1)$ in the domain of concentration of $\nu^{N,L}_n(y)$, and
the sum over $\Zz^d$ in (\ref{rho-infty-c}) remains bounded as $N,L\to\infty$. Within the terms belonging to $n\propto N$, $L|y|=O(N^{1/d-1/2})$, and the whole integral tends to 1, which explains Eq.~(\ref{rho0-macr-cycles}).

Let
\bea\label{F(x)bis}
G\left[n,\{n_l\}_1^p\right](x)
=\prod_{l=1}^p\int_\Lambda \d x_l \int W_{x_l x_l}^{n_l\beta}(\d\omega_l)
e^{-\beta U(\omega_l)}\left(\prod_{1\leq l'<l\leq p}e^{-\beta  U(\omega_{l'},\omega_l)}\right)
\nonumber\\
L^d\int W_{0x}^{n\beta}(\d\omega) e^{-\beta U(\omega)}\,e^{-\beta \sum_{l=1}^p U(\omega,\omega_l)}
\eea
and for $n<N$
\be
G^N_{n}(x)=\sum_{p=1}^{N-n}\frac{1}{p!}\sum_{n_1,\dots,n_p\geq 1:\sum n_l=N-n}\,\frac{1}{\prod_{l=1}^p n_l} G\left[n,\{n_l\}_1^p\right](x)
=L^d\int W_{0x}^{n\beta}(\d\omega) e^{-\beta U(\omega)} Q_{N-n,L}(\omega).
\ee
Setting $Q_{0,L}(\omega)=1$, the right-hand side of this equation defines also $G^N_N(x)$. With the help of
$G^N_{n}(x)$ the equations (\ref{sigma(x)-with-rho_n}) and (\ref{rho_0-with-rho_n}) take on the form
\be\label{sigma-rho-with-G}
\langle x|\sigma^{N,L}_1|0\rangle=\sum_{n=1}^N \rho^{N,L}_n\,\frac{G^N_{n}(x)}{G^N_{n}(0)},\qquad
\rho^{N,L}_0=\sum_{n=1}^N \rho^{N,L}_n\,\frac{\int_\Lambda\d x\, G^N_{n}(x)}{L^d\, G^N_{n}(0)}.
\ee
Also, with
\be\label{G-short}
G\left[\{n_l\}_1^p\right]=\prod_{l=1}^p\int_\Lambda \d x_l \int W_{x_l x_l}^{n_l\beta}(\d\omega_l)
e^{-\beta U(\omega_l)}\left(\prod_{1\leq l'<l\leq p}e^{-\beta  U(\omega_{l'},\omega_l)}\right)
\ee
and ($n_0=n$)
\bea\label{G-0-to-p-x=0}
G\left[\{n_l\}_0^p\right]\equiv G\left[n,\{n_l\}_1^p\right]\equiv G\left[n,\{n_l\}_1^p\right](0)
=\prod_{l=0}^p\int_\Lambda \d x_l \int W_{x_l x_l}^{n_l\beta}(\d\omega_l)
e^{-\beta U(\omega_l)}\left(\prod_{0\leq l'<l\leq p}e^{-\beta  U(\omega_{l'},\omega_l)}\right)
\nonumber\\
=L^d \int W_{00}^{n_0\beta}(\d\omega_0)e^{-\beta U(\omega_0)}
\prod_{l=1}^p\int_\Lambda \d x_l \int W_{x_l x_l}^{n_l\beta}(\d\omega_l)
e^{-\beta U(\omega_l)}\left(\prod_{0\leq l'<l\leq p}e^{-\beta  U(\omega_{l'},\omega_l)}\right)
\eea
the partition function still can be written as
\be\label{QNL-average-1-to-p}
Q_{N,L}=\sum_{p=1}^{N}\frac{1}{p!}\sum_{n_1,\dots,n_p\geq 1:\sum n_l=N} \frac{1}{\prod_{l=1}^pn_l}\ G\left[\{n_l\}_1^p\right]
\ee
or, setting $G^N_n(0)=G^N_n$,
\be\label{QNL-with-G}
Q_{N,L}=
\frac{1}{N}G[N]+\frac{1}{N}\sum_{n=1}^{N-1}\sum_{p=1}^{N-n}\frac{1}{p!}\sum_{n_1,\dots,n_p\geq 1:\sum n_l=N-n}\,\frac{1}{\prod_{l=1}^p n_l}G\left[n,\{n_l\}_1^p\right]
=\frac{1}{N}\sum_{n=1}^N G^N_n,
\ee
the respective analogues of Eqs.~(\ref{QNL-1}) and (\ref{QNL-2}). Comparison of Eqs.~(\ref{F(x)bis}) and (\ref{G-0-to-p-x=0}) shows that $G\left[n,\{n_l\}_1^p\right](x)$ is continuous at $x=0$. This is a consequence of translation invariance.

Throughout the paper the numbering of the cycles may run from 1 to $p$ or from 0 to $p$ if cycle 0 is to be distinguished. In the latter case we understand $n_0=n$, and thus $G\left[\{n_l\}_0^p\right](x)= G\left[n,\{n_l\}_1^p\right](x)$.

The proof of Theorem~\ref{first-step} relies on the following lemma.

\begin{lemma}\label{lemma-G}
Suppose that $\hat{u}$ exists and is integrable. Then
\bea\label{G[nl](x)}
\lefteqn{
G\left[\{n_l\}_0^p\right](x)=
\exp\left\{-\frac{\beta\hat{u}(0)N(N-1)}{2L^d}\right\}
 \sum_{\{\alpha^k_j\in\Nn_0|1\leq j<k\leq N\}}\
\prod_{1\leq j<k\leq N}\frac{1}{\alpha^k_j !}
}
\nonumber\\
&&
\prod_{r=1}^{\alpha^k_j} \frac{-\beta}{L^d} \sum_{z^k_{j,r}\in\Zz^d\setminus\{0\}}\hat{u}\left(\frac{z^k_{j,r}}{L}\right) \int_0^1\d t^k_{j,r}\
\left[\delta_{Z^0_1,0}\sum_{z\in\Zz^d} \exp\left\{-\frac{\pi \lambda_\beta^2}{L^2}\sum_{q=1}^{n_0-1}\int_0^1\left[z+Z_q(t)\right]^2\d t\right\}\cos \left(\frac{2\pi}{L}z\cdot x\right)
 \right.
 \nonumber\\
&&
\left. \prod_{l=1}^p \delta_{Z^l_1,0}
 \sum_{z\in\Zz^d} \exp\left\{-\frac{\pi \lambda_\beta^2}{L^2}\sum_{q\in C_l}\int_0^1\left[z+Z_q(t)\right]^2\d t\right\}
\right].
\eea
There is a summation with respect to $\alpha^k_j$ for every pair $1\leq j<k\leq N$, and the quantity in the square brackets is under all these summations and integrals. Furthermore,
\be
C_l=\{N_{l-1}+1,N_{l-1}+2,\dots,N_l\}\quad\mbox{where}\quad N_l=\sum_{l'=0}^l n_{l'}\quad (N_p=N)
\ee
and for $q\in C_l$
\bea\label{Zqt}
Z_q(t)=-\sum_{j=1}^{q-1}\sum_{k=q}^{N_l}\sum_{r=1}^{\alpha^{k}_{j}}{\bf 1}_{\{  t^{k}_{j,r}\geq t\}} z^{k}_{j,r} +\sum_{j=q}^{N_l}\sum_{k=N_l+1}^{N}\sum_{r=1}^{\alpha^{k}_{j}}{\bf 1}_{\{  t^{k}_{j,r}\geq t\}} z^{k}_{j,r}\nonumber\\
-\sum_{j=1}^{q}\sum_{k=q+1}^{N_l}\sum_{r=1}^{\alpha^{k}_{j}}{\bf 1}_{\{  t^{k}_{j,r}<t\}} z^{k}_{j,r} +\sum_{j=q+1}^{N_l}\sum_{k=N_l+1}^{N}\sum_{r=1}^{\alpha^{k}_{j}}{\bf 1}_{\{  t^{k}_{j,r}<t\}} z^{k}_{j,r}\ .
\eea
In particular,
\be\label{Z^l_1}
Z^l_1:= Z_{N_{l-1}+1}(0)=-\sum_{j=1}^{N_{l-1}} \sum_{k\in C_l} \sum_{r=1}^{\alpha^{k}_{j}}z^{k}_{j,r}+\sum_{j\in C_l} \sum_{k=N_l+1}^N \sum_{r=1}^{\alpha^{k}_{j}}z^{k}_{j,r}\ .
\ee
\end{lemma}
Preceding the proof of this lemma complementary information will be given in Section 3.

\vspace{5pt}\noindent
The second step of the proof is done in

\begin{theorem}\label{second-step}
Consider $N$ identical bosons on a $d\geq 3$-torus of side $L$ at inverse temperature $\beta$ that interact via a pair potential $u:\Rr^d\to\Rr$ such that
\begin{itemize}
\item[(i)]
$u\geq 0$ and $\hat{u}\geq 0$,
\item[(ii)]
$u(x)=O\left(|x|^{-\eta}\right)$ with some $\eta>d$ as $x\to\infty$, $\hat{u}$ is integrable and $\int \hat{u}(x)x^2\d x<\infty$.
\end{itemize}
There exists a temperature-dependent positive number $\zeta_c(\beta)$ such that if $\rho>\zeta_c(\beta)/\lambda_\beta^d$ then
\be\label{infinite-cycles}
\sum_{n=1}^\infty \rho_n=\frac{\zeta_c(\beta)}{\lambda_\beta^d}
\quad\mbox{and}\quad
\rho-\sum_{n=1}^\infty \rho_n=\lim_{\varepsilon\downarrow 0}\,\lim_{N,L\to\infty, N/L^d=\rho}\,\sum_{n\geq \varepsilon N}\rho^{N,L}_n=\rho_0>0.
\ee
\end{theorem}
The inequality $\rho_0>0 $ is already the corollary of the two theorems.
Theorem~\ref{second-step} is also based on Lemma~\ref{lemma-G}, but a key element of the proof is another lemma. Below $\zeta$ is Riemann's zeta function.

\begin{lemma}\label{bounds-to-free-energy}
Let $F_{N,L}$ denote the free energy and
\be
f(\rho,\beta)
=\lim_{N,L\to\infty, N/L^d=\rho}L^{-d} F_{N,L}
=- \lim_{N,L\to\infty, N/L^d=\rho}\ \frac{1}{\beta L^d}\ln Q_{N,L}
\ee
the free energy density for an integrable superstable pair potential $u$ [R2, R3], and let $F^0_{N,L}$ and $f^0(\rho,\beta)$ be the same quantities for the ideal Bose gas. Then
\be\label{FNL-upper-lower}
\frac{C_\Lambda[u]}{L^d}N(N-1)-BN+F^0_{N,L}
\leq F_{N,L}
\leq \frac{\|u\|_1}{2L^d}N(N-1)+\frac{2^{d/2-1}\zeta(d/2)\|u\|_1}{\lambda_\beta^d}N+F^0_{N,L}
\ee
and
\be\label{f-upper-lower}
C[u] \rho^2 - B\rho + f^0(\rho,\beta)
\leq f(\rho,\beta)
\leq
\frac{\|u\|_1}{2}\rho^2 + \frac{2^{d/2-1}\zeta(d/2)\|u\|_1}{\lambda_\beta^d}\rho + f^0(\rho,\beta)
\ee
where $B>0$ and $0<C_\Lambda[u],C[u]\leq \hat{u}(0)/2$.
If $u$ is bounded, for $B$ one can substitute $u_L(0)/2$ in the first line and $u(0)/2$ in the second line. For a positive-type $u$, $C_\Lambda[u]=C[u]=\hat{u}(0)/2$. Thus, if both $u\geq 0$ and $\hat{u}\geq 0$ then
\be\label{bounds-for-u-pos-postype}
\frac{\hat{u}(0)}{2} \rho^2 - \frac{u(0)}{2}\rho + f^0(\rho,\beta)
\leq f(\rho,\beta)
\leq
\frac{\hat{u}(0)}{2}\rho^2 + \frac{2^{d/2-1}\zeta(d/2)\hat{u}(0)}{\lambda_\beta^d}\rho + f^0(\rho,\beta).
\ee
\end{lemma}

The bounds (\ref{bounds-for-u-pos-postype}) explain why we are able to show (\ref{infinite-cycles}) if both $u$ and $\hat{u}$ are nonnegative. By
subtracting the mean-field contribution $\hat{u}(0)N(N-1)/2L^d$ from the potential energy the interaction loses its superstability but remains stable; meanwhile, the condition for BEC is unchanged. In this regard the system becomes similar to the noninteracting gas, and comparison with it facilitates the proof. The result is also similar.

A more fundamental reason why $u\geq 0$ and $\hat{u}\geq 0$ can help the proof of (\ref{infinite-cycles}) is as follows.
If $\hat{u}\geq 0$, at high densities the distribution of particles in classical ground states is uniform, while if $\hat{u}$ has a negative part, there is some structural order [Su5]. Also, at moderate densities the ground state can be highly degenerate for $\hat{u}\geq 0$, as it was shown for $\hat{u}$ having a compact support [Su13]. The uniform distribution in the ground state most probably excludes any stable crystalline phase at positive temperatures. Moreover, $u\geq 0$ excludes classical condensation. On the contrary, if $u$ or $\hat{u}$ has a negative part, classical condensation or crystallization may take place at a higher temperature than that of the expected appearance of infinite cycles, so the latter should be proven in a restricted ensemble: in a liquid (the liquid helium) or in a crystal [Uel1].

It is worth noting how the conditions on the pair potential strengthen as we advance in the proof. For Lemma~\ref{lemma-G} only the existence and integrability of $\hat{u}$ is necessary. For Theorem~\ref{first-step} we need $\hat{u}\geq 0$ and $\int \hat{u}(x) x^2 \d x<\infty$, and for Theorem~\ref{second-step} we have to ask $u\geq 0$ as well.

The paper is organized as follows. In Section 2 we present the two-step proof for the ideal Bose gas. This result is not new, it appeared in papers [Su1] and [Su2], but the proof is new and its analogue will be applicable to the interacting gas.
Sections 3, 4 and 5 contain some explaining remarks about Lemma~\ref{lemma-G}, the proof of the lemma and the proof of Theorem~\ref{first-step}, respectively. In a first reading Section 4 can be skipped.
Sections 6 and 7 are devoted to the proof of Lemma~\ref{bounds-to-free-energy} and Theorem~\ref{second-step}.
The physical meaning of the permutation cycles is not obvious, a possible interpretation is given in Section 8. Section 9 ends the article with a survey of the long history of the research on BEC.

\newpage
\newsec{Two-step proof for the ideal Bose gas}

If there is no interaction, $G\left[\{n_l\}_0^p\right]$ becomes
\be
G^0\left[\{n_l\}_0^p\right]=\prod_{l=0}^p q_{n_l}
\ee
where
\be\label{qn}
q_n=\sum_{z\in\Zz^d}\exp\left\{-\frac{\pi n \lambda_\beta^2}{L^2}z^2\right\}=\frac{L^d}{n^{d/2}\lambda_\beta^d}\sum_{z\in\Zz^d}\exp\left\{-\frac{\pi L^2}{n\lambda_\beta^2}z^2\right\},
\ee
the one-particle partition function at inverse temperature $n\beta$, a monotone decreasing function of $n$ bounded below by 1. Thus, the canonical partition function on the $d$-torus of side $L$ is
\be\label{Q0NL}
Q^0_{N,L}=\frac{1}{N}\sum_{n=1}^N q_n Q^0_{N-n,L}.
\ee
Together with the initial condition $Q^0_{0,L}=1$, Eq.~(\ref{Q0NL}) defines recursively $Q^0_{N,L}$. From this equation, which could be obtained without referring to $G\left[\{n_l\}_0^p\right]$, one can easily reproduce most of the results about cycle percolation (the appearance of infinite permutation cycles) and its connection with BEC, obtained in [Su1,~2].

With the help of the single-particle energies $\epsilon_k=\hbar^2k^2/2m_0$, $k\in(2\pi/L)\Zz^d$, the canonical partition function can still be written as
\be\label{Q0N}
Q^0_{N,L}=\sum_{\sum_{k\neq 0} n_k\leq N}e^{-\beta\sum n_k\epsilon_k}=\sum_{M=0}^N \widehat{Q}^0_M,
\ee
where
\be
\widehat{Q}^0_M=\sum_{\sum_{k\neq 0} n_k=M}e^{-\beta\sum n_k\epsilon_k},
\ee
and the summations run over sets $\{n_k\}_{k\in(2\pi/L)\Zz^d\setminus\{0\}}$ of nonnegative integers. This shows that $Q^0_{N,L}>Q^0_{N-1,L}$, a crucial property for the proof of cycle percolation [Su1], which can also be obtained from (\ref{Q0NL}).

\begin{lemma}
Let $A_{-1}=0$, $A_0=1$,
$a_1,a_2,\ldots$ arbitrary numbers, and for $N\geq 1$ define recursively $A_N$ by
\be\label{A}
A_N=\frac{1}{N}\sum_{n=1}^N a_nA_{N-n}.
\ee
Then
\be\label{deltaA}
A_N-A_{N-1}=\frac{1}{N}\sum_{n=1}^N(a_n-1)(A_{N-n}-A_{N-n-1}).
\ee
\end{lemma}

\vspace{5pt}
\noindent
{\em Proof.} This follows by a simple computation. $\quad\Box$

\vspace{5pt}
Now if $a_n>1$, then $A_N>A_{N-1}$ can be proved by induction; and, because $q_n>1$, this applies to $Q_{N,L}^0$. Writing Eq.~(\ref{deltaA}) for
$Q^0_{N,L}-Q^0_{N-1,L}=\widehat{Q}^0_N$, keeping only the $n=1$ term and iterating one obtains
\be
\widehat{Q}^0_N> \frac{1}{N}(q_1-1)\widehat{Q}^0_{N-1}>\cdots> \frac{(q_1-1)^N}{N!}\quad(N\geq 1).
\ee
With $\widehat{Q}^0_0=1$ one then concludes that for fixed $L$,
\be
\lim_{N\to\infty}Q^0_{N,L}>e^{q_1-1}.
\ee
The limit is finite and is easy to compute from the middle member of (\ref{Q0N}). Because the restriction $\sum_{k\neq 0} n_k\leq N$  drops, the multiple sum factorizes. Using $\beta\epsilon_k=\pi\lambda_\beta^2 z^2/L^2$,
\be
\lim_{N\to\infty}Q^0_{N,L}=\prod_{z\in\Zz^d\setminus\{0\}}\left[1-e^{-\pi(\lambda_\beta/L)^2z^2}\right]^{-1} =\exp\left\{\sum_{n=1}^\infty\frac{q_n-1}{n}\right\}\sim e^{\zeta(1+d/2)(L/\lambda_\beta)^d}\quad (L/\lambda_\beta\gg 1).
\ee
Here $\zeta(x)=\sum_{n=1}^\infty n^{-x}$. The multiplier of $L^d$ in the exponent is $-\beta$ times the infinite-volume free energy density for the density $\rho$ above its critical value,
\be\label{f0-limit}
f^0(\rho,\beta)=-\frac{\zeta(1+d/2)}{\beta\lambda_\beta^d},\quad \rho\geq \rho^0_c(\beta)=\frac{\zeta(d/2)}{\lambda_\beta^d}.
\ee
Some simple properties of the Wiener measure are useful:
\be\label{norms-P-W}
\int P^\beta_{xy}(\d\omega)
=\lambda_\beta^{-d}e^{-\pi(y- x)^2/\lambda_\beta^2},\quad
\int W^{n\beta}_{0x}(\d\omega)
= \frac{1}{\lambda_{n\beta}^d}\sum_{z\in\Zz^d}e^{-\pi (x+L z)^2/\lambda_{n\beta}^2},
\ee
\be
\int_\Lambda \d x \int W^{n\beta}_{xx}(\d\omega)=L^d \int W^{n\beta}_{00}(\d\omega)
=\frac{L^d}{\lambda_{n\beta}^d}\sum_{z\in\Zz^d}e^{-\pi L^2 z^2/\lambda_{n\beta}^2}=q_n
\ee
and
\be\label{open-trajectory}
\int_\Lambda \d x \int W^{n\beta}_{0x}(\d\omega)=\int_\Lambda \d x\sum_{z\in\Zz^d}\int P^\beta_{0,x+Lz}(\d\omega)=\int_{\Rr^d}\d x\int P^\beta_{0,x}(\d\omega)=1.
\ee
With them
\be\label{int-GNn(x)--GNn(0)}
\frac{\int_\Lambda\d x\, G^N_{n}(x)}{L^d G^N_{n}(0)}=\frac{1}{q_n}.
\ee
The first step of the proof of BEC is as follows.
\begin{theorem}\label{thm-ideal-1}
In the ideal Bose gas there is BEC if and only if there exists a $c>0$ such that
\be\label{cond-P0-limit}
\lim_{N,L\to\infty, N/L^d=\rho} \sum_{n>cN^{2/d}}\rho^{N,L}_n>0.
\ee
\end{theorem}

\vspace{5pt}
\noindent
{\em Proof.} (i) Suppose first that (\ref{cond-P0-limit}) holds true. From Eqs.~(\ref{sigma-rho-with-G}) and (\ref{int-GNn(x)--GNn(0)})
\be\label{cond-fraction-ideal}
\rho^{N,L}_0
=\sum_{n=1}^N\frac{\rho^{N,L}_n}{q_n} \geq\sum_{n>cN^{2/d}}\frac{\rho^{N,L}_n}{q_n}
> \frac{\sum_{n>cN^{2/d}}\rho^{N,L}_n}{\sum_{z\in\Zz^d}\exp\{-\pi c(\rho^{1/d}\lambda_\beta)^2 z^2\}},
\ee
where we used the monotonic decrease of $q_n$. Taking the limit we find $\rho_0>0$.

\vspace{10pt}\noindent
(ii) Suppose now that for any $c>0$,
$
\lim_{N,L\to\infty, N/L^d=\rho} \sum_{n>cN^{2/d}}\rho^{N,L}_n=0.
$
We have
\bea
\rho^{N,L}_0=\sum_{n\leq cN^{2/d}}\frac{\rho^{N,L}_n}{q_n} +\sum_{n>cN^{2/d}}\frac{\rho^{N,L}_n}{q_n}
\leq \frac{\sum_{n\leq cN^{2/d}}\rho^{N,L}_n}{\sum_{z\in\Zz^d}\exp\{-\pi c(\rho^{1/d}\lambda_\beta)^2 z^2\}}
+\sum_{n>cN^{2/d}}\rho^{N,L}_n
\nonumber\\
\leq \frac{\rho}{\sum_{z\in\Zz^d}\exp\{-\pi c(\rho^{1/d}\lambda_\beta)^2 z^2\}}
+\sum_{n>cN^{2/d}}\rho^{N,L}_n
\eea
because $q_n>1$. Taking the limit,
\be
\rho_0\leq \frac{\rho}{\sum_{z\in\Zz^d}\exp\{-\pi c(\rho^{1/d}\lambda_\beta)^2 z^2\}}
\ee
for $c$ arbitrarily small, therefore $\rho_0=0$. $\quad\Box$

The second step can be performed in many different ways. We present one which will have its analogue in the case of interacting bosons.

\begin{theorem}
If $\rho\leq \zeta(d/2)/\lambda_\beta^d$ then in the infinite system there are only finite cycles. If $\rho> \zeta(d/2)/\lambda_\beta^d$ then
\be\label{ideal-concise}
\rho_0=\rho-\sum_{n=1}^\infty \rho_n
 =\lim_{\varepsilon\downarrow0}\lim_{N,L\to\infty,N/L^d=\rho}\sum_{n>\varepsilon N}\rho^{N,L}_n>0.
\ee
\end{theorem}

\vspace{5pt}
\noindent
{\em Proof.}
Note first that with $N$ and $L$ increasing $Q^0_{N,L}=\exp\{-\beta F^0_{N,L}\}=\exp\{-\beta L^d [f^0(\rho,\beta)+o(1)]\}$. Furthermore, if $n=o(N)$ then
\be
\frac{Q^0_{N-n,L}}{Q^0_{N,L}}=e^{\beta n\frac{F^0_{N,L}/L^d-F^0_{N-n,L}/L^d}{(n/L^d)}}
=
e^{\beta n[\partial f^0(\rho,\beta)/\partial\rho+o(1)]}.
\ee
In $\partial f^0(\rho,\beta)/\partial\rho$ we recognise the chemical potential computed in the canonical ensemble.
Let $g_N=o(N^{2/d})$ be any monotone increasing sequence of integers that tends to infinity.
Applying the second form of $q_n$,
\be
\frac{1}{N}\sum_{n=1}^{g_N}q_n \frac{Q^0_{N-n,L}}{Q^0_{N,L}}
=\frac{1}{\rho\lambda_\beta^d}\sum_{n=1}^{g_N}\frac{1}{n^{d/2}}\left[1+O\left(e^{-\pi L^2/(n\lambda_\beta^2)}\right)\right]
e^{\beta n[\partial f^0(\rho,\beta)/\partial\rho+o(1)]}.
\ee
We cut the sum in two parts, $\sum_{n=1}^M+\sum_{n=M+1}^{g_N}$ and take $\lim_{M\to\infty}\lim_{N,L\to\infty, N/L^d=\rho}$ resulting
\bea\label{lim1}
\lim_{N,L\to\infty, N/L^d=\rho}\frac{1}{N}\sum_{n=1}^{g_N}q_n \frac{Q^0_{N-n,L}}{Q^0_{N,L}}
&=&
\frac{1}{\rho\lambda_\beta^d}\sum_{n=1}^\infty \frac{\exp\{\beta n\partial f^0(\rho,\beta)/\partial\rho\}}{n^{d/2}}
\nonumber\\
&+&
\lim_{M\to\infty}\lim_{N,L\to\infty, N/L^d=\rho}\frac{1}{N}\sum_{n=M+1}^{g_N}q_n \frac{Q^0_{N-n,L}}{Q^0_{N,L}}.
\eea
The infinite sum contains the contribution of all the finite cycles. The rest comes from cycles whose length although diverges with $N$, but the divergence is slower than $N^{2/d}$, and the double limit yields zero:
\[
\lim_{M\to\infty}\lim_{N,L\to\infty, N/L^d\rho}\frac{1}{N}\sum_{n=M+1}^{g_N}q_n \frac{Q^0_{N-n,L}}{Q^0_{N,L}}
=\frac{1}{\rho\lambda_\beta^d}\lim_{M\to\infty}\lim_{N,L\to\infty, N/L^d\rho}\sum_{n=M+1}^{g_N}\frac{1}{n^{d/2}} \frac{Q^0_{N-n,L}}{Q^0_{N,L}}=0.
\]
Next, fix any $c>0$ and choose any monotone increasing sequence $h_N$ of integers such that $h_N>c N^{2/d}$ and $h_N=o(N)$. Then
\be\label{lim2}
\lim_{N,L\to\infty, N/L^d=\rho}\frac{1}{N}\sum_{n=c N^{2/d}}^{h_N}q_n \frac{Q^0_{N-n,L}}{Q^0_{N,L}}
=0
\ee
because $q_n Q^0_{N-n,L}/Q^0_{N,L}=O(1)$. Combining Eqs.~(\ref{lim1}) and (\ref{lim2}),
\be\label{1=limit0}
1=\frac{1}{\rho\lambda_\beta^d}\sum_{n=1}^\infty \frac{\exp\{\beta n\partial f^0(\rho,\beta)/\partial\rho\}}{n^{d/2}}+
\lim_{\varepsilon\downarrow 0}\lim_{N,L\to\infty, N/L^d=\rho}\frac{1}{N}\sum_{n=\varepsilon N}^N\frac{Q^0_{N-n,L}}{Q^0_{N,L}}.
\ee
We used also that $q_n\to 1$ if $n$ increases faster than $N^{2/d}$.
Now $f^0$ is a negative strictly monotone decreasing function of $\rho$ below the critical density
$\zeta(d/2)/\lambda_\beta^d$. Therefore, for $\rho\leq\zeta(d/2)/\lambda_\beta^d$, $\varepsilon>0$ and $n\geq \varepsilon N$
\be
\frac{Q^0_{N-n,L}}{Q^0_{N,L}}\sim e^{-\beta L^d\left[f^0((1-n/N)\rho,\beta)-f^0(\rho,\beta)\right]}
\leq e^{-\beta L^d\left[f^0((1-\varepsilon)\rho,\beta)-f^0(\rho,\beta)\right]}
\ee
decays exponentially fast with the volume, $N^{-1}\sum_{n=\varepsilon N}^N Q^0_{N-n,L}/Q^0_{N,L}$ also decays at this rate and turns Eq.~(\ref{1=limit0}) into
\be
\frac{1}{\rho\lambda_\beta^d}\sum_{n=1}^\infty \frac{\exp\{\beta n\partial f^0(\rho,\beta)/\partial\rho\}}{n^{d/2}}=1\qquad (\rho\leq\zeta(d/2)/\lambda_\beta^d).
\ee
This equation determines $\partial f^0(\rho,\beta)/\partial\rho$ and assigns to it a non-positive value.
At $\rho=\zeta(d/2)/\lambda_\beta^d$ the sum over $n$ and $\partial f^0(\rho,\beta)/\partial\rho$ reach their maxima, $\zeta(d/2)$ and zero, respectively.
Therefore, if $\rho\lambda_\beta^d>\zeta(d/2)$, Eq.~(\ref{1=limit0}) can hold only with a positive contribution from macroscopic cycles. For this conclusion we are not obliged to further analyse the second term on the right-hand side of Eq.~(\ref{1=limit0}). Still, it must be detached from zero exactly when the first term sinks below 1, and this is easy to see.
If $\rho>\zeta(d/2)/\lambda_\beta^d$ then $\partial f^0(\rho,\beta)/\partial\rho\equiv 0$. At the same time, if
$(1-n/N)\rho\geq \zeta(d/2)/\lambda_\beta^d$, then
$f^0((1-n/N)\rho,\beta)=f^0(\rho,\beta)=-\frac{\zeta(1+d/2)}{\beta\lambda_\beta^d}$,
and $Q^0_{N-n,L}/Q^0_{N,L}\to 1$.
(In the case of periodic boundary conditions there is no surface/edge/corner correction to the limit of $Q^0_{N-n,L}/Q^0_{N,L}$.) So (\ref{1=limit0}) becomes
\[
\frac{\zeta(d/2)}{\rho\lambda_\beta^d}+\lim_{\varepsilon\downarrow 0}\lim_{N,L\to\infty, N/L^d=\rho}\frac{1}{N}\sum_{n=\varepsilon N}^{(1-\zeta(d/2)/\rho\lambda_\beta^d)N}\frac{Q^0_{N-n,L}}{Q^0_{N,L}}
=\frac{\zeta(d/2)}{\rho\lambda_\beta^d}+\left(1-\frac{\zeta(d/2)}{\rho\lambda_\beta^d}\right)=1\quad (\rho>\zeta(d/2)/\lambda_\beta^d).\quad\Box
\]

Finally, we recall some more results obtained in [Su2]. First, the expected number of infinite cycles that contain at least a fraction $x$ of the total number of particles is
$\ln \frac{\rho_0}{x\rho}$ for any $ x<\rho_0/\rho$, cf. Eq.~(44) of [Su2]. This number can be arbitrarily large if $x$ is sufficiently small. For $m \geq \ln \rho/\rho_0$ the expected number of infinite cycles of density between
$e^{-(m+1)}\rho$ and $e^{-m}\rho$ is
\[\ln\frac{e^{m+1}\rho_0}{\rho}-\ln\frac{e^{m}\rho_0}{\rho} = 1.\]
The intervals $[e^{-(m+1)}, e^{-m})$ are disjoint, their number is infinite and on average there belongs one infinite cycle to each interval.

Second, the limit shape of partitions of $N$ in Vershik's sense [V] can be inferred from Eqs.~(26), (27) and (44) of [Su2]. Let $r_k(\lambda)$ denote the number of elements of length $k$ in the partition $\lambda$ of $N$. Then the limit measure over the set of partitions is singular,
\be
\lim_{N,L\to\infty, N/L^d=\rho}\frac{r_k(\lambda)}{N}= \frac{1}{k}\frac{\rho_k}{\rho}
=\left\{
\begin{array}{lll}
\frac{z^k}{k^{d/2+1}\rho\lambda_\beta^d} & \mbox{if}& \rho\lambda_\beta^d\leq \zeta(d/2)\\
\frac{1}{k^{d/2+1}\rho^0_c(\beta)\lambda_\beta^d}=\frac{1}{k^{d/2+1}\zeta(d/2)} & \mbox{if} & \rho\lambda_\beta^d> \zeta(d/2)
\end{array}
\right.
\ee
with probability one, where $z$ is the solution of
\[
\sum_{n=1}^\infty\frac{z^n}{n^{d/2}}=\rho\lambda_\beta^d.
\]
Thus, with the scaling factor $a=1$ the limit shape for the finite elements of the partitions is
\be
\tilde{\varphi}_\lambda(t):=\frac{a}{N}\sum_{k\geq at}r_k(\lambda)\to\left\{
\begin{array}{lll}
\frac{1}{\rho\lambda_\beta^d}\sum_{k\geq t}\frac{z^k}{k^{d/2+1}} & \mbox{if}& \rho\lambda_\beta^d\leq \zeta(d/2)\\
\frac{1}{\zeta(d/2)}\sum_{k\geq t}\frac{1}{k^{d/2+1}} & \mbox{if} & \rho\lambda_\beta^d> \zeta(d/2).
\end{array}
\right.
\ee
Furthermore, when $\rho\lambda_\beta^d> \zeta(d/2)$, with the scaling sequence $a_N=N\rho^{N,L}_0/\rho$ we obtain the limit shape for the macroscopic elements of the partitions : dropping the prefactor $\rho^{N,L}_0/\rho$,
\be
\tilde{\varphi}^{\rm macr}_\lambda(t)\propto\sum_{k\geq (\rho^{N,L}_0/\rho)Nt}r_k(\lambda)\to \left(\ln\frac{1}{t}\right)_+
\ee
 with probability one.

\newsec{Remarks to Lemma~\ref{lemma-G}}\label{remarks-lemma-G}

\begin{remark}
In the proof we will obtain
$$\int_0^1\d t^k_{j,\alpha^k_j}\int_0^{t^k_{j,\alpha^k_j}}\d t^k_{j,\alpha^k_j-1} \cdots\int_0^{t^k_{j,2}}\d t^k_{j,1}$$
and replace it with the symmetric
$$\frac{1}{\alpha^k_j!}\int_0^1\d t^k_{j,1}\,\d t^k_{j,2}\cdots \d t^k_{j,\alpha^k_j}\ .$$
This can be done because in the formula (\ref{Zqt}) the times $t^{k}_{j,r}$ are compared to $t$, not to each other, therefore any permutation of $t^{k}_{j,1},\dots,t^{k}_{j,\alpha^k_j}$ rearranges only the sums with respect to $r$.
\end{remark}

\begin{remark}
The exponents in Eq.~(\ref{G[nl](x)}) can be expanded yielding
\bea\label{exprewritten}
\lefteqn{\sum_{z\in\Zz^d} \exp\left\{-\frac{\pi \lambda_\beta^2}{L^2}\sum_{q\in C_l}\int_0^1\left[z+Z_q(t)\right]^2\d t\right\}}
\nonumber\\
&&=\exp\left\{-\frac{\pi n_l \lambda_\beta^2}{L^2}\left[\overline{\left(Z^l_{^\cdot}\right)^2}-\overline{Z^l_{^\cdot}}^2\right]\right\}
\sum_{z\in\Zz^d}\exp\left\{-\frac{\pi n_l \lambda_\beta^2}{L^2}\left(z+\overline{Z^l_{^\cdot}}\right)^2\right\}
\eea
where
\be\label{intZq}
\overline{Z^l_{^\cdot}}=\frac{1}{n_l}\sum_{q\in C_l}\int_0^1 Z_q(t)\d t,\quad \overline{\left(Z^l_{^\cdot}\right)^2}=\frac{1}{n_l}\sum_{q\in C_l}\int_0^1 Z_q(t)^2\d t.
\ee
Performing the time integral and interchanging the order of summations,
\bea\label{intZqbis}
&&n_l\overline{Z^l_{^\cdot}}=\sum_{q\in C_l}\int_0^1 Z_q(t)\d t=-\sum_{j=1}^{N_{l-1}}\sum_{k\in C_l}\sum_{r=1}^{\alpha^{k}_{j}}(k-N_{l-1}-1+t^{k}_{j,r})z^{k}_{j,r}
\nonumber\\
&&-\sum_{\{ j<k\}\subset C_l}(k-j)\sum_{r=1}^{\alpha^{k}_{j}}z^{k}_{j,r}
+\sum_{j\in C_l}\sum_{k=N_l+1}^N \sum_{r=1}^{\alpha^{k}_{j}}(j-N_{l-1}-1+t^{k}_{j,r})z^{k}_{j,r}\nonumber\\
&&\equiv -\sum_{k\in C_l}\sum_{j=1}^{k-1}\sum_{r=1}^{\alpha^{k}_{j}}\left(k-N_{l-1}-1+t^{k}_{j,r}\right)z^{k}_{j,r} +\sum_{j\in C_l}\sum_{k=j+1}^N\sum_{r=1}^{\alpha^{k}_{j}}\left(j-N_{l-1}-1+t^{k}_{j,r}\right)z^{k}_{j,r}.\nonumber\\
\eea
The second form is derived differently, see the proof of the lemma, but the identity is evident if we write
\[
k-j=(k-N_{l-1}-1+t^{k}_{j,r})-(j-N_{l-1}-1+t^{k}_{j,r}).
\]
The formula for $\overline{\left(Z^l_{^\cdot}\right)^2}$ is lengthier and not very illuminating. It can be obtained by integration with respect to $t$, but there is still another way described in the proof of the lemma. The special cases needed in the proof of the theorems will be given there.
\end{remark}

\vspace{5pt}
\begin{remark}\label{D_l^2=0}
$\overline{\left(Z^l_{^\cdot}\right)^2}-\overline{Z^l_{^\cdot}}^2=0$ if and only if $\overline{\left(Z^l_{^\cdot}\right)^2}=\overline{Z^l_{^\cdot}}^2=0$ and this holds if and only if
\be\label{in-l-cycle}
\alpha^k_j=0 \quad\mbox{whenever}\quad \{j,k\}\cap C_l\neq\emptyset.
\ee
Equation (\ref{in-l-cycle}) implies $Z_q(t)\equiv 0$ and hence $\overline{\left(Z^l_{^\cdot}\right)^2}=\overline{Z^l_{^\cdot}}^2=0$.
For a proof in the other direction one must inspect Eqs.~(\ref{Zqt}), (\ref{Z^l_1}) and (\ref{intZq}). $\overline{\left(Z^l_{^\cdot}\right)^2}-\overline{Z^l_{^\cdot}}^2=0$ if and only  if $Z_q(t)$ is the same constant for all $q\in C_l$ for almost all $t\in [0,1]$. The times $ t^{k}_{j,r}$ occurring in (\ref{Zqt}) are almost surely positive, so there is an interval above 0 in which $Z_{N_{l-1}+1}(t)=Z_{N_{l-1}+1}(0)=Z^l_1$; therefore $Z_q(t)=Z^l_1$ ($=0$ because of $\delta_{Z^l_1,0}$)
for all $q\in C_l$ for almost all $t\in [0,1]$. This however needs the condition (\ref{in-l-cycle}).

A physical interpretation is the following. $\hbar(2\pi/L)Z_q(t)$ may be considered as the shift due to interactions of the momentum of the $q$th particle at "time" $t$ compared to its value in the ideal gas. The shift does not fluctuate if the interaction does not fluctuate, like when the particle is exposed only to a constant external field. This is precisely what (\ref{in-l-cycle}) means, cf. the next remark.
\end{remark}

\vspace{5pt}
\begin{remark}\label{constraint}
$Z^l_1=0$ will be seen as a consequence of the closing of the $l$th trajectory. If $x=0$, this holds for all the trajectories; if $x\neq 0$, the trajectories $l=1,\dots,p$ are closed, but the zeroth trajectory is open. Nonetheless, $Z^0_1=0$, because the general formula (\ref{Z^l_1}) is valid for $l=0$ and because $\sum_{l=0}^p Z^l_1\equiv 0$. ($Z^0_1$ must be zero, otherwise $G\left[n,\{n_l\}_1^p\right](x)$ would not be continuous at $x=0$.)
The constraint is absent in the mean-field approximation which consists in keeping a single term,
$\alpha^k_j= 0$ for all $1\leq j<k\leq N$. For $x=0$ this yields the partition function of the mean-field Bose gas,
\be
Q_{N,L}^{\rm m.f.}= \exp\left\{-\frac{\beta\rho\hat{u}(0)(N-1)}{2}\right\}
\sum_{p=1}^N
\frac{1}{p!}\sum_{n_1,\dots,n_p\geq 1:\sum_1^p n_l=N}\ \prod_{l=1}^p\frac{1}{n_l}
\sum_{z\in\Zz^d}\exp\left\{-\frac{\pi n_l \lambda_\beta^2}{L^2}z^2\right\},
\ee
differing from that of the noninteracting gas only in the first factor.
One can go beyond the  mean-field approximation and still avoid the constraint. $Z^l_1$ depends only on $z^k_{j,r}$ where one and only one of $j$ and $k$ is in
$C_l,$
i.e., in the $l$th cycle. A meaningful completion of the mean-field result is to add terms in which the cycles are decoupled by retaining only $\alpha^k_j=0$ if $j$ and $k$ are in different cycles. The {\em cycle-decoupling model} thus obtained plays a central role in the proof of Theorem\ref{second-step}.

To go even further,
\be
A_l:=\sum_{j=1}^{N_{l-1}}\sum_{k\in C_l}\alpha^k_j+\sum_{j\in C_l}\sum_{k=N_l+1}^N\alpha^k_j\geq 2
\ee
is necessary to satisfy $Z^l_1=0$ with nonzero vectors. Such an equation cannot stand alone: if $\alpha^k_j>0$, one of $j$ and $k$ is in another cycle $l'$, implying a coupled equation $Z^{l'}_1=0$. Suppose that $l_1<\cdots<l_s$ ($s\geq 2$) label a maximal set of coupled permutation cycles, giving rise to $s$ homogeneous linear equations. Clearly, $s\leq \frac{1}{2}\sum_{i=1}^s A_{l_i}$, the number of involved pairs $(j,k)$ at $\alpha^k_j$ different times, i.e., the total number of variables in the $s$ equations. Explicitly, these variables are
\[
\bigcup_{1\leq i'<i\leq s}\left\{z^k_{j,r}|j\in C_{l_{i'}},\ k\in C_{l_i},\ r=1,\dots,\alpha^k_j\right\}.
\]
In general, if $j$ is in cycle $l'$ and $k$ is in cycle $l>l'$, $z^k_{j,r}$ appears in two equations with different signs: as $+z^k_{j,r}$ in $Z^{l'}_1=0$ and as $-z^k_{j,r}$ in $Z^{l}_1=0$. For a given set $\{\alpha^k_j\}$ the system of equations may not be soluble with all $z^k_{j,r}$ nonzero. Such an $\{\alpha^k_j\}$ is illicit and is discarded by $\prod_{l=0}^p \delta_{Z^l_1,0}$. On the other hand, if $\{\alpha^k_j\}$ is a valid set and if the number of linearly independent equations is $K_{\{\alpha^k_j\}}$ then $\prod_{l=0}^p \delta_{Z^l_1,0}$ makes $K_{\{\alpha^k_j\}}$ sums over $\Zz^d\setminus\{0\}$ collapse into a single vector leaving a factor $ (L^{-d})^{K_{\{\alpha^k_j\}}}$ uncompensated by sums. This appears as a loss of weight in $Q_{N,L}$ compared to the mean-field and other contributions with uncoupled cycles.

A systematic study of the constraint can be done with the help of graph theory. Consider a graph ${\cal G}_{\{\alpha^k_j\}}$ of $p+1$ vertices numbered from 0 to $p$ with
$$
b_{l'l}=\sum_{j\in C_{l'}}\sum_{k\in C_l}\alpha^k_j
$$
edges between the vertices $l'$ and $l>l'$. The edge set is valid if to every edge one can assign a nonzero vector so that at any vertex $l$ the sum of the vectors on the incident edges $(l,l_1),(l,l_2),\dots$, taken with minus sign if $l_i<l$ and with plus sign if $l_i>l$, is zero.
A solution to this graph problem together with examples is presented in the proof of Lemma~\ref{lemma-G}.
Here is the result:

\noindent
A set $\{\alpha^k_j\}$ is compatible with the constraint $\prod_{l=0}^p \delta_{Z^l_1,0}$ if and only if every maximal connected subgraph of ${\cal G}_{\{\alpha^k_j\}}$ is either a single vertex or a merger through vertices or along edges of circular graphs of arbitrary length.

A simple argument shows (see Proposition~\ref{K=V-1}) that for a connected merger graph of $V$ vertices the number $K$ of independent equations for the edge variables is $V-1$, whatever be the number of edges. Let the $\sum_{0\leq l'<l\leq p}b_{l'l}$
edges of ${\cal G}_{\{\alpha^k_j\}}$ couple the $p+1$ vertices into $m_{\{\alpha^k_j\}}$ maximal connected subgraphs of $V_1,\dots,V_{m_{\{\alpha^k_j\}}}$ vertices, respectively. The $i$th one contributes to $K_{\{\alpha^k_j\}}$ with $V_i-1$, so
\be\label{K-sum-V}
K_{\{\alpha^k_j\}}=\sum_{i=1}^{m_{\{\alpha^k_j\}}}(V_i-1)=p+1-m_{\{\alpha^k_j\}}.
\ee
The cycles can be coupled in so many different ways that the coupling always overcompensates the loss and results in a decrease of the free energy density. This will be seen in the proof of Theorem~\ref{second-step}.
\end{remark}

\begin{remark}
Asymptotically, for $L$ so large that the Riemann integral-approximating sum $L^{-d}\sum_{z\in\Zz^d\setminus\{0\}}\hat{u}(z/L)$ can be replaced with $\int_{\Rr^d} \hat{u}(x) \d x$, by the substitutions $x^{k}_{j,r}=z^{k}_{j,r}/L$, $X^l_1=Z^l_1/L$ and $X_q(t)=Z_q(t)/L$ one obtains
\bea\label{G[nl](x)-asymp}
G\left[\{n_l\}_0^p\right](x)
=
\exp\left\{-\frac{\beta\hat{u}(0)N(N-1)}{2L^d}\right\}
\sum_{\{\alpha^k_j\in\Nn_0|1\leq j<k\leq N\}}\
\Delta_{\{\alpha^k_j\},\{n_l\}} \left(L^{-d}\right)^{K_{\{\alpha^k_j\}}}
 \prod_{1\leq j<k\leq N}
\frac{ \left(-\beta\right)^{\alpha^k_j}}{\alpha^k_j !}
\nonumber\\
\prod_{r=1}^{\alpha^k_j}\int_0^1\d t^k_{j,r} \int \d x^k_{j,r}\ \hat{u}\left(x^k_{j,r}\right)
\left[
\delta(X^0_1,\dots,X^p_1) \sum_{z\in\Zz^d} \exp\left\{-\frac{\pi \lambda_\beta^2}{L^2}\sum_{q=1}^{n_0-1}\int_0^1\left[z+LX_q(t)\right]^2\d t\right\}\cos \left(\frac{2\pi}{L}z\cdot x\right)
 \right.
\nonumber\\
\left. \prod_{l=1}^p \sum_{z\in\Zz^d}
\exp\left\{-\frac{\pi \lambda_\beta^2}{L^2}\sum_{q\in C_l}\int_0^1\left[z+LX_q(t)\right]^2\d t\right\}
\right].
\eea
The factor $(L^{-d})^{K_{\{\alpha^k_j\}}}$ appears explicitly because
\bea
\lefteqn{
\left[ \prod_{1\leq j<k\leq N}\prod_{r=1}^{\alpha^k_j}
\frac{1}{L^d} \sum_{z^k_{j,r}\in\Zz^d\setminus\{0\}}\hat{u}\left(\frac{z^k_{j,r}}{L}\right)\right] \prod_{l=0}^p \delta_{Z^l_1,0}
}  \nonumber\\
&&\sim\left(L^{-d}\right)^{K_{\{\alpha^k_j\}}}
\int \hat{u}\left(x^k_{j,r}\right) \delta(X^0_1,\dots,X^p_1)\prod_{1\leq j<k\leq N}\prod_{r=1}^{\alpha^k_j}\d x^k_{j,r}
\eea
where $\delta(X^0_1,\dots,X^p_1)$
restricts the multiple integral to a $d\left(\sum_{j<k}\alpha^k_j-K_{\{\alpha^k_j\}}\right)$-dimensional manifold on which $X^0_1=\cdots=X^p_1=0$.
Moreover, $\Delta_{\{\alpha^k_j\},\{n_l\}}=1$ if ${\cal G}_{\{\alpha^k_j\}}$ is a merger graph and is zero otherwise.
As in Eqs.~(\ref{exprewritten}) and (\ref{intZq}),
\bea\label{exp-asymp}
\lefteqn{
\sum_{z\in\Zz^d}\exp\left\{-\frac{\pi \lambda_\beta^2}{L^2}\sum_{q\in C_l}\int_0^1\left[z+LX_q(t)\right]^2\d t\right\}   }
\nonumber\\
&&=\exp\left\{-\pi n_l \lambda_\beta^2\left[\overline{\left(X^l_{^\cdot}\right)^2}-\overline{X^l_{^\cdot}}^2\right]\right\}
\sum_{z\in\Zz^d}\exp\left\{-\frac{\pi n_l \lambda_\beta^2}{L^2}\left(z+L\overline{X^l_{^\cdot}}\right)^2\right\}
\eea
and
\be\label{exp-asymp-expanded}
\overline{X^l_{^\cdot}}=\frac{1}{n_l}\sum_{q\in C_l}\int_0^1 X_q(t)\d t,\quad   \overline{\left(X^l_{^\cdot}\right)^2}
=\frac{1}{n_l}\sum_{q\in C_l}\int_0^1 X_q(t)^2\d t
\ee
with the replacements $\overline{X^l_{^\cdot}}=\overline{Z^l_{^\cdot}}/L$, $\overline{(X^l_{^\cdot})^2}=\overline{(Z^l_{^\cdot})^2}/L^2$.
\end{remark}

\newsec{Proof of Lemma~\ref{lemma-G}}

The main steps of the proof are as follows.

\vspace{5pt}
\noindent
(1) Discrete-time approximation: the time integrals of the pair potentials are replaced by their Riemann approximating sum. Then the Wiener measure depends only on a finite sequence of values that the trajectory assumes in the selected instants, and appears as a product of a finite number of (periodized) Gaussians.\\
(2) Each Gaussian and each Boltzmann factor is Fourier-expanded, making possible the integration with respect to the spatial variables. This introduces relations among the Fourier variables and permits to eliminate those associated with the Gaussians, i.e., with the kinetic energy.\\
(3) One takes the limit that restores the continuous time.

\vspace{5pt}
\noindent
First we prove the formula for $G\left[\{n_l\}_1^p\right]$, cf. Eq.~(\ref{G-short}).

\subsection{Discrete-time approximation}

\noindent
The interval $[0,\beta]$ is divided into $m$ subintervals of length $\beta/m$.  Let
\be\label{xl-km+i}
x^l_{km+i}=\omega_l((km+i)\beta/m),\qquad l=1,\ldots
,p,\quad k=0,\ldots,n_l-1,\quad i=1,\ldots,m,\qquad x^l_{n_l m}\equiv x^l_0=x_l.
\ee
Thus, the spatial integration variables will be $x^l_1,\dots, x^l_{n_l m}$ for $l=1\dots,p$. Then, with the notation
\be
E_m(x)=\exp\left\{-\frac{\beta}{m} u_L(x)\right\}
\ee
the discrete-time approximation of $G\left[\{n_l\}_1^p\right]$ is
\bea
G_m\left[\{n_l\}_1^p\right]= \int_{\Lambda^{mN}} \prod_{l=1}^p\prod_{j=1}^{n_l m}\d x^l_j\int\prod_{l=1}^p\prod_{j=1}^{n_l m} W^{\beta/m}_{x^l_{j-1}x^l_j}(\d\omega^l_j) \prod_{i=1}^m \prod_{l=1}^p\ \prod_{0\leq j<k\leq n_l-1}E_m(x^l_{km+i}-x^l_{jm+i})\nonumber\\
\prod_{i=1}^m\, \prod_{1\leq l'<l\leq p}\prod_{j=0}^{n_{l'}-1}\prod_{k=0}^{n_{l}-1} E_m(x^l_{km+i}-x^{l'}_{jm+i}).
\eea

\subsection{ Fourier expansions}

Let us start with the Wiener measure.
\be
\int W^{\beta/m}_{xy}(\d\omega)=\sum_{z\in\Zz^d}\int P^{\beta/m}_{x,y+Lz}(\d\omega) =\lambda_{\beta/m}^{-d}\sum_{z\in\Zz^d}e^{-\pi(x-y+Lz)^2/\lambda_{\beta/m}^2} =\frac{1}{L^d}\sum_{z\in\Zz^d}e^{-\pi\lambda_{\beta/m}^2z^2/L^2} e^{\i\frac{2\pi}{L}z\cdot (x-y)}.
\ee
The rightmost form, obtained via the Poisson summation formula, is the Fourier representation we need. Denoting the dual vector of $x^l_{j}- x^l_{j-1}$ by $v^l_{j,j-1}$ ($j=1,\dots,n_lm$),
\bea\label{prodW}
\int\prod_{j=1}^{n_l m} W^{\beta/m}_{x^l_{j-1}x^l_j}(\d\omega^l_j) =\frac{1}{L^{dn_lm}} \sum_{v^l_{10},v^l_{21},\ldots,v^l_{n_lm,n_lm-1}\in\Zz^d} \exp\left\{-\frac{\pi\lambda_{\beta}^2}{mL^2}\sum_{j=1}^{n_lm}(v^l_{j,j-1})^2\right\} \nonumber\\
\times \exp\left\{\i\frac{2\pi}{L}\sum_{j=1}^{n_lm}v^l_{j,j-1}\cdot(x^l_j-x^l_{j-1})\right\}.
\eea
Now
\be
\sum_{j=1}^{n_lm}v^l_{j,j-1}\cdot(x^l_j-x^l_{j-1})=\sum_{j=1}^{n_lm}(v^l_{j,j-1}-v^l_{j+1,j})\cdot x^l_j \quad \mbox{where}\quad  v^l_{n_lm+1,n_lm}\equiv v^l_{10}.
\ee
Introduce
\be
v^l_j=v^l_{j,j-1}-v^l_{j+1,j}\ ,\quad j=1,\ldots,n_lm.
\ee
There are only $n_lm-1$ linearly independent differences,
\be\label{sumvlj}
\sum_{j=1}^{n_lm} v^l_j=0.
\ee
The new variables $\{v^l_j\}_{j=1}^{n_lm-1}$ together with $v^l_{10}$ form a linearly independent set. Utilizing them,
\be
v^l_{j,j-1}=v^l_{10}+\sum_{j'=j}^{n_lm}v^l_{j'},\qquad j=1,\ldots,n_lm,
\ee
\be
\sum_{j=1}^{n_lm}(v^l_{j,j-1})^2= \sum_{j=1}^{n_1m}(v^l_{10}+\sum_{j'=j}^{n_lm}v^l_{j'})^2,
\ee
\be
\sum_{j=1}^{n_lm}v^l_{j,j-1}\cdot(x^l_j-x^l_{j-1})=\sum_{j=1}^{n_lm}v^l_j\cdot x^l_j.
\ee
Substituting the last two expressions into Eq.~(\ref{prodW}) and replacing the summation variables $v^l_{j+1,j}$ with $v^l_j$ for $j=1,\dots,n_lm-1$ while keeping $v^l_{10}$,
\bea\label{Wxj-1xj}
\int\prod_{j=1}^{n_l m} W^{\beta/m}_{x^l_{j-1}x^l_j}(\d\omega^l_j) =\frac{1}{L^{dn_lm}}
\sum_{v^l_{10}\in\Zz^d}\ \sum_{v^l_{1},v^l_{2},\ldots,v^l_{n_lm-1}\in\Zz^d} \exp\left\{-\frac{\pi\lambda_{\beta}^2}{mL^2}\sum_{j=1}^{n_lm}(v^l_{10}+\sum_{j'=j}^{n_lm}v^l_{j'})^2\right\}
\nonumber\\
\prod_{j=1}^{n_l m}\exp\left\{\i\frac{2\pi}{L} x^l_{j}\cdot v^l_{j}\right\}.
\eea
The Fourier expansion of the Boltzmann factors is
\be\label{Ell'}
E_m(x^l_{km+i}-x^{l'}_{jm+i})=\sum_{z_{l'j}^{lk}(i)\in\Zz^d}\hat{E}_m\left(z_{l'j}^{lk}(i)\right) \exp\left\{\i\frac{2\pi}{L}z_{l'j}^{lk}(i)\cdot (x^l_{km+i}-x^{l'}_{jm+i}) \right\}.
\ee
Here
\be\label{hatEmz}
\hat{E}_m(z)=\frac{1}{L^d}\int_\Lambda \exp\left\{-{\mathrm i}\frac{2\pi}{L}z\cdot x\right\}\exp\left\{-\frac{\beta}{m}u_L(x)\right\} \d x =\delta_{z,0} -\frac{\beta}{mL^d}\ \hat{u}(z/L)+\frac{1}{L^d}O(1/m^2)
\ee
with
\be
\hat{u}(z/L)=\int \exp\left\{-\i\frac{2\pi}{L}z\cdot x\right\} u(x) \d x.
\ee
In Eq.~(\ref{hatEmz}) after the Taylor-expansion of $\exp\{-(\beta/m)u_L(x)\}$ we substituted (\ref{u_L-def}). The summation over $\Zz^d$ combined with the integral over $\Lambda$ is equivalent to integrating over the whole space.
Notice that $\hat{E}_m(z)=\hat{E}_m(-z)=\hat{E}_m(z)^*$. This holds true
because $u_L(x)=u_L(-x)$. Similarly, $\hat{u}(z/L)$ and also the remainder are real even functions of $z$. Some short notations will be useful:
\be
\calZ_{l}=\left(\Zz^d\right)^{mn_l(n_l-1)/2}, \quad
\calZ_{<}=\left(\Zz^d\right)^{m\sum_{1\leq l'<l\leq p}n_{l'}n_l}, \quad
\calZ=\left(\Zz^d\right)^{mN(N-1)/2}.
\ee
The elements of these sets are considered as sets of vectors from $\Zz^d$,
and the notation
\be\label{E^Z}
\hat{E}_m^Z=\prod_{z\in Z}\hat{E}_m(z)
\ee
will be used for $Z\in\calZ_{l}$, $\calZ_{<}$, $\calZ$.
Now
\be
\prod_{i=1}^m\ \prod_{0\leq j<k\leq n_l-1} E_m(x^l_{km+i}-x^l_{jm+i}) =\sum_{Z\in\calZ_{l}}\hat{E}_m^Z
\prod_{i=1}^m \exp\left\{\i\frac{2\pi}{L}\sum_{k=1}^{n_l-1}\sum_{j=0}^{k-1} z^{lk}_{lj}(i)\cdot(x^l_{km+i}-x^l_{jm+i})\right\}
\ee
which, after substituting
\be
\sum_{k=1}^{n_l-1}\sum_{j=0}^{k-1} z^{lk}_{lj}(i)\cdot(x^l_{km+i}-x^l_{jm+i})=\sum_{k=0}^{n_l-1} x^l_{km+i}\cdot  \left(\sum_{j=0}^{k-1}z^{lk}_{lj}(i) - \sum_{j=k+1}^{n_l-1}z^{lj}_{lk}(i)\right)
\ee
results in
\be\label{Eml}
\prod_{i=1}^m\ \prod_{0\leq j<k\leq n_l-1} E_m(x^l_{km+i}-x^l_{jm+i}) =\sum_{Z\in\calZ_{l}}\hat{E}_m^Z\prod_{i=1}^m \prod_{k=0}^{n_l-1} \exp\left\{\i\frac{2\pi}{L} x^l_{km+i}\cdot  \left(\sum_{j=0}^{k-1}z^{lk}_{lj}(i) - \sum_{j=k+1}^{n_l-1}z^{lj}_{lk}(i)\right)\right\}.
\ee
Furthermore,
\be
\sum_{1\leq l'<l\leq p}\sum_{k=0}^{n_l-1}\sum_{j=0}^{n_{l'}-1}z^{lk}_{l'j}(i)\cdot \left(x^l_{km+i}-x^{l'}_{jm+i}\right) =\sum_{l=1}^p\sum_{k=0}^{n_l-1}x^l_{km+i}\cdot\left(\sum_{l'=1}^{l-1}\sum_{j=0}^{n_{l'}-1}z^{lk}_{l'j}(i) -\sum_{l'=l+1}^p\sum_{j=0}^{n_{l'}-1}z^{l'j}_{lk}(i)\right),
\ee
so
\bea\label{Eml'ljk}
\lefteqn{\prod_{i=1}^m\, \prod_{1\leq l'<l\leq p}\prod_{j=0}^{n_{l'}-1}\prod_{k=0}^{n_{l}-1} E_m(x^l_{km+i}-x^{l'}_{jm+i})}
\nonumber\\
&&=\sum_{Z\in\calZ_{<}}\hat{E}_m^Z\prod_{i=1}^m\prod_{l=1}^p\prod_{k=0}^{n_l-1} \exp\left\{\i\frac{2\pi}{L} x^l_{km+i}\cdot \left(\sum_{l'=1}^{l-1}\sum_{j=0}^{n_{l'}-1}z^{lk}_{l'j}(i) -\sum_{l'=l+1}^p\sum_{j=0}^{n_{l'}-1}z^{l'j}_{lk}(i)\right)\right\}.
\eea
From Eqs.~(\ref{Wxj-1xj}), (\ref{Eml}) and (\ref{Eml'ljk}) one can collect the multiplier of $x^l_{km+i}$.
If we write $G_m\left[\{n_l\}_1^p\right]$ as
\be
G_m\left[\{n_l\}_1^p\right]=\int_{\Lambda^{mN}} \prod_{l=1}^p\prod_{j=1}^{n_lm}\d x^l_j\  G\left[\left\{(x^l_j)_{j=1}^{n_lm},n_l\right\}_{l=1}^p\right]
\ee
then
\bea\label{Gxlj-nl}
\lefteqn{
G\left[\left\{(x^l_j)_{j=1}^{n_lm},n_l\right\}_{l=1}^p\right]=\sum_{Z\in\calZ}\hat{E}_m^Z}
\nonumber\\
&&\prod_{l=1}^p
\left\{\sum_{v^l_{10}\in\Zz^d}\ \sum_{v^l_{1},v^l_{2},\ldots,v^l_{n_lm-1}\in\Zz^d} \left[e^{-\frac{\pi\lambda_{\beta}^2}{mL^2}\sum_{j=1}^{n_lm}(v^l_{10}+\sum_{j'=j}^{n_lm}v^l_{j'})^2}
\prod_{k=0}^{n_l-1}\prod_{i=1}^m\frac{1}{L^d} e^{\i\frac{2\pi}{L}x^l_{km+i}\cdot(v^l_{km+i}- z^l_{km+i})}\right]\right\}
\nonumber\\
\eea
where
\be\label{zlkm+i}
z^l_{km+i}=-\sum_{j=0}^{k-1}z^{lk}_{lj}(i)+\sum_{j=k+1}^{n_l-1}z^{lj}_{lk}(i)
-\sum_{l'=1}^{l-1}\sum_{j=0}^{n_{l'}-1}z^{lk}_{l'j}(i) +\sum_{l'=l+1}^p\sum_{j=0}^{n_{l'}-1}z^{l'j}_{lk}(i).
\ee
The multiple sum (\ref{Gxlj-nl}) is absolutely convergent and can be integrated term by term. In each term there is factorization according to the spatial variables $x^l_{km+i}$, so the integration of the complex units can be made separately.
Integration with respect to $x^l_j$ in $\Lambda$ makes $v^l_j$ coincide with $z^l_j$ for all $j$
and turns Eq.~(\ref{sumvlj}) into
\be\label{sum-zlj}
\sum_{j=1}^{n_lm}z^l_j=0.
\ee
Thus,
\be\label{Gm1}
G_m\left[\{n_l\}_1^p\right]=\sum_{Z\in\calZ}\hat{E}_m^Z\prod_{l=1}^p \delta_{\sum_{j=1}^{n_lm}z^l_j,0} \sum_{z\in\Zz^d} \exp\left\{-\frac{\pi\lambda_{\beta}^2}{mL^2}\sum_{j=1}^{n_lm}(z+\sum_{j'=j}^{n_lm}z^l_{j'})^2\right\}.
\ee
Finally, defining
\be
Z^l_j=\sum_{j'=j}^{n_lm}z^l_{j'},\qquad \overline{Z^l_{^\cdot}}=\frac{1}{n_lm}\sum_{j=1}^{n_lm}Z^l_j,\qquad \overline{\left(Z^l_{^\cdot}\right)^2}=\frac{1}{n_lm}\sum_{j=1}^{n_lm}(Z^l_j)^2,
\ee
where we use the same notations for the averages as for their $m\to\infty$ limit in (\ref{exprewritten}), Eq.~(\ref{Gm1}) becomes
\bea\label{Gm2}
\lefteqn{
G_m\left[\{n_l\}_1^p\right] =  \sum_{Z\in\calZ}\hat{E}_m^Z \prod_{l=1}^p \delta_{Z^l_1,0} \sum_{z\in\Zz^d} \exp\left\{-\frac{\pi\lambda_{\beta}^2}{mL^2}\sum_{j=1}^{n_lm}(z+Z^l_j)^2\right\}
} \nonumber\\
&&=\sum_{Z\in\calZ}\hat{E}_m^Z \prod_{l=1}^p \delta_{Z^l_1,0} \exp\left\{-\frac{\pi n_l\lambda_{\beta}^2}{L^2} \left[\overline{\left(Z^l_{^\cdot}\right)^2}-\overline{Z^l_{^\cdot}}^2\right]\right\}\sum_{z\in\Zz^d} \exp\left\{-\frac{\pi n_l\lambda_\beta^2}{L^2}\left(z+\overline{Z^l_{^\cdot}}\right)^2\right\}.
\eea
A consequence of $Z^l_1=0$ is that $\overline{\left(Z^l_{^\cdot}\right)^2}-\overline{Z^l_{^\cdot}}^2=0$ if and only if $Z^l_j=0$ for $j=1,\ldots,n_lm$ which holds if and only if $z^l_j=0$ for $j=1,\ldots,n_lm$. (Note: $Z^l_j=-\sum_{j'=1}^{j-1}z^l_{j'}$ as well, cf. Eq.~(\ref{sum-zlj}).) This is the finite-$m$ equivalent of Remark~\ref{D_l^2=0} in Section 3.
Two identities in the individual $z$ vectors are noteworthy because they hold for every $i$:
\be\label{iden-fixed-l-i}
\sum_{k=0}^{n_l-1}z^l_{km+i}
=-\sum_{l'=1}^{l-1}\sum_{j=0}^{n_{l'}-1}\sum_{k=0}^{n_l-1}z^{lk}_{l'j}(i) +\sum_{l'=l+1}^p\sum_{j=0}^{n_{l'}-1}\sum_{k=0}^{n_l-1}z^{l'j}_{lk}(i),
\ee
i.e. the intra-cycle contribution vanishes from the sum on the left-hand side, and
\be
\sum_{l=1}^p \sum_{k=0}^{n_l-1}z^l_{km+i}=0.
\ee
Summing this latter with respect to $i$ yields $\sum_{l=1}^p Z^l_1\equiv 0$.

At this point it is useful to change the notation and number the particles continuously from 1 to $N$. Particle $lk$, the $k$th particle (starting from 0) of the $l$th cycle will carry the number $N_{l-1}+k+1$ when counted continuously, so the identities new$\equiv$old are
\be\label{z-continuous-number}
z^{N_{l-1}+k+1}_{N_{l'-1}+j+1}(i)\equiv z^{lk}_{l'j}(i)\ \mbox{or}\ z^q_{j'}(i)\equiv z^{l,q-N_{l-1}-1}_{l',j'-N_{l'-1}-1}(i),
\quad z_{N_{l-1}+k+1}(i)\equiv z^l_{km+i}\ \mbox{or}\ z_q(i)\equiv z^l_{(q-N_{l-1}-1)m+i}
\ee
for $q\in C_l$ and $j'\in C_{l'}$.
The new notation is better suited to $\hat{E}_m^Z$ which is independent of the cycle structure. Also, the expression (\ref{zlkm+i}) is replaced with the cycle-independent
\be\label{zji}
z_q(i)=-\sum_{j=1}^{q-1} z^q_{j}(i) + \sum_{k=q+1}^N z^{k}_q(i), \qquad q=1,\ldots,N, \quad i=1,\ldots,m,
\ee
and the expanded form of $Z^l_j$, $\overline{Z^l_{^\cdot}}$ and $\overline{\left(Z^l_{^\cdot}\right)^2}$ in terms of the individual $z$ vectors becomes more transparent. For $q\in C_l$ one finds
\bea\label{Zlqi}
\lefteqn{Z^l_{(q-N_{l-1}-1)m+i}=\sum_{i'=i}^m\sum_{k=q}^{N_l}z_{k}(i')+\sum_{i'=1}^{i-1}\sum_{k=q+1}^{N_l}z_{k}(i')}
\\
&&=\sum_{i'=i}^m\left[-\sum_{j=1}^{q-1}\sum_{k=q}^{N_l}z^{k}_{j}(i') +\sum_{j=q}^{N_l}\sum_{k=N_l+1}^Nz^{k}_{j}(i')\right]
+\sum_{i'=1}^{i-1}\left[-\sum_{j=1}^{q}\sum_{k=q+1}^{N_l}z^{k}_{j}(i') +\sum_{j=q+1}^{N_l}\sum_{k=N_l+1}^N z^{k}_{j}(i')\right].
\nonumber
\eea
In particular,
\be\label{Zl1new}
Z^l_1=\sum_{i=1}^m\left[-\sum_{j=1}^{N_{l-1}}\sum_{k\in C_l}z^{k}_{j}(i) +\sum_{j\in C_l}\sum_{k=N_l+1}^N z^{k}_{j}(i)\right].
\ee
Moreover,
\bea\label{Zldot}
\lefteqn{\overline{Z^l_{^\cdot}}=\frac{1}{n_l}\sum_{q\in C_l}\sum_{i=1}^m \left(q-N_{l-1}-1+\frac{i}{m}\right) z_q(i)}\nonumber\\
&&=\frac{1}{n_l}\sum_{q\in C_l}\left[-\sum_{j=1}^{q-1}\sum_{i=1}^m\left(q-N_{l-1}-1+\frac{i}{m}\right)\,z^{q}_{j}(i) +\sum_{k=q+1}^{N}\sum_{i=1}^m\left(q-N_{l-1}-1+\frac{i}{m}\right)\,z^{k}_{q}(i)\right]
\nonumber\\
\eea
and
\bea\label{Zldotsquare}
\overline{\left(Z^l_{^\cdot}\right)^2}&=&\frac{1}{n_l}\sum_{k,k'=0}^{n_l-1}\ \sum_{i,i'=1}^m \min\left\{k+\frac{i}{m},k'+\frac{i'}{m}\right\} z^l_{km+i}\cdot z^l_{k'm+i'}
\nonumber\\
&=&\frac{1}{n_l}\sum_{q,q'\in C_l}\ \sum_{i,i'=1}^m\left(\min\left\{q+\frac{i}{m},q'+\frac{i'}{m}\right\}-N_{l-1}-1\right) z_{q}(i)\cdot z_{q'}(i').
\eea
The last formula becomes complete after (\ref{zji}) is substituted in it.

\newpage
\subsection{The case $x\neq 0$}

For the zeroth cycle we have to insert
\[
L^d \int W_{0x}^{n\beta}(\d\omega_0)e^{-\beta U(\omega_0)}
\prod_{l=1}^ p e^{-\beta  U(\omega_0,\omega_l)}.
\]
The formulas or parts of formulas not involving cycle 0 are unchanged. There are $nm-1$ new integration variables $x^0_j=\omega_0(j\beta/m)$, $j=1,\dots,nm-1$ completed with the fixed $x^0_0=0$ and $x^0_{nm}=x$. The new Fourier variables associated with differences of position vectors are: $v_{j,j-1}$ for $x^0_j-x^0_{j-1}$ ($j=1,\dots,nm$), $z^{0k}_{0j}(i)$ for $x^0_{km+i}-x^0_{jm+i}$ ($0\leq j<k\leq n-1$, $i=1,\dots,m$) and $z^{lk}_{0j}(i)$ for $x^l_{km+i}-x^0_{jm+i}$ ($l=1,\dots,p$, $k=0,\dots,n_l-1$, $j=0,\dots,n-1$, $i=1,\dots,m$). We have
\be
\sum_{j=1}^{nm}v^0_{j,j-1}\cdot(x^0_j-x^0_{j-1})=\sum_{j=1}^{nm-1}(v^0_{j,j-1}-v^0_{j+1,j})\cdot x^0_j+v^0_{nm,nm-1}\cdot x
=\sum_{j=1}^{nm-1}v^0_j\cdot x^0_j+v^0_{nm}\cdot x,
\ee
the second equality defining $v^0_j$. In contrast to (\ref{sumvlj}), their sum is nonzero,
\be
\sum_{j=1}^{nm} v^0_j=v^0_{10}.
\ee
Defining $v^0_{nm+1,nm}=0$,
\be
v^0_{j,j-1}=\sum_{j'=j}^{nm}v^0_{j'},\quad j=1,\dots,nm,
\ee
so
\be
\sum_{j=1}^{nm}\left(v^0_{j,j-1}\right)^2=\sum_{j=1}^{nm}\left(\sum_{j'=j}^{nm} v^0_{j'}\right)^2
\ee
and
\be
L^d\prod_{j=1}^{nm}\int W^{\beta/m}_{x^0_{j-1}x^0_j}(\d\omega_0)=\sum_{v^0_1,\dots,v^0_{nm}\in\Zz^d}e^{-\frac{\pi\lambda_\beta^2}{mL^2}
\sum_{j=1}^{nm}\left(\sum_{j'=j}^{nm} v^0_{j'}\right)^2}
\left(\prod_{j=1}^{nm-1}\frac{1}{L^d}e^{\i\frac{2\pi}{L}v^0_j\cdot x^0_j}\right)
\ e^{\i\frac{2\pi}{L}v^0_{nm}\cdot x}.
\ee
Equation~(\ref{zlkm+i}) extends to $l=0$ and thus defines $z^0_j$. Moreover, the identity (\ref{iden-fixed-l-i}) also holds true. It is because of this latter that the formula (\ref{Z^l_1}) is valid also for $l=0$ with the consequence that $\sum_{l=0}^p Z^l_1\equiv 0$ which, together with $Z^1_1=\cdots=Z^p_1=0$ implies $Z^0_1=0$. Including the $z$-vectors associated with the Boltzmann factors, in the argument of the complex units $x^0_j$ is scalar-multiplied with $v^0_j-z^0_j$. Integrating with respect to $x^0_j$ over $\Lambda$ for $j=1,\dots,nm-1$ makes $v^0_j$ coincide with $z^0_j$, leaving behind
\bea
L^d\int_{\Lambda^{nm-1}}\prod_{j=1}^{nm-1}\d x^0_j \prod_{j=1}^{nm}\int W^{\beta/m}_{x^0_{j-1}x^0_j}(\d\omega_0)
\prod_{i=1}^m\prod_{0\leq j<k\leq n-1} E_m\left(x^0_{km+i}-x^0_{jm+i}\right) \prod_{l=1}^p\prod_{j=1}^{n-1}\prod_{k=1}^{n_l-1}E_m\left(x^l_{km+i}-x^0_{jm+i}\right)
\nonumber\\
=\sum_Z \hat{E}_m^Z\sum_{v^0_{nm}\in\Zz^d}
e^{-\frac{\pi\lambda_\beta^2}{mL^2}
\sum_{j=1}^{nm}\left(v^0_{nm}+\sum_{j'=j}^{nm-1} z^0_{j'}\right)^2}\ e^{\i\frac{2\pi}{L}x\cdot (v^0_{nm}-z^0_{nm})}
=
\sum_Z \hat{E}_m^Z\sum_{z\in\Zz^d}
e^{-\frac{\pi\lambda_\beta^2}{mL^2}
\sum_{j=1}^{nm}\left(z+\sum_{j'=j}^{nm} z^0_{j'}\right)^2}\ e^{\i\frac{2\pi}{L}x\cdot z}
\nonumber\\
=\sum_Z \hat{E}_m^Z\sum_{z\in\Zz^d}
e^{-\frac{\pi\lambda_\beta^2}{mL^2}
\sum_{j=1}^{nm}\left(z+Z^0_j\right)^2}\ e^{\i\frac{2\pi}{L}x\cdot z}.\hspace{8mm}
\eea
Above $Z=\{z^{0k}_{0j}(i)\}_{j,k,i}\cup\{z^{lk}_{0j}(i)\}_{l,j,k,i}$, each vector running over $\Zz^d$. Because the expression is real, we can take the real part and find the cosine appearing in the statement of the lemma.

\newpage
\subsection{The limit of continuous time}
There is no difference between $x=0$ or $x\neq 0$. We describe only the first case.
The result of the limit when $m$ tends to infinity can be seen on the simplest example, that of $N=2$.
Because in this case there is a single pair, the notation can be simplified by writing $z(i)$ instead of $z^2_1(i)$.
Now $\calZ=(\Zz^d)^m$ whose elements are $Z=\{z(1),\ldots,z(m)\}$. There are two partitions of 2: $p=1$, $n_1=2$ and $p=2$, $n_1=n_2=1$.

\vspace{10pt}
\noindent
{\bf One two-particle trajectory}

\vspace{10pt}
\noindent
When $p=1$ then $l=1$, $N_{l-1}=0$, $N=N_1=n_1=2$, so from (\ref{zji}) and (\ref{Zl1new})
\be
z_1(i)=z(i),\quad z_2(i)=-z(i), \quad Z^1_1\equiv 0.
\ee
Moreover,
\be
\overline{Z^1_\cdot}=\frac{1}{2}\sum_{i=1}^m\left[ -\left(1+\frac{i}{m}\right)z(i) +\frac{i}{m}z(i)\right] = - \frac{1}{2}\sum_{i=1}^m z(i),
\ee
\bea
\overline{\left(Z^1_\cdot\right)^2}=\frac{1}{2}\sum_{j,j'=1}^2\sum_{i,i'=1}^m \left(\min\left\{j+\frac{i}{m},j'+\frac{i'}{m}\right\}-1\right) z_j(i)\cdot z_{j'}(i') \nonumber\\
=\frac{1}{2}\left(\sum_{i=1}^m z(i)\right)^2 - \frac{1}{2}\sum_{i,i'=1}^m\frac{|i-i'|}{m} z(i)\cdot z(i')\nonumber\\
= 2 \overline{Z^1_\cdot}^2  - \frac{1}{2}\sum_{i,i'=1}^m\frac{|i-i'|}{m} z(i)\cdot z(i').
\eea
Thus,
\be
\overline{\left(Z^1_\cdot\right)^2}- \overline{Z^1_\cdot}^2= \sum_{i,j=1}^m \left(\frac{1}{4}-\frac{|i-j|}{2m}\right)z(i)\cdot z(j).
\ee

\vspace{10pt}
\noindent
{\bf Two one-particle trajectories}

\vspace{10pt}
\noindent
In this case
\[
Z^1_1=-Z^2_1=\sum_{i=1}^m z(i)
\]
\[
\overline{Z^1_\cdot}=-\overline{Z^2_\cdot}=\sum_{i=1}^m\frac{i}{m} z(i)
\]
\[
\overline{(Z^1_\cdot)^2}=\overline{(Z^2_\cdot)^2}=\sum_{i,j=1}^m  \min\left\{\frac{i}{m},\frac{j}{m}\right\}z(i)\cdot z(j).
\]

If $z(i)\neq 0$ precisely for $\alpha$ values of $i$, say, for $i_1<i_2<\cdots<i_\alpha$ then everywhere one can replace $\sum_{i=1}^m$ with $\sum_{r=1}^\alpha$ and $i$ in the summand with $i_r$. Because $z(i_r)$ is a summation variable over $\Zz^d\setminus\{0\}$, for its labelling the value of $i_r$ is unimportant and the notation $z_r$ can be used for it. All this permits to rewrite
\be
G_m\left[\{n_l\}_1^p\right]=\sum_{Z\in(\Zz^d)^m}  \left(\prod_{l=1}^p \delta_{Z^l_1,0}\right)\hat{E}_m^Z\ f_{\left[\{n_l\}_1^p\right]}\left(\{i/m\}_{i=1}^m,Z\right)
\ee
where
\be
f_{[2]}\left(\{i/m\}_{i=1}^m,Z\right)=\exp\left\{-\frac{\pi\lambda_\beta^2}{L^2}\sum_{i,j=1}^m\left(\frac{1}{2}-\frac{|i-j|}{m}\right)z(i)\cdot z(j)\right\}
\sum_{z\in\Zz^d}\exp\left\{-\frac{2\pi\lambda_\beta^2}{L^2}\left(z-\frac{1}{2}\sum_{i=1}^m z(i)\right)^2\right\}
\ee
and
\bea
f_{[1,1]}\left(\{i/m\}_{i=1}^m,Z\right)
=
\exp\left\{-\frac{2\pi\lambda_\beta^2}{L^2} \sum_{i,j=1}^m\left[\min\left\{\frac{i}{m},\frac{j}{m}\right\}-\frac{ij}{m^2}\right]z(i)\cdot z(j)\right\}\nonumber\\
\left[\sum_{z\in\Zz^d}\exp\left\{-\frac{\pi\lambda_\beta^2}{L^2}\left(z+\sum_{i=1}^m \frac{i}{m}z(i)\right)^2\right\}\right]^2
\eea
as
\bea
G_m\left[\{n_l\}_1^p\right]
&=&
\sum_{z(1),\dots,z(m)\in\Zz^d}\left(\prod_{l=1}^p \delta_{Z^l_1,0}\right) \hat{E}_m(z(1))\cdots\hat{E}_m(z(m))
f_{\left[\{n_l\}_1^p\right]}\left(\{i/m\}_{i=1}^m,Z\right)
\nonumber\\
&=&
\sum_{\alpha=0}^m \hat{E}_m(0)^{m-\alpha} \sum_{z_1,\dots,z_\alpha\in\Zz^d\setminus\{0\}}\left(\prod_{l=1}^p \delta_{Z^l_1,0}\right) \hat{E}_m(z_1) \cdots \hat{E}_m(z_\alpha)
\nonumber\\
&\times&
\sum_{1\leq i_1<\cdots <i_\alpha\leq m}\,f_{\left[\{n_l\}_1^p\right]}\left(\{i_r/m\}_{r=1}^\alpha,\{z_r\}_{r=1}^\alpha\right).\nonumber\\
\eea
Now
\bea
\sum_{1\leq i_1<\cdots <i_\alpha\leq m}\, f_{\left[\{n_l\}_1^p\right]}\left(\{i_r/m\}_{r=1}^\alpha,\{z_r\}_{r=1}^\alpha\right) =\sum_{i_\alpha=\alpha}^m\ \sum_{i_{\alpha-1}=\alpha-1}^{i_\alpha-1}\cdots \sum_{i_1=1}^{i_2-1}f_{\left[\{n_l\}_1^p\right]}\left(\{i_r/m\}_{r=1}^\alpha,\{z_r\}_{r=1}^\alpha\right)\nonumber\\
=m^\alpha\, \frac{1}{m}\sum_{i_\alpha=\alpha}^m\ \frac{1}{m}\sum_{i_{\alpha-1}=\alpha-1}^{i_\alpha-1}\cdots \frac{1}{m}\sum_{i_1=1}^{i_2-1}f_{\left[\{n_l\}_1^p\right]}\left(\{i_r/m\}_{r=1}^\alpha,\{z_r\}_{r=1}^\alpha\right)\nonumber\\
=m^\alpha \frac{1}{m}\sum_{t_\alpha=\alpha/m}^1\ \frac{1}{m}\sum_{t_{\alpha-1}=(\alpha-1)/m}^{t_\alpha-1/m}\cdots \frac{1}{m}\sum_{t_1=1/m}^{t_2-1/m}f_{\left[\{n_l\}_1^p\right]}\left(\{t_r\}_{r=1}^\alpha,\{z_r\}_{r=1}^\alpha\right)
\nonumber\\
\eea
where each $t_r=i_r/m$ varies by steps $1/m$. From Eq.~(\ref{hatEmz}) we substitute $\hat{E}_m(z)$, dropping the $O(1/m^2)$ term which disappears in the $m\to\infty$ limit. This gives
\bea
\lefteqn{
G_m\left[\{n_l\}_1^p\right]=\sum_{\alpha=0}^m \left(1-\frac{\beta \hat{u}(0)}{mL^d}\right)^{m-\alpha}\left(\frac{-\beta}{L^d}\right)^\alpha  \sum_{z_1,\dots,z_\alpha\in\Zz^d\setminus\{0\}}\left(\prod_{l=1}^p \delta_{Z^l_1,0}\right)\hat{u}(z_1/L) \cdots \hat{u}(z_\alpha/L)
}\nonumber\\
&&\times\frac{1}{m}\sum_{t_\alpha=\alpha/m}^1\ \frac{1}{m}\sum_{t_{\alpha-1}=(\alpha-1)/m}^{t_\alpha-1/m}\cdots \frac{1}{m}\sum_{t_1=1/m}^{t_2-1/m}f_{\left[\{n_l\}_1^p\right]}\left(\{t_r\}_{r=1}^\alpha,\{z_r\}_{r=1}^\alpha\right)
\nonumber\\
\eea
The summand is bounded above by
\[
\left(\sum_{z\in\Zz^d} \exp\left\{-\frac{\pi\lambda_\beta^2}{L^2}z^2\right\}\right)^2
\frac{1}{\alpha !}\left(\frac{\beta \sum_{z\neq 0}|\hat{u}(z/L)|}{L^d}\right)^\alpha
\]
whose sum over $\alpha$ extended to infinity is convergent. So the dominated convergence theorem applies and yields
\bea
G\left[\{n_l\}_1^p\right]
=e^{-\beta\hat{u}(0)/L^d}
\sum_{\alpha=0}^\infty \left(\frac{-\beta}{L^d}\right)^\alpha  \sum_{z_1,\dots,z_\alpha\in\Zz^d\setminus\{0\}}\left(\prod_{l=1}^p \delta_{Z^l_1,0}\right)\hat{u}(z_1/L) \cdots \hat{u}(z_\alpha/L)
\nonumber\\
\times\int_0^1\d t_\alpha\int_0^{t_\alpha}\d t_{\alpha-1}\cdots\int_0^{t_2}\d t_1 f_{\left[\{n_l\}_1^p\right]}\left(\{t_r\}_{r=1}^\alpha,\{z_r\}_{r=1}^\alpha\right).
\nonumber\\
\eea
This ends the proof of the lemma for $N=2$. Note that for $p=2$ the constraint $\sum_{r=1}^\alpha z_r=0$ acts both on $\alpha$ and on the set of variables $z_1,\dots,z_\alpha$: for $\alpha=1$ the sum over $\Zz^d\setminus \{0\}$ is empty and for $\alpha>1$ only $\alpha-1$ variables can be chosen freely from $\Zz^d\setminus\{0\}$, meaning that these terms are of order $L^{-d}$.

For a general $N$, consider first the limit of the exponents in Eq.~(\ref{Gm2}). In the first line
\be
\frac{1}{m}\sum_{j=1}^{n_lm}(z+Z^l_j)^2=\sum_{k=0}^{n_l-1}\frac{1}{m}\sum_{i=1}^m (z+Z^l_{km+i})^2 =\sum_{q\in C_l}\frac{1}{m}\sum_{i=1}^m (z+Z^l_{(q-N_{l-1}-1)m+i})^2.
\ee
Substituting for $Z^l_{(q-N_{l-1}-1)m+i}$ the expression (\ref{Zlqi}) and keeping only the nonzero $z$ vectors,
\be
\sum_{i'=i}^m z^{k}_{j}(i')=\sum_{r=1}^{\alpha^{k}_{j}}{\bf 1}_{\left\{i^{k}_{j,r}/m\geq i/m\right\}}z^{k}_{j}\left(i^{k}_{j,r}\right), \qquad \sum_{i'=1}^{i-1} z^{k}_{j}(i')=\sum_{r=1}^{\alpha^{k}_{j}}{\bf 1}_{\left\{i^{k}_{j,r}/m< i/m\right\}}z^{k}_{j}\left(i^{k}_{j,r}\right).
\ee
When $m$ tends to infinity
\[
i/m\to t,\quad i^{k}_{j,r}/m\to t^{k}_{j,r},\quad  z^{k}_{j}\left(i^{k}_{j,r}\right)\to z^{k}_{j,r},\quad Z^l_{(q-N_{l-1}-1)m+i}\to Z_q(t).
\]
These together yield $\int_0^1[z+Z_q(t)]^2\d t$
as shown in Eqs.~(\ref{G[nl](x)}) and (\ref{Zqt}). For the $m\to\infty$ limit of (\ref{Zldot}) first we rewrite it as
\be
\overline{Z^l_{^\cdot}}
=\frac{1}{n_l}\sum_{q\in C_l}\left[-\sum_{j=1}^{q-1}\sum_{r=1}^{\alpha^q_j}\left(q-N_{l-1}-1+\frac{i^q_{j,r}}{m}\right)\,z^{q}_{j}(i^q_{j,r}) +\sum_{k=q+1}^{N}\sum_{r=1}^{\alpha^k_q}\left(q-N_{l-1}-1+\frac{i^k_{q,r}}{m}\right)\,z^{k}_{q}(i^k_{q,r})\right].
\ee
Interchanging the order of summations with respect to $q$ and $j$ and to $q$ and $k$ and letting $m$ go to infinity one obtains the second form of $\overline{Z^l_{^\cdot}}$, see Eq.~(\ref{intZqbis}). In particular,
\be
\int_0^1 Z_q(t)\d t=
-\sum_{j=1}^{q-1}\sum_{r=1}^{\alpha^q_j}\left(q-N_{l-1}-1+t^q_{j,r}\right)\,z^{q}_{j,r} +\sum_{k=q+1}^{N}\sum_{r=1}^{\alpha^k_q}\left(q-N_{l-1}-1+t^k_{q,r}\right)\,z^{k}_{q,r}.
\ee
The $m\to\infty$ limit of (\ref{Zldotsquare}) can be computed with the help of (\ref{zji}). The lengthy formulas can be found in [Su12]. We will need only $\overline{\left(X^0_{^\cdot}\right)^2}=\overline{\left(Z^0_{^\cdot}\right)^2}/L^2$ and present it in the proof of Theorem~\ref{first-step}.

The last point to check is that the $m\to\infty$ limit could indeed be taken under the summation signs. The cycle-dependent part of the summand of $G_m\left[\{n_l\}_1^p\right]$,
\be
f_{\left[\{n_l\}_1^p\right]}\left(\{t^k_{j,,r}\}_{r=1}^{\alpha^k_j},\{z^k_{j,r}\}_{r=1}^{\alpha^k_j}\right)= \prod_{l=1}^p \delta_{Z^l_1,0}\exp\left\{-\frac{\pi n_l\lambda_\beta^2}{L^2}\left[\overline{\left(Z^l_{^\cdot}\right)^2}-\overline{Z^l_{^\cdot}}^2\right]\right\}\sum_{z\in\Zz^d} \exp\left\{-\frac{\pi n_l\lambda_\beta^2}{L^2}\left(z+\overline{Z^l_{^\cdot}}\right)^2\right\}
\ee
is bounded as
\be\label{fbound}
|f_{\left[\{n_l\}_1^p\right]}|\leq \prod_{l=1}^p \sum_{z\in\Zz^d} \exp\left\{-\frac{\pi n_l\lambda_\beta^2}{L^2}z^2\right\}.
\ee
Therefore, in $G_m\left[\{n_l\}_1^p\right]$ for each pair $j<k$ we can apply the dominated convergence theorem to the sum over $\alpha^k_j$.

\subsection{Merger graphs: Analysis of the constraint}

We start with a class of graphs slightly different from the one presented under Remark \ref{constraint} of Section~\ref{remarks-lemma-G} but interesting in its own right. A graph of $V$ vertices and $E$ edges in this class corresponds to a system of $V$ homogeneous linear equations for $E$ variables, each appearing with coefficient 1, that has a solution in which all the variables take a nonzero integer value.

\begin{definition}\label{merge-gen1}
1.  A (merger) generator is a circle of even length, two odd circles with a common vertex, or two odd circles joining through a vertex the opposite endpoints of a linear graph. 2. The generators are mergers;  merging two mergers through one or more vertices and/or along one or more edges provides a merger. A graph composed of two disconnected mergers is a merger. 3. Given a merger, a set of generators whose merging provides the graph is called a covering. A covering is minimal if each of its elements contributes to the merger with at least one edge not covered by the other generators. A max-min covering is a minimal covering that contains the largest number of generators.
\end{definition}

As an example, the Petersen graph is a merger obtained by merging five "washtub" hexagons. The external edges are covered by three, the middle ones by two, the internal edges by one of the hexagons. Bipartite graphs, suitable subgraphs of the triangular lattice, complete graphs of more than three vertices are also mergers. The definition extends to multigraphs whose construction then involves also two-circles. In general the max-min covering is not unique, but the number of its constituting generators is uniquely determined because of the maximal property.

\begin{proposition}\label{merger1}
A graph is a merger if and only if to every edge one can assign a nonzero number in such a way that at every vertex the sum of the numbers assigned to the incident edges is zero. The numbers as variables form a manifold whose dimension is greater than or equal to the number of generators in the max-min coverings.
\end{proposition}

\noindent{\em Proof.} To mark the edges of a generator one can use a single and only a single variable denoted by $x_i$ for the $i$th generator of a merger. To the edges of a circle of length $2n$ one assigns $x_i$ and $-x_i$ in alternation. The same can be done with two odd circles sharing a vertex. In the case of two odd circles linked by a linear graph one assigns $x_i$ to those two edges of one of the circles that join the vertex of degree 3, and $-x_i$ and $x_i$ in alternation to the remaining edges of the same circle.
One then continues by alternating $-2x_i$ and $2x_i$ on the edges of the linear graph until reaching the second circle whose edges can again be marked by $x_i$ and $-x_i$ in a proper alternation. When merging, the generators carry their numbers. The number on a multiply covered edge is the sum of the numbers of the covering edges while one keeps the original number for the singly covered edges. Changing some $x_{i}$ in case of an accidental cancellation the sums on the edges are nonzero and the constraint remains satisfied.
The number of free variables cannot be smaller than the number of elements in max-min coverings, but it can be larger.
By construction, if the free variables are chosen to be integers, the others also take integers values.

In the opposite direction the proof goes by noting first that a graph does not contain any merger generator as a subgraph if and only if each of its maximal connected components is without circles or contains a single circle of odd length. The edges of such graphs cannot be marked in the required manner because either they have a vertex of degree 1 or they are unions of disjoint odd circles. Let $\cal G$ be any graph with properly marked edges. Thus, it has subgraphs which are merger generators. We can proceed by successive demerging. Let us choose a generator $g$ in $\cal G$ and select one of its edges denoted by $e$. Let $x$ be the number assigned to $e$. Prepare an image $g'$ of $g$ outside $\cal G$ and assign $-x$ to the image $e'$ of $e$. This uniquely determines the numbering of the other edges of $g'$ in such a way that the constraint is satisfied. Add the number on every edge of $g'$ to the number on its pre-image in $g$. As a result, the new number on $e$ is zero. Dropping all the edges from $\cal G$ whose new number is zero
and dropping also the vertices that become isolated we obtain a new graph having at least one edge less than $\cal G$ while the total numbering still satisfies the constraint. In a finite number of steps we can empty $\cal G$ which is, therefore, a merger. $\Box$

Now we define the class of graphs that we need for this paper. The vertices of a graph in this class correspond to permutation cycles, its edges represent the nonzero $\alpha^k_j$ that allow the solution of all the equations $Z^l_1=0$, cf. (\ref{Z^l_1}), with each variable taking a nonzero integer vector value.
We shall refer to these graphs as mergers generated by circles. The forthcoming discussion is restricted to them.

\begin{definition}\label{merge-gen2}
1. A (merger) generator is a circle of any (even or odd) length $n\geq 2$ with $n$ different positive integers assigned to the vertices in an arbitrary order. 2. The generators are mergers. Merging two mergers through all their vertices that carry the same number and optionally along some of the edges whose endpoints are common in the two mergers provides a merger. A graph composed of two disconnected mergers with disjoint vertex-numbering is a merger. 3. Given a merger, a set of generators whose merging provides the graph is called a covering. A covering is minimal if each of its element contributes to the merger with at least one edge not covered by the other generators. A max-min covering is a minimal covering that contains the largest number of generators.
\end{definition}

\begin{proposition}\label{merger2}
A graph whose vertices $\{1,2,\dots\}$ carry different positive integers $l_1,l_2,\dots$ is a merger if and only if to every edge one can assign a nonzero vector in such a way that at any vertex $i$ the sum of the vectors on the incident edges $(i,j)$ taken with minus sign if $l_j<l_i$ and with plus sign if $l_j>l_i$, is zero.
If $N_I$ and $M$ denote the number of linearly independent vectors and the number of generators in the max-min coverings, respectively, then $N_I\geq M$.
\end{proposition}

\noindent
In Proposition \ref{K=V-1} we shall give the precise value of $N_I$.

\vspace{5pt}
\noindent{\em Proof.} First let us see how to assign a vector to the edges of a generator. Let $l_i$ be the number carried by the $i$th vertex of a $n$-circle, where $i=1,\dots,n$ label the clockwise consecutive vertices. Let $x_{12}, x_{23},\dots,x_{n-1,n}, x_{n1}$ denote the $n$ edge variables that must assume a suitable value. The equation to be solved at vertex $i$  ($n+1\equiv 1$) is one of
\[
(1)\ \ x_{i-1,i}+x_{i,i+1}=0\quad \mbox{if   $l_{i-1},l_{i+1}>l_i$},\quad (2)\ \  -x_{i-1,i}+x_{i,i+1}=0\quad \mbox{if   $l_{i-1}<l_i<l_{i+1}$},
\]
\[
(3)\ \ x_{i-1,i}-x_{i,i+1}=0\quad \mbox{if   $l_{i-1}>l_i>l_{i+1}$},\quad (4)\ \  -x_{i-1,i}-x_{i,i+1}=0\quad \mbox{if   $l_{i-1},l_{i+1}<l_i$}.
\]
It is seen that whatever be the choice of, say, $x_{12}$, the other variables must take the same value with plus or minus sign. So the solution, if any, is a one-dimensional manifold.
To be definite, let $l_1$ be the smallest number. Take an arbitrary nonzero vector $v$ and set $x_{12}=v$. The equation at vertex 1 is (1), it is solved with $x_{n1}=-v$. We must prove that going around the circle there is no "frustration", all the equations can be solved. Call $i$ a source if $l_i<l_{i-1},l_{i+1}$ and a sink if $l_i>l_{i-1},l_{i+1}$. It is helpful to imagine an arrow on every edge, pointing towards the larger-numbered vertex. The problem is soluble because the number of sinks equals the number of sources and there is at least one source. Passing a source $-v$ changes to $v$ while solving equation (1), stays $v$ until the next sink and solves equations of the type (2), passing the sink it changes to $-v$ while solving equation (4), stays $-v$ and solves equations of the type (3), and so on.
By merging the circles of a covering as written down in Proposition~\ref{merger1} provides the solution for the merger graph.

The proof in the opposite direction goes again by successive demerging as described above provided we can show that no other vertex-numbered graph than those defined as mergers can be edge-marked in the required manner. Suppose there is such a finite graph. It must contain at least one circle, since linear graphs, tree graphs obviously cannot satisfy the condition at vertices of degree 1. Demerging successively all the circles, what remains is nonempty and cannot be marked -- a contradiction.  $\Box$

As a matter of fact, the numbering can be dropped from the definition because the vertices of any merger of circular graphs can be labelled {\em a posteriori} and in an arbitrary order with different $l_1,l_2,\dots$, still a proper assignment of nonzero vectors to the edges is possible.

\vspace{10pt}\noindent
{\em Example 1.}  A multigraph of two vertices and $n>1$ edges is a merger of $n-1$ two-circles.  Let $x_1,\dots,x_n$ be the edge variables. The equations at the two vertices are $\pm(x_1+\cdots+x_n)=0$. Their general solution with nonzero vectors is $x_1=v_1$, $x_2=-v_1+v_2$,\dots, $x_{n-1}=-v_{n-2}+v_{n-1}$, $x_n=-v_{n-1}$, a $(n-1)$-dimensional manifold. On the other hand, for $n$ even the collection of every second 2-circles constitutes a max-min covering. For $n$ odd we need one more circle. So $N_I=n-1$ and $M=n/2$ or $(n+1)/2$.

\noindent
{\em Example 2.} Consider the complete 4-graph of vertices 1, 2, 3, 4 with $l_i=i$. A max-min covering is for example the three circles $(123)$, $(234)$ and $(134)$. If prior to merging
\begin{eqnarray*}
z^2_1=x,\quad z^3_1=-x,\quad z^3_2=x,\quad\mbox{for $(123)$}\\
z^3_2=y,\quad z^4_2=-y,\quad z^4_3=y\quad\mbox{for $(234)$}\\
z^3_1=z,\quad z^4_1=-z,\quad z^4_3=z\quad\mbox{for $(134)$}\\
\end{eqnarray*}
then after merging the edges of the tetrahedron will carry
\begin{equation*}
z^2_1=x,\quad z^3_1=-x+z,\quad z^3_2=x+y,\quad
z^4_1=-z,\quad z^4_2=-y,\quad z^4_3=y+z\\
\end{equation*}
which solve the four equations
\[
z^2_1+z^3_1+z^4_1=0,\quad -z^2_1+z^3_2+z^4_2=0,\quad -z^3_1-z^3_2+z^4_3=0,\quad -z^4_1-z^4_2-z^4_3=0.
\]
A covering which is minimal but not max-min is the two 4-circles $(1234)$, $(1243)$. Using them we would get only two independent variables. A covering which is not minimal is obtained by merging e.g. $(1234)$ to the above three triangles with a fourth variable $v$. This only changes $x$ to $x'=x+v$ and $z$ to $z'=z+v$ without increasing the number of free variables. Now $N_I=M$.

\noindent
{\em Example 3.} Consider a part of the  triangular lattice composed of four small triangles, three with top up and one in the middle with top down. Let the V=6 vertices be numbered from 1 to 6 along the big outer triangle. This is a merger graph in both senses, generated by $M= 3$ elements in a max-min covering: according to Definition~\ref{merge-gen1}, by the rhombi $(1246)$, $(2346)$ and $(2456)$, and according to Definition~\ref{merge-gen2}, by the triangles $(126)$, $(234)$ and $(456)$.
We focus on mergers of the second kind.
Let the E=9 edge variables form the vector $Z$ with  $Z=(z^2_1\,  z^3_2\,  z^4_3\,  z^5_4\,  z^6_5\,  z^6_1\,  z^4_2\,  z^6_2\,  z^6_4)^T$ and let $A$ be the $V\times E$ matrix of the coefficients in the system of equations $AZ=0$ ($A$ is the incidence matrix of the directed graph with arrows towards the larger-numbered vertices).
One can check that ${\rm rank} A=5$, so 5 variables take a unique value once for the $N_I=4$ free variables a nonzero value is chosen. Thus, $N_I>M$. This can also be seen by choosing first a value for the variables associated with the three generators, as it was done in the proof of the propositions above. For the triangles, let $z^2_1=z^6_2=-z^6_1=x$, $z^3_2=z^4_3=-z^4_2=y$, $z^5_4=z^6_5=-z^6_4=z$. We can add any $v$ to $z^4_2$ so that the new value is $-y+v$; if we change simultaneously $z^6_2$ into $x-v$ and $z^6_4$ into $-z+v$, the four vectors $x$, $-y+v$, $x-v$ and $-z+v$ are linearly independent and
the three equations at vertices 2, 4 and 6, namely
\[
-z^2_1+z^3_2+z^4_2+z^6_2=0,\qquad -z^4_2-z^4_3+z^5_4+z^6_4=0,\qquad -z^6_1-z^6_2-z^6_4-z^6_5=0
\]
remain satisfied. The equations at the 2-degree vertices 1, 3 and 5 are unchanged.

\begin{proposition}\label{K=V-1}
Let $\cal G$ be a connected merger of circles having $V$ vertices and $E$ edges. Let $K$ denote the maximal number of linearly independent equations among the $V$ equations for the edge variables. Then $K=V-1$ independently of $E$, so the number of free variables is $N_I=E-V+1$. Moreover, any $V-1$ equations are linearly independent.
\end{proposition}

\noindent{\em Proof.} We proceed by induction according to the number of vertices. For $V=2$ the claim holds true, see Example 1. Suppose that it holds for $V=n$ and let $A_n$ denote the matrix of the coefficients in the $n$ equations. By the induction hypothesis $K\equiv{\rm rank} A_n=n-1$. If the vertex $n+1$ is coupled to any vertex $i\leq n$ with at most a single edge then
\[
A_{n+1}=\begin{pmatrix}
A_n & {\rm diag}\left(b_1^{n+1},\ldots,b_n^{n+1}\right)\\
\phantom{.}&\phantom{.}\\
0      & -b_1^{n+1}\ \ldots\ -b_n^{n+1}
\end{pmatrix}
\]
where
\[
b_i^{n+1}=\left\{\begin{array}{ll}
1 & \mbox{if there is an edge between vertices $i$ and $n+1$}\\
0 & \mbox{otherwise,}
\end{array}
\right.
\]
and at least two of them is 1. In the general case when vertex $n+1$ is coupled to vertex $i$ with $r_i$ edges, the diagonal matrix is replaced by a block-diagonal one where the $i$th block is
\[
B_i=(b^{n+1}_{i,1}\ldots b^{n+1}_{i,r_i}), \qquad i=1,\dots,n.
\]
At the same time in the last line we have $-b^{n+1}_{1,1}\ \ldots\ -b^{n+1}_{n,r_n}$ with at least two nonzero elements.
Now in $A_{n+1}$ the first $n$ lines are linearly independent, and replacing any one with the last line the $n$ lines are still linearly independent. On the other hand, the sum of the $n+1$ lines gives zero. So ${\rm rank}A_{n+1}=n$. $\quad\Box$

\vspace{5pt}
\noindent
{\em Example 4.}
For plane graphs and graphs consisting of the vertices and edges of convex polyhedra $N_I$ is related to the Euler characteristic $V+F-E=2$. Comparison with it shows that $N_I=F-1$ where $F$ is the number of faces (the exterior face included for plane graphs). Example 2 is a tetrahedron with $N_I=F-1=3=M$. In the case of Example 3 the general solution can also be obtained by choosing freely a variable for all the four triangles and then merge them.
The procedure applies to any connected plane graph: demerge it into faces, assign to the edges of the circle around each interior face a value as was shown in the proof of Proposition~\ref{merger2}, and merge the faces. This is an alternative way to solve the associated system of equations.
For edge-connected plane graphs $M<N_I$ because to cover the edges joining a vertex of even degree greater than 2 half of the faces incident on that vertex is enough. For the square lattice $M/N_I\to 1/2$. If the degree is odd, more faces are necessary. For the honeycomb lattice $M/N_I\to 3/4$.

\noindent
{\em Example 5.} There is another set of circle-generated merger graphs for which the general solution of the system of equations is easily obtained: the complete $n$-graphs. In this case $N_I=n(n-1)/2-n+1=(n-1)(n-2)/2$, the number of edges of the complete $(n-1)$-graph. The variables of this latter, $z^k_j$ for $1\leq j<k\leq n-1$ are free; then with
\[
z^n_l=\sum_{j=1}^{l-1}z^l_j-\sum_{k=l+1}^{n-1}z^k_l\qquad l=1,\dots,n-1
\]
all the equations will be satisfied. (For the complete 4-graph this offers the alternative to choose first $z^2_1=x$, $z^3_2=y$, $z^3_1=z$ and conclude with $z^4_1=-x-z$, $z^4_2=x-y$, $z^4_3=y+z$.) Note that for the complete $n$-graphs $N_I=M$: the triangles $(j,k,n)$ with $j<k<n$ form a max-min covering and $M=(n-1)(n-2)/2$ indeed. The edges $(j,k)$ are covered by a single triangle and each edge $(l,n)$ is covered by $n-2$ triangles.

In general, finding $N_I$ edges that can carry free variables or, equivalently, a unimodular $(V-1)\times(V-1)$ submatrix of the incidence matrix (whose columns belong to a possible set of dependent variables) may be difficult. Instead, one can easily find $N_I$ circles that constitute a covering, and assign a free variable to each circle. The variable is then equipped with edge-dependent signs on the edges of the circle. After merging the $N_I$ circles, on every edge we obtain the sum of the signed circle-variables belonging to the circles incident on that edge, and the linear equations at each vertex will be satisfied with nonzero integer vectors.

\newpage
\newsec{Proof of Theorem~\ref{first-step}}

Introduce
\be\label{tildeGn(x)}
\tilde{G}\left[n,\{n_l\}_1^p\right](x)=\exp\left\{\frac{\beta\hat{u}(0)N(N-1)}{2L^d}\right\}G\left[n,\{n_l\}_1^p\right](x),
\quad \tilde{G}^N_{n}(x)=\exp\left\{\frac{\beta\hat{u}(0)N(N-1)}{2L^d}\right\}G^N_{n}(x)
\ee
and substitute in the first formula the asymptotic form Eq.~(\ref{exp-asymp}). This yields ($n_0=n$)
\bea\label{tildeGn(x)-asymp}
\tilde{G}\left[n,\{n_l\}_1^p\right](x)
=
\sum_{\{\alpha^k_j\in\Nn_0|1\leq j<k\leq N\}}
\Delta_{\{\alpha^k_j\},\{n_l\}_0^p}\ \left(L^{-d}\right)^{K_{\{\alpha^k_j\}}}
\prod_{1\leq j<k\leq N}
\frac{\left(-\beta\right)^{\alpha^k_j}}{\alpha^k_j !} \prod_{r=1}^{\alpha^k_j}
 \int_0^1\d t^k_{j,r} \int\d x^k_{j,r}\ \hat{u}\left(x^k_{j,r}\right)
\nonumber\\
\left[\delta(X^0_1,\dots,X^p_1)
\exp\left\{-\pi n \lambda_\beta^2\left[\overline{\left(X^0_{^\cdot}\right)^2}-\overline{X^0_{^\cdot}}^2\right]\right\}
f_n\left(x;L\overline{X^0_{^\cdot}}\right)
\prod_{l=1}^p \exp\left\{-\pi n_l \lambda_\beta^2\left[\overline{\left(X^l_{^\cdot}\right)^2}-\overline{X^l_{^\cdot}}^2\right]\right\}
f_{n_l}\left(0;L\overline{X^l_{^\cdot}}\right)\right]
\nonumber\\
\eea
where
\be
f_{n_l}\left(0;L\overline{X^l_{^\cdot}}\right)=
\sum_{z\in\Zz^d}\exp\left\{-\frac{\pi n_l\lambda_\beta^2}{L^2}(z+L\overline{X^l_{^\cdot}})^2\right\}
\ee
and
\bea\label{fnx}
f_n\left(x;L\overline{X^0_{^\cdot}}\right)
&=&
\sum_{z\in\Zz^d}\exp\left\{-\frac{\pi n\lambda_\beta^2}{L^2}(z+L\overline{X^0_{^\cdot}})^2\right\} \cos \left[\frac{2\pi}{L}z\cdot x\right]
\nonumber\\
&=&
\frac{L^d}{\lambda_{n\beta}^d}\sum_{z\in\Zz^d}\exp\left\{-\frac{\pi(x+Lz)^2}{n\lambda_\beta^2}\right\}
\cos\left[2\pi\overline{X^0_{^\cdot}} \cdot (x+Lz)\right].
\eea
Equations (\ref{sigma-rho-with-G}) can be replaced with
\be\label{sigma-rho-with-tildeG}
\langle x|\sigma^{N,L}_1|0\rangle=\sum_{n=1}^N \rho^{N,L}_n\,\frac{\tilde{G}^N_{n}(x)}{\tilde{G}^N_{n}(0)},\qquad
\rho^{N,L}_0=\sum_{n=1}^N \rho^{N,L}_n\,\frac{\int_\Lambda\d x\, \tilde{G}^N_{n}(x)}{L^d\, \tilde{G}^N_{n}(0)}.
\ee
Let us start by proving a lemma. The first form of $f_n\left(x;L\overline{X^0_{^\cdot}}\right)$ suggests the evaluation
\[
f_n\left(x;L\overline{X^0_{^\cdot}}\right)\sim L^d
\int \exp\left\{-\pi n\lambda_\beta^2(y+\overline{X^0_{^\cdot}})^2\right\} \cos 2\pi y\cdot x \d y
=\frac{L^d}{\lambda_{n\beta}^d}\exp\left\{-\frac{\pi x^2}{n\lambda_\beta^2}\right\} \cos 2\pi \overline{X^0_{^\cdot}}\cdot x.
\]
However, this gives the right result only when $n\lambda_\beta^2/L^2\to 0$. To cover all the cases when $n$ tends to infinity we need detailed estimates.

\begin{lemma}
For $x$ fixed, when $N, L\to\infty$, $N/L^d=\rho>0$ we have the following asymptotic forms for $f_n\left(x;L\overline{X^0_{^\cdot}}\right)$.

\begin{itemize}
\item[(i)]
 If $n\lambda_\beta^2/L^2\to 0$ then
\be\label{to-0}
f_n\left(x;L\overline{X^0_{^\cdot}}\right)
=
\frac{L^d}{\lambda_{n\beta}^d}\exp\left\{-\frac{\pi x^2}{n\lambda_\beta^2}\right\}\left[\cos2\pi\overline{X^0_{^\cdot}}\cdot x
+o(1)\right].
\ee
\item[(ii)]
If $n\lambda_\beta^2/L^2\to \infty$ then
\bea\label{to-infty}
f_n\left(x;L\overline{X^0_{^\cdot}}\right)
&=&
\exp\left\{-\frac{\pi n\lambda_\beta^2}{L^2}\{L\overline{X^0_{^\cdot}}\}^2\right\}
\nonumber\\
&\times&
\left[\left(1+\sum_{z:\, \max|z_i|=1}\exp\left\{-\frac{\pi n\lambda_\beta^2}{L^2}z\cdot (z+2\{L\overline{X^0_{^\cdot}}\})\right\}   \right)\cos2\pi\overline{X^0_{^\cdot}}\cdot x + o(1)   \right]
\eea
where
$\{L\overline{X^0_{^\cdot}}\}$ is the fractional part of $L\overline{X^0_{^\cdot}}$, each component of which is bounded in modulus by 1/2.
\item[(iii)]
If $n\lambda_\beta^2/L^2=c>0$ then
\be\label{=c}
f_n\left(x;L\overline{X^0_{^\cdot}}\right)=\sum_{z\in\Zz^d}\exp\left\{-\pi c(z+\{L\overline{X^0_{^\cdot}}\})^2\right\} \left[\cos2\pi\overline{X^0_{^\cdot}}\cdot x + o(1)\right].
\ee
\end{itemize}
\end{lemma}

\vspace{10pt}
\noindent
{\em Proof.} (i) From the second form of $f_n\left(x;L\overline{X^0_{^\cdot}}\right)$,
\bea\label{fn-for-lemma}
f_n\left(x;L\overline{X^0_{^\cdot}}\right)
&=&
\frac{L^d}{\lambda_{n\beta}^d}
\exp\left\{-\frac{\pi x^2}{n\lambda_\beta^2}\right\}
\nonumber\\
&\times& \left[\cos2\pi\overline{X^0_{^\cdot}}\cdot x
+\sum_{z\in\Zz^d\setminus\{0\}}\exp\left\{-\frac{\pi\left[(x+Lz)^2-x^2\right]}{n\lambda_\beta^2}\right\}\cos2\pi\overline{X^0_{^\cdot}} \cdot (x+Lz)\right].
\nonumber\\
\eea
Because for $z\in\Zz^d$
\[
|x+Lz|^2-x^2\geq L^2 z^2\left(1-2|x|/L\right),
\]
the sum over $\Zz^d\setminus\{0\}$ can be bounded above in modulus by
\[
\sum_{z\in\Zz^d\setminus\{0\}}\exp\left\{-\frac{\pi L^2 z^2}{n\lambda_\beta^2}(1-2|x|/L)\right\}
\]
which tends to zero as $L^2/(n\lambda_\beta^2)$ goes to infinity.

\vspace{5pt}\noindent
(ii) We use the first form of $f_n\left(x;L\overline{X^0_{^\cdot}}\right)$. With $L\overline{X^0_{^\cdot}}=[L\overline{X^0_{^\cdot}}]+\{L\overline{X^0_{^\cdot}}\}$, a decomposition into integer and fractional parts,
\bea\label{largest-n}
f_n\left(x;L\overline{X^0_{^\cdot}}\right)=\sum_{z\in\Zz^d}\exp\left\{-\frac{\pi n\lambda_\beta^2}{L^2}(z+\{L\overline{X^0_{^\cdot}}\})^2\right\}
\cos \frac{2\pi}{L}\left(z-[L\overline{X^0_{^\cdot}}]\right)\cdot x      \hspace{5.5cm}
\nonumber\\
=\exp\left\{-\frac{\pi n\lambda_\beta^2}{L^2}\{L\overline{X^0_{^\cdot}}\}^2\right\}
\left[\cos\frac{2\pi}{L}[L\overline{X^0_{^\cdot}}]\cdot x+\sum_{z\in\Zz^d\setminus\{0\}}\exp\left\{-\frac{\pi n\lambda_\beta^2}{L^2}
z\cdot (z+2\{L\overline{X^0_{^\cdot}}\})\right\}   \cos\frac{2\pi}{L}\left(z-[L\overline{X^0_{^\cdot}}]\right)\cdot x   \right].
\nonumber\\
\eea
$z\cdot (z+2\{L\overline{X^0_{^\cdot}}\})\geq 0$, and the sum restricted to  $\max_{1\leq i\leq d}|z_i|\geq 2$ is $o(1)$. The sum over $z$ with $\max|z_i|=1$ can give a contribution of order 1 if $|\{L\overline{X^0_{^\cdot}}\}_i|=1/2$ for one or more components of $\{L\overline{X^0_{^\cdot}}\}$. For these $z$
\[
\cos\frac{2\pi}{L}\left(z-[L\overline{X^0_{^\cdot}}]\right)\cdot x
= \cos\frac{2\pi}{L}\left(L\overline{X^0_{^\cdot}}-\{L\overline{X^0_{^\cdot}}\}-z\right)\cdot x
= \cos2\pi\overline{X^0_{^\cdot}}\cdot x + O(L^{-1}).
\]
However, $|\{L\overline{X^0_{^\cdot}}\}_i|=1/2$ can only be if the times $t^k_{j,r}$ entering $L\overline{X^0_{^\cdot}}$ (only for $j\leq n$, $k>n$, cf. Eq.~(\ref{intZqbis}))
take value from a zero-measure subset of $[0,1]^{M}$ where $M=\sum_{j=1}^n\sum_{k=n+1}^N\alpha^k_j$. Because $f_n\left(x;L\overline{X^0_{^\cdot}}\right)$ appears in $\tilde{G}\left[n,\{n_l\}_1^p\right](x)$ under integrals over $t^k_{j,r}$, the sum over  $z$ with $\max|z_i|=1$ can be dropped.

\vspace{5pt}\noindent
(iii) Looking at the first line of (\ref{largest-n}) it is seen that the summand has a summable majorizing function. Therefore the asymptotic approximation can be done under the summation sign: for any $z$ fixed,
$\cos \frac{2\pi}{L}\left(z-[L\overline{X^0_{^\cdot}}]\right)\cdot x=\cos 2\pi\overline{X^0_{^\cdot}}\cdot x + O(L^{-1})$.
$   \Box$

\vspace{5pt}
The essential information provided by the above lemma is that for large systems the $x$-dependence of $f_n\left(x;L\overline{X^0_{^\cdot}}\right)$ and, hence, of $\tilde{G}\left[n,\{n_l\}_1^p\right](x)$ is in the factors
$\exp\left\{-\frac{\pi x^2}{n\lambda_\beta^2}\right\}$ and $\cos2\pi\overline{X^0_{^\cdot}}\cdot x$. From Eqs.~(\ref{to-0})-(\ref{=c})
\be\label{f_n(0)-asymp}
f_n\left(0;L\overline{X^0_{^\cdot}}\right)=
\left\{\begin{array}{lll}
\frac{L^d}{\lambda_{n\beta}^d}\,[1+o(1)]&\mbox{if}& n\lambda_\beta^2/L^2\to 0\\
e^{-\pi n\lambda_\beta^2\{L\overline{X^0_{^\cdot}}\}^2/L^2}\left[1
+o(1)\right]&\mbox{if}&n\lambda_\beta^2/L^2\to\infty\\
\sum_{z\in\Zz^d}e^{-\pi c(z+\{L\overline{X^0_{^\cdot}}\})^2}&\mbox{if}&n\lambda_\beta^2/L^2=c.
\end{array}\right.
\ee
Comparison with (\ref{to-0})-(\ref{=c}) shows that
\[
f_n\left(x;L\overline{X^0_{^\cdot}}\right)=
\left\{\begin{array}{lll}
f_n\left(0;L\overline{X^0_{^\cdot}}\right)\exp\left\{-\frac{\pi x^2}{n\lambda_\beta^2}\right\}\left[\cos2\pi\overline{X^0_{^\cdot}}\cdot x+o(1)\right]
&\mbox{if}& n=O(1) \\
f_n\left(0;L\overline{X^0_{^\cdot}}\right)\left[\cos2\pi\overline{X^0_{^\cdot}}\cdot x+o(1)\right]
&\mbox{if}&n\to\infty.\\
\end{array}\right.
\]
However, because $\exp\{-\pi x^2/(n\lambda_\beta^2)\}\to 1$ as $n\to\infty$,
\be\label{fn-unified}
f_n\left(x;L\overline{X^0_{^\cdot}}\right)
=
f_n\left(0;L\overline{X^0_{^\cdot}}\right)
\exp\left\{-\frac{\pi x^2}{n\lambda_\beta^2}\right\}\left[\cos2\pi\overline{X^0_{^\cdot}}\cdot x+o(1)\right]
\ee
holds true in all the cases. This can be substituted into Eq.~(\ref{tildeGn(x)-asymp}).

To write down $\tilde{G}^N_n(x)/\tilde{G}^N_n(0)$ we separate the contribution of the $p$ closed trajectories,
\bea\label{HN-n}
H_{N-n}
=\sum_{p=1}^{N-n}\frac{1}{p!}\sum_{n_1,\dots,n_p\geq 1:\sum_1^p n_l=N-n}\frac{1}{\prod_1^p n_l}  \hspace{9cm}
\nonumber\\
\sum_{\{\alpha^k_j\in\Nn_0|n+1\leq j< k\leq N\}} \left(\prod_{n+1\leq j<k\leq N}\frac{\left(-\beta\right)^{\alpha^k_j}}{\alpha^k_j !}\right)
\int_0^1 \prod_{n+1\leq j<k\leq N}\prod_{r=1}^{\alpha^k_j}\d t^k_{j,r}
\int \prod_{n+1\leq j<k\leq N} \prod_{r=1}^{\alpha^k_j}\d x^k_{j,r}\hat{u}\left(x^k_{j,r}\right)
\nonumber\\
\left[\delta(X^0_1,\dots,X^p_1)\,\Delta_{\{\alpha^k_j\},\{n_l\}_0^p}\, \left(L^{-d}\right)^{K_{\{\alpha^k_j\}}}
\prod_{l=1}^p
e^{-\pi n_l \lambda_\beta^2\left[\overline{\left(X^l_{^\cdot}\right)^2}-\overline{X^l_{^\cdot}}^2\right]}
f_{n_l}\left(0;L\overline{X^l_{^\cdot}}\right)\right].
\nonumber\\
\eea
$H_{N-n}$ depends on all the variables that connect the cycles $l=1,\dots,p$ to cycle 0, i.e. on
\[
\{\alpha^k_j,x^k_{j,r},t^k_{j,r}| j\leq n, k\geq n+1, r\leq \alpha^k_j\}.
\]
Let
\bea\label{Theta}
\lefteqn{
\Theta^N_n(y)
=
\sum_{\{\alpha^k_j\in\Nn_0|1\leq j\leq n,\, j+1\leq k\leq N\}} \left(\prod_{ j=1}^n\prod_{k=j+1}^N\frac{\left(-\beta\right)^{\alpha^k_j}}{\alpha^k_j !}\right)
\int_0^1 \prod_{ j=1}^n\prod_{k=j+1}^N\prod_{r=1}^{\alpha^k_j}\d t^k_{j,r}    }
\nonumber\\
&& \int \prod_{j=1}^n\prod_{k=j+1}^N \prod_{r=1}^{\alpha^k_j}\d x^k_{j,r}\, \hat{u}\left(x^k_{j,r}\right) H_{N-n}
\exp\left\{-\pi n \lambda_\beta^2\left[\overline{\left(X^0_{^\cdot}\right)^2}-y^2\right]\right\} f_n\left(0;Ly\right) \delta(y-\overline{X^0_{^\cdot}}),
\eea
i.e. the integrations over $\{t^k_{j,r},\,x^k_{j,r}|j\leq n, k\geq j+1, r\leq \alpha^k_j\}$ are restricted to values that yield $\overline{X^0_{^\cdot}}=y$.
Then
\be\label{L^d F^N_n(x)}
\tilde{G}^N_n(x)
=
\exp\left\{-\frac{\pi x^2}{n\lambda_\beta^2}\right\}
\int_{\Rr^d}\Theta^N_n(y)[\cos (2\pi y\cdot x)+o(1)] \d y.
\ee
Introducing
\be\label{nu}
\nu^{N,L}_n(y)=\frac{\Theta^N_n(y)}{\tilde{G}^N_n(0)}=\frac{\Theta^N_n(y)}{\int\Theta^N_n(y')\d y'},
\ee
the ratio $\tilde{G}^N_n(x)/\tilde{G}^N_n(0)$ takes the form
\be\label{Fnx/Fn0}
\frac{\tilde{G}^N_n(x)}{\tilde{G}^N_n(0)}
=
\exp\left\{-\frac{\pi x^2}{n\lambda_\beta^2}\right\}\int_{\Rr^d}[\cos(2\pi y\cdot x)+o(1)]\,\nu^{N,L}_n(y)\d y \,.
\ee
$\int\nu^{N,L}_n(y)\d y=1$, but $\nu^{N,L}_n(y)$ can be negative for some values of $y$. Note also that $\nu^{N,L}_n(y)=\nu^{N,L}_n(-y)$ and, if $u$ is spherically symmetric then $\nu^{N,L}_n$ has a cubic symmetry that tends to spherical as $L$ inreases.

\subsection{ Condition for ODLRO}\label{cond-ODLRO}

The general concept of off-diagonal long-range order was introduced by C. N. Yang [Y]. In the present context it means that
$\sigma_1(x)$ does not go to zero as $x$ goes to infinity.
Now with
\be
\lim_{N,L\to\infty,N/L^d=\rho}\frac{\tilde{G}^N_{n}(x)}{\tilde{G}^N_{n}(0)}\leq \exp\left\{-\frac{\pi x^2}{n\lambda_\beta^2}\right\}\leq 1
\ee
and
\be\label{cycle-percolation}
\rho-\sum_{n=1}^\infty \rho_n
=
\lim_{M\to\infty} \lim_{N,L\to\infty,N/L^d=\rho}\sum_{n=M+1}^N\rho^{N,L}_n
\ee
one obtains
\bea
\sigma_1(x)
&=&
\sum_{n=1}^\infty \rho_n \lim_{N,L\to\infty,N/L^d=\rho}\frac{\tilde{G}^N_{n}(x)}{\tilde{G}^N_{n}(0)}
+\lim_{M\to\infty}\lim_{N,L\to\infty,N/L^d=\rho}\sum_{n=M+1}^N \rho^{N,L}_n \frac{\tilde{G}^N_{n}(x)}{\tilde{G}^N_{n}(0)}
\nonumber\\
&\leq&
 \sum_{n=1}^\infty \rho_n \exp\left\{-\frac{\pi x^2}{n\lambda_\beta^2}\right\} + \rho-\sum_{n=1}^\infty \rho_n.
\eea
This is the upper bound in (\ref{cond1}).
If in the infinite system the full density arises from particles in finite cycles, i.e. $\rho=\sum_{n=1}^\infty \rho_n$, then
\be\label{sigma1-from-finite n}
\sigma_1(x) \leq
\sum_{n=1}^\infty \rho_n \exp\left\{-\frac{\pi x^2}{n\lambda_\beta^2}\right\},
\ee
the infinite sum decays to zero as $x$ goes to infinity. Thus, there is no ODLRO and no BEC either:
\be
\rho^{N,L}_0\leq \sum_{n=1}^N \rho^{N,L}_n \frac{1}{L^d}\int_\Lambda\exp\left\{-\frac{\pi x^2}{n\lambda_\beta^2}\right\}\d x \rightarrow 0.
\ee
In the absence of interaction the inequalities hold with equality.

On the other hand, if $\rho-\sum_{n=1}^\infty \rho_n>0$,
the infinite-volume Gibbs measure assigns a positive weight to infinite cycles. When $N$, $L$ and $n$ tend to infinity then for $x$ fixed $\exp\left\{-\frac{\pi x^2}{n\lambda_\beta^2}\right\}$ tends to 1, and
\be\label{Phi}
\frac{\tilde{G}^N_n(x)}{\tilde{G}^N_n(0)}\to\Phi(x)=\int_{\Rr^d}\cos (2\pi y\cdot x)\nu(\d y).
\ee
For the existence of ODLRO $\nu$ must contain a positive multiple of $\delta_0$: $\Phi$ being nonnegative, in the absence of an atomic component at 0 there could be no other atomic component either, so $\nu$ would define a continuous measure, and by the Riemann-Lebesgue lemma $\Phi(x)$ would go to zero when $x$ goes to infinity.
For the ideal Bose gas it is easy to check that $G^N_{n}(x)/G^N_{n}(0)=f_n(x;0)/f_n(0;0)\to 1$ whenever $n\to\infty$, i.e. $\nu=\delta_0$.
In principle, $\nu$ and $\Phi$ may depend on how fast $n$ increases with $N$. It will be seen that $\nu^{N,L}_n$ converges weakly to $\delta_0$ by concentrating to a $1/\sqrt{n}$-neighborhood of the origin, so the rate of convergence depends only on $n$.

Let now $g_N<N$ be any monotone increasing sequence that tends to infinity.
Clearly,
\[
\sigma_1(x)\geq \lim_{N,L\to\infty,N/L^d=\rho}\sum_{n=g_N}^N \rho^{N,L}_n \frac{\tilde{G}^N_{n}(x)}{\tilde{G}^N_{n}(0)}.
\]
For $N$ large enough, $n\geq g_N$ and $y$ in a $1/\sqrt{n}$-neighborhood of the origin $\cos (2\pi y\cdot x)\geq 1-x^2/(g_N \lambda^2)$ with some constant $\lambda^2$, therefore
$
\tilde{G}^N_{n}(x)/\tilde{G}^N_{n}(0)\geq \exp\{-\pi x^2/(g_N\lambda_\beta^2)\}(1-x^2/(g_N \lambda^2)),
$
implying
$
\sigma_1(x)\geq  \lim_{N,L\to\infty,N/L^d=\rho}\sum_{n=g_N}^N \rho^{N,L}_n.
$
Because this holds true for any $g_N\to\infty$, we obtain the lower bound
\be\label{lower-bound-to-sigma1}
\sigma_1(x)\geq  \rho-\sum_{n=1}^\infty \rho^{N,L}_n.
\ee

To see the concentration of $\nu^{N,L}_n$ we must analyse the large-$n$ behavior of $\overline{X^0_{^\cdot}}$ and $\overline{\left(X^0_{^\cdot}\right)^2}$. From Eq.~(\ref{intZqbis}) for $l=0$
\be\label{avX_0}
n\overline{X^0_{^\cdot}}= -\sum_{k=2}^n\sum_{j=1}^{k-1}(k-j)\sum_{r=1}^{\alpha^{k}_{j}}x^{k}_{j,r}
+\sum_{j=1}^{n}\sum_{k=1}^{N-n}\sum_{r=1}^{\alpha^{+k}_{j}}\left(j-1+t^{+k}_{j,r}\right)x^{+k}_{j,r}
=: nY^0_0 +nY^+_0.
\ee
In the new notation we restart the count of the particles outside the zeroth cycle. Now cycle 0 is distinguished without making distinction among the different partitions of $N-n$. Similarly, combining Eq.~(\ref{Zldotsquare}) with  (\ref{zji}) and taking the $m\to\infty$ limit we find, cf. [Su12],
\bea\label{avZl-2}
n\overline{\left(X^0_{^\cdot}\right)^2}
&=&\sum_{1\leq j<k\leq n}\ \sum_{1\leq j'<k'\leq n}\sum_{r=1}^{\alpha^k_j}\sum_{r'=1}^{\alpha^{k'}_{j'}}A^{k'j'r'}_{kjr}x^k_{j,r}\cdot x^{k'}_{j',r'}
\nonumber\\
&+&\sum_{j=1}^{n}\sum_{k=1}^{N-n} \sum_{j'=1}^{n}\sum_{k'=1}^{N-n}
\sum_{r=1}^{\alpha^{+k}_j}\sum_{r'=1}^{\alpha^{+k'}_{j'}}A^{+k'j'r'}_{+kjr} x^{+k}_{j,r}\cdot x^{+k'}_{j',r'}
-2\sum_{1\leq j<k\leq n}\sum_{j'=1}^{n}\sum_{k'=1}^{N-n} \sum_{r=1}^{\alpha^k_j}\sum_{r'=1}^{\alpha^{+k'}_{j'}}A^{+k'j'r'}_{kjr} x^k_{j,r}\cdot x^{+k'}_{j',r'}
\nonumber\\
\eea
where
\be
A^{+k'j'r'}_{+kjr}=\min\left\{j-1+t^{+k}_{j,r},\,j'-1+t^{+k'}_{j',r'}\right\},
\ee
\be
A^{+k'j'r'}_{kjr}=\left\{\begin{array}{lll}
k-j&\mbox{if}&k<j'\\
j'-j+t^{+k'}_{j',r'}-t^{k}_{j,r}&\mbox{if}&j< j'<k\\
0&\mbox{if}&j'<j\ ,
\end{array}\right.
\ee
\be
A^{k'j'r'}_{kjr}=\left\{\begin{array}{lll}
0&\mbox{if}&j'<k'< j<k\ \ \mbox{or}\ \ j<k< j'<k'\\
k-j&\mbox{if}&j'< j<k< k'\\
k'-j'&\mbox{if}&j< j'<k'< k\\
k-j'+t^k_{j,r}-t^{k'}_{j',r'}&\mbox{if}&j<j'< k<k'\\
k'-j+t^{k'}_{j',r'}-t^{k}_{j,r}&\mbox{if}&j'< j<k'< k\ .\\
\end{array}\right.
\ee
If two indices coincide, there may appear a correction in modulus not larger than 1, coming from the times $t^k_{j,r}, t^{+k}_{j,r}$, but there is no such correction to $A^{kjr}_{kjr}=k-j$.

In the analysis below we treat $n$ as an independent variable and $N={\cal N}(n)$ strictly increasing through jumps. The fastest increase of $n$ that we shall consider here is $n\sim\varepsilon N$ with some $0<\varepsilon<1$, so the slowest increase for ${\cal N}$ is ${\cal N}(n)=\lfloor n/\varepsilon\rfloor$. Accordingly, $L$ must satisfy ${\cal N}(n)/L^d=\rho$.

Because $\overline{X^0_{^\cdot}}$ and $\overline{\left(X^0_{^\cdot}\right)^2}$ depend on all the variables $\{\alpha^k_j,\,t^k_{j,r},\,x^k_{j,r}|j\leq n, k\geq j+1, r\leq \alpha^k_j\}$, it is advantageous to rearrange the summations in $\Theta^N_n$. Let
\be\label{vector-alpha}
~\alpha^0_n=\{\alpha^k_j|1\leq j< k\leq n\},\quad ~\alpha^+_n=\{\alpha^{+k}_j|1\leq j\leq n,\, 1\leq k\leq {\cal N}(n)-n\},\quad ~\alpha_n=\{~\alpha^0_n,\,~\alpha^+_n\}.
\ee
Let, moreover,
\be\label{alpha-norms}
\|~\alpha^0_n\|=\sum_{k=2}^n\sum_{j=1}^{k-1} j\  \alpha^k_{k-j},\quad \|~\alpha^+_n\|=\sum_{k=1}^{{\cal N}(n)-n}\sum_{j=1}^n \ j\alpha^{+k}_j,
\quad \|~\alpha_n\|=\|~\alpha^0_n\|+\|~\alpha^+_n\|.
\ee
Then
\be
\Theta^N_n(y)=\sum_{A,B=0}^\infty\,\sum_{~\alpha^0_n:\|~\alpha^0_n\|=A}\,\sum_{~\alpha^+_n:\|~\alpha^+_n\|=B}\vartheta_{~\alpha_n}(y)
\ee
where $\vartheta_{~\alpha_n}(y)$ is defined by Eq.~(\ref{Theta}).
For most of what follows we shall consider the infinite sequences $~\alpha^0=\{\alpha^k_j|1\leq j<k<\infty\}$ and $~\alpha^+=\{\alpha^{+k}_j| j, k\in \Nn_+\}$ as given beforehand, and then $~\alpha^0_n$ and $~\alpha^+_n$ are their restrictions to a finite number of elements as shown in Eq.~(\ref{vector-alpha}).
Since $t^k_{j,r}$ and $t^{+k}_{j,r}$ play a minor role, they can also be considered as given beforehand.

$\tilde{G}^N_n(0)=\int\Theta^N_n(y)\d y$ written as a sum over $~\alpha_n$ makes it possible to examine the contribution of the different terms and to decide which of them dominate the sum. In order to better distinguish between the different cases we provisionally strengthen the condition on $\hat{u}$ by assuming that its support is compact, implying also $\int\hat{u}(x) x^2 \d x<\infty$.

\vspace{3pt}
\noindent
(i) If $~\alpha$ contains a finite number of nonzero elements then $\|~\alpha_n\|$ attains a constant at a finite $n$, implying
$\overline{X^0_{^\cdot}}\to 0$. We shall refer to these sequences as rationals,
while those with infinitely many positive elements will be referred to as irrationals.
With all the $x^k_{j,r}$ and $x^{+k}_{j,r}$ on the support of $\hat{u}$ (that we suppose in the sequel) $n\overline{X^0_{^\cdot}}$ and $n\overline{\left(X^0_{^\cdot}\right)^2}$ are bounded, so $|\overline{X^0_{^\cdot}}|=O(1/n)$ and
\[
n\left[\overline{\left(X^0_{^\cdot}\right)^2}-\overline{X^0_{^\cdot}}^2\right]
= n\overline{\left(X^0_{^\cdot}\right)^2}-O(1/n)=O(1).
\]

\noindent
(ii) $\overline{X^0_{^\cdot}}$ tends surely to zero also for some irrational sequences.
If  $\|~\alpha_n\|\to\infty$ but
\be\label{cond-sure-convergence}
\frac{\|~\alpha_n\|}{n}
=\frac{1}{n}\left[ \sum_{k=2}^n\sum_{j=1}^{k-1} j\  \alpha^k_{k-j}
+\sum_{k=1}^{{\cal N}(n)-n}\sum_{j=1}^n \ j\alpha^{+k}_j\right] \to 0
\ee
then $\overline{X^0_{^\cdot}}$ goes to zero, although the decay is slower than $1/n$. An example is $~\alpha^+=~0$, $\alpha^k_{k-1}=1$ if $k$ is a prime and $\alpha^k_j=0$ otherwise. Concerning the asymptotic form of $n\left[\overline{\left(X^0_{^\cdot}\right)^2}-\overline{X^0_{^\cdot}}^2\right]$ more information is necessary and will be obtained later.

\vspace{3pt}
\noindent
(iii) To enlarge further the family of irrational sequences yielding $\overline{X^0_{^\cdot}}\to 0$ and also to get some insight into the asymptotic form of $\overline{\left(X^0_{^\cdot}\right)^2}-\overline{X^0_{^\cdot}}^2$ we use probabilistic arguments.
The vectors $x^k_{j,r}$ and $x^{+k}_{j,r}$ can be considered as identically distributed zero-mean random variables with the common probability density $|\hat{u}(\cdot)|/\|\hat{u}\|_1$. Viewed as such, $n\overline{X^0_{^\cdot}}$ is the result of a finite number of steps of a symmetric random walk in $\Rr^d$.
The vectors $x^k_{j,r}$ are independent among themselves and from  $x^{+k}_{j,r}$, but the latter are weakly dependent because their sum vanishes,
\be\label{x+1,11}
X^0_1=\sum_{j=1}^n\sum_{k=1}^{N-n}\sum_{r=1}^{\alpha^{+k}_j}x^{+k}_{j,r}=0.
\ee
One vector, e.g. $x^{+1}_{1,1}$ can be expressed with the others, yielding
\be\label{Y+0-indept}
n Y^+_0=\sum_{j=1}^{n}\sum_{k=1}^{N-n}\sum_{r=1}^{\alpha^{+k}_{j}}\left(j-1+t^{+k}_{j,r}-t^{+1}_{1,1}\right)x^{+k}_{j,r}.
\ee
The vectors appearing here with a nonzero coefficient are already independent. The product measures for finite $N, L$ form a consistent family and can be extended into a unique probability measure $\Pp$ in the measurable space $\left((\Rr^d)^\infty, \Sigma\right)$, where $\Sigma$ is the smallest $\sigma$-algebra containing all sets that depend only on a finite number of $x^k_{j,r}$, $x^{+k}_{j,r}$. In a short-hand notation $\tilde{G}^N_n(0)$ can be rewritten as
\be
\tilde{G}^{{\cal N}(n)}_n(0)=
\sum_{~\alpha_n} \frac{(-\beta \|\hat{u}\|_1)^{~\alpha_n}}{~\alpha_n!} \int \d t_{~\alpha_n}\,
\Ee\left[H_{{\cal N}(n)-n}\exp\left\{-\pi n \lambda_\beta^2\left[\overline{\left(X^0_{^\cdot}\right)^2}-\overline{X^0_{^\cdot}}^2\right]\right\}
f_n\left(0;L\overline{X^0_{^\cdot}}\right) \right].
\ee
Now $\overline{X^0_{^\cdot}}\to 0$
is a tail event, so it occurs with probability 0 or 1.
In the probabilistic setting the convergence of $Y^0_0$ to zero is an instance of the strong law of large numbers and it holds with probability one e.g. for the sequence
\be\label{intracycle}
\alpha^k_{k-j}=\left\{\begin{array}{ll}
a & {\rm if}\ j\leq j_0, k=2, 3,\dots\\
0 & {\rm otherwise}
\end{array}\right.
\ee
where $a, j_0\geq 1$. This can be seen by expressing $Y^0_0$ as an average of independent zero-mean random variables,
\be\label{lambda-u}
Y^0_0=\frac{1}{n}\sum_{k=2}^n\xi_k\quad\mbox{with}\quad \xi_k=-\sum_{j=1}^{k-1}j\sum_{r=1}^{\alpha^k_{k-j}}x^k_{k-j,r}\quad\mbox{and}\quad
\Ee[\xi_k^2]=\frac{1}{\lambda_u^2} \sum_{j=1}^{k-1}j^2\alpha^k_{k-j}.
\ee
Here the length $\lambda_u$ is defined by $1/\lambda_u^2=\int|\hat{u}(x)|x^2\d x/\|\hat{u}\|_1$.
The condition
\be\label{conditionF}
\sum_{k=1}^\infty \frac{1}{k^2}\Ee[\xi_k^2]
<\infty
\ee
implies that with probability one $\sum_{k=1}^\infty k^{-1}\xi_k$ is convergent and $n^{-1}\sum_{k=1}^n\xi_k\to 0$ [Fe], and is satisfied by (\ref{lambda-u}). So for $~\alpha^0$ defined by Eq.~(\ref{intracycle}) $\|~\alpha^0_n\|\propto n$ and $Y^0_0\to 0$ almost surely. The latter is true also if in (\ref{intracycle})
$k$ is restricted to a positive- or zero-density subset of $\Nn_+$. We shall return to this case later. In these examples $\xi_k$ is a bounded sequence (for $x^k_{k-j,r}\in\supp \hat{u}$). The condition (\ref{conditionF}) may hold also if in (\ref{lambda-u}) $\xi_k$ is unbounded, e.g., if
\be\label{alpha-in}
\alpha^k_{k-j}=\left\{\begin{array}{lll}
a & {\rm if} & j\leq  k^\theta, k=2, 3,\dots\\
0 & {\rm if} & j>k^\theta
\end{array}\right.
\ee
provided that $\theta<1/3$. In this case $\|~\alpha^0_n\|/n$ diverges as $n^{2\theta}$, still $Y^0_0\to 0$ almost surely.

To apply the strong law of large numbers to $Y^+_0$, we extend the summation with respect to $k$ up to infinity and introduce
\[
\eta_j=\sum_{k=1}^\infty\sum_{r=1}^{\alpha^{+k}_j} \left(j-1+t^{+k}_{j,r}-t^{+1}_{1,1}\right)x^{+k}_{j,r},\quad V_n=\frac{1}{n}\sum_{j=1}^n\eta_j.
\]
Then
\[
\Ee[\eta_j^2]=\frac{1}{\lambda_u^2}\sum_{k=1}^\infty\sum_{r=1}^{\alpha^{+k}_j} \left(j-1+t^{+k}_{j,r}-t^{+1}_{1,1}\right)^2,
\]
and
\[
\sum_{j=1}^\infty\frac{1}{j^2}\Ee[\eta_j^2]=\frac{1}{\lambda_u^2}\sum_{k=1}^\infty\sum_{r=1}^{\alpha^{+k}_j}
\frac{ \left(j-1+t^{+k}_{j,r}-t^{+1}_{1,1}\right)^2}{j^2}
\leq \frac{1}{\lambda_u^2}\sum_{k=1}^\infty\alpha^{+k}_j<\infty
\]
if $\alpha^{+k}_j>0$ only for a finite number of pairs $(j,k)$, i.e. if $~\alpha^+$ is rational. This guaranties the sure convergence of $V_n$ and also of $Y^+_0$ to zero, but the result is not new.

In Eq.~(\ref{lambda-u}) we wrote $Y^0_0$ as $1/n$ times the resulting vector of $n$ steps of a random walk. If the step lengths are bounded then the averaged distancing from the origin must decay as $1/\sqrt{n}$. The rest of the proof is devoted to show that in the dominant contribution to the partition function the step lengths are indeed bounded.

\vspace{5pt}
\noindent
(iv) More systematically, we can find irrational sequences giving rise to $\overline{X^0_{^\cdot}}\to 0$ a.s. as follows. Note first that
$\Pp(\overline{X^0_{^\cdot}}\to y)=0$ if $y\neq 0$: $\overline{X^0_{^\cdot}}\to y$ is also a tail event, and because $\Ee[\overline{X^0_{^\cdot}}]=0$, $\Pp(\overline{X^0_{^\cdot}}\to y)= 1$ is impossible .
This implies that for the given $~\alpha$ the asymptotic probability distribution is continuous outside 0, but it is not excluded that also $\Pp(\overline{X^0_{^\cdot}}\to 0)=0$, e.g. $\overline{X^0_{^\cdot}}$ is symmetrically distributed about 0. On the other hand, $\Ee\left[\overline{X^0_{^\cdot}}^2\right]\to 0$ already implies $\Pp(\overline{X^0_{^\cdot}}\to 0)=1$. However, as shown below, arguing with $\Ee\left[\overline{X^0_{^\cdot}}^2\right]\to 0$ will not significantly extend the result of points (i)-(iii).
\[
\Ee\left[\overline{X^0_{^\cdot}}^2\right]
=\Ee\left[\left(Y^0_0\right)^2\right] + \Ee\left[\left(Y^+_0\right)^2\right]
\]
where
\be\label{EY0-EY+}
\Ee\left[\left|Y^0_0\right|^2\right]=\frac{1}{n^2\lambda_u^2} \sum_{k=2}^n \sum_{j=1}^{k-1}j^2\alpha^k_{k-j},\quad
\Ee\left[\left|Y^+_0\right|^2\right]=\frac{1}{n^2\lambda_u^2}\sum_{j=1}^{n}\sum_{k=1}^{{\cal N}(n)-n}\ \sum_{r=1}^{\alpha^{+k}_{j}}\left(j-1+t^{+k}_{j,r}-t^{+1}_{1,1}\right)^2.
\ee
Hence, $\overline{X^0_{^\cdot}}^2$ goes to zero with probability one if
\be\label{E[.]->0}
\frac{1}{n^2}\sum_{k=2}^n \sum_{j=1}^{k-1} j^2 \alpha^k_{k-j}\to 0
\quad\mbox{and}\quad
\frac{1}{n^2} \sum_{k=1}^{{\cal N}(n)-n}\sum_{j=1}^{n}\ \sum_{r=1}^{\alpha^{+k}_{j}}\left(j-1+t^{+k}_{j,r}-t^{+1}_{1,1}\right)^2\to 0.
\ee
Equation (\ref{E[.]->0}) can hold if $\|~\alpha_n\|/n$ does not go to zero or even tends to infinity. For example, the first condition is satisfied for $~\alpha^0$ defined by (\ref{intracycle}) or (\ref{alpha-in}); thus, one can reproduce the result obtained from the strong law of large numbers. With (\ref{intracycle}), $\|~\alpha_n\|/n\sim 1$ but
\[\Ee\left[\left(Y^0_0\right)^2\right]\propto \frac{1}{n^2}\sum_{k=2}^n \sum_{j=1}^{\min\{k-1, j_0\}} j^2 \alpha^k_{k-j}=O\left(1/n\right);\]
with (\ref{alpha-in}), $\|~\alpha_n\|/n\sim n^{2\theta}$
but
\[\Ee\left[\left(Y^0_0\right)^2\right]\propto \frac{1}{n^2}\sum_{k=2}^n \sum_{j=1}^{\min\{k-1, k^\theta\}} j^2 \alpha^k_{k-j}=O\left(\frac{1}{n^{1-3\theta}}\right)\to 0\]
if $\theta<1/3$.

The second condition in (\ref{E[.]->0}) is also fulfilled by some irrational sequences.
Suppose first that ${\cal N}(n)/n^2\to 0$ as $n\to\infty$, for example, $n=\varepsilon N$. If for any $k$
\[
\alpha^{+k}_{j}=\left\{\begin{array}{lll}
a & {\rm if} & j\leq  j_0\\
0 & {\rm if} & j>j_0.
\end{array}\right.
\]
then
\be\label{alpha+k_j}
\frac{1}{n^2} \sum_{k=1}^{{\cal N}(n)-n}\ \sum_{j=1}^{n}\ \sum_{r=1}^{\alpha^{+k}_{j}}\left(j-1+t^{+k}_{j,r}-t^{+1}_{1,1}\right)^2
\leq \frac{j_0(j_0+1)(2j_0+1)a}{6}\ \frac{{\cal N}(n)-n}{n^2}\to 0.
\ee
If ${\cal N}(n)/ n^2$ does not go to zero, let $\{k_i\}$ be a lacunary sequence such that $n^{-2}\sum_{i:\ k_i\leq {\cal N}(n)-n}1\to 0$. Then for
\[
\alpha^{+k}_{j}=\left\{\begin{array}{ll}
a & {\rm if}\ j\leq  j_0,\ k\in\{k_i\}\\
0 & {\rm otherwise}
\end{array}\right.
\]
the condition is satisfied, $Y^+_0\to 0$ with probability one. This is new but irrelevant.
The statistical weight of $~\alpha$ -- rational or irrational, producing $\overline{X^0_{^\cdot}}\to 0$ or not -- is decreased by a factor $\left(L^{-d}\right)^{K_{~\alpha}}$ in which $K_{~\alpha}$ (depending only on $\alpha^k_j$ where $j$ and $k$ are in different cycles) is
at least as large as the number of cycles which the zeroth cycle is coupled with, cf. Proposition~\ref{K=V-1}.
This suggests that the asymptotic contribution of each cycle to $K_{~\alpha}$ is finite; or, in graph language, the infinite graph ${\cal G}_{~\alpha}$ is almost surely of finite degree, whether or not there exist infinite cycles. Later we shall see that more is true: $~\alpha^+$ (and its analogues for cycles $1,\dots,p$) must be rational.

\noindent
(v) $\Ee\left[\overline{(X^0_{^\cdot})^2}\right]\geq \Ee\left[\overline{X^0_{^\cdot}}^2\right]$, therefore $\Ee\left[\overline{(X^0_{^\cdot})^2}\right]\to 0$ implies $\Pp(\overline{X^0_{^\cdot}}\to 0)=1$, but this will not be new.
Substituting $x^{+1}_{1,1}$ from Eq.~(\ref{x+1,11}) into Eq.~(\ref{avZl-2}) and taking the expectation value,
\bea
\Ee\left[\overline{(X^0_{^\cdot})^2}\right]
=
\frac{1}{n\lambda_u^2}\left[ \sum_{k=2}^n\sum_{j=1}^{k-1}j\alpha^k_{k-j}
+ \sum_{k=1}^{{\cal N}(n)-n}\sum_{j=2}^n\ \sum_{r=1}^{\alpha^{+k}_j}\left(j-1+t^{+k}_{j,r}-t^{+1}_{1,1}\right)
\right.
\nonumber\\
\left.
+\sum_{k=1}^{{\cal N}(n)-n}\ \sum_{r=1}^{\alpha^{+k}_1}\left(t^{+k}_{1,r}+t^{+1}_{1,1}-2\min\left\{t^{+k}_{1,r},t^{+1}_{1,1}\right\}\right)\right]
\nonumber\\
=
\frac{1}{n\lambda_u^2}\left[ \sum_{k=2}^n\sum_{j=1}^{k-1}j\alpha^k_{k-j}+ \sum_{k=1}^{{\cal N}(n)-n}\sum_{j=1}^n\ \sum_{r=1}^{\alpha^{+k}_j}\left|j-1+t^{+k}_{j,r}-t^{+1}_{1,1}\right| \right]
\approx
\frac{1}{n\lambda_u^2}\|~\alpha_n\|.
\eea
Thus, $\Ee\left[\overline{(X^0_{^\cdot})^2}\right]\to 0$ is equivalent to (\ref{cond-sure-convergence}) which was the condition for $\overline{X^0_{^\cdot}}\to 0$ surely.

\vspace{3pt}
\noindent
(vi) To summarize, if $~\alpha$ is rational, the result is deterministic, $|\overline{X^0_{^\cdot}}|=O(1/n)$ and $\overline{\left(X^0_{^\cdot}\right)^2}=O(1/n)$. If $~\alpha$ is irrational and $\|~\alpha_n\|/n\to 0$, the result is still deterministic, both $\overline{X^0_{^\cdot}}$ and $\overline{\left(X^0_{^\cdot}\right)^2}$ go surely to zero. We found also other irrational $~\alpha$ with non-decaying or even diverging $\|~\alpha_n\|/n$
that allow for an $\overline{X^0_{^\cdot}}$ almost surely converging to zero. However, as we shall see, while $\|~\alpha_n\|\sim n$ provides the leading-order contribution, no $~\alpha$ with $\|~\alpha_n\|$ diverging faster than $n$ contributes asymptotically to $\nu^{N,L}_n$.

By inspecting
\bea\label{variance}
\lefteqn{
\Ee\left[\overline{\left(X^0_{^\cdot}\right)^2}-\overline{X^0_{^\cdot}}^2\right]
=
\frac{1}{n\lambda_u^2}
\left[\sum_{k=2}^n\sum_{j=1}^{k-1}j\alpha^k_{k-j}\left(1-\frac{j}{n}\right)
\right. }\nonumber\\
&&+
\left.
\sum_{k=1}^{{\cal N}(n)-n}\sum_{j=1}^n\sum_{r=1}^{\alpha^{+k}_j}\left|j-1+t^{+k}_{j,r}-t^{+1}_{1,1}\right|\left(1-\frac{\left|j-1+t^{+k}_{j,r}-t^{+1}_{1,1}\right|}{n}\right)\right]
\eea
one observes that, unless $\alpha^k_{k-j}> 0$ for $k$ and $j$ of order $n$, $\Ee\left[\overline{\left(X^0_{^\cdot}\right)^2}-\overline{X^0_{^\cdot}}^2\right]$ is of the same order as $\Ee\left[\overline{\left(X^0_{^\cdot}\right)^2}\right]$.
The exception is illustrated by the following example. Let $~\alpha^+_n=~0$, $k_0\geq 1$, $\alpha^{k}_{k-j}=1$ for $n-k_0\leq j<k\leq n$ and $\alpha^k_{k-j}=0$ otherwise. Then $\|~\alpha^0_n\|=\sum_{n-k_0\leq j<k\leq n}j$ is of order $n$, so both $\Ee\left[\overline{\left(X^0_{^\cdot}\right)^2}\right]$ and $\Ee\left[\overline{X^0_{^\cdot}}^2\right]$ are of order 1, but their difference computed from the first line of (\ref{variance}) is of order $1/n$. Later we return to this example, here we note only that $~\alpha^0_n$ is exceptional also because it is not the restriction of some infinite $~\alpha^0$, since $~\alpha^0_n\to ~0$ as $n\to\infty$.

Focusing first on the "regular" case $\Ee\left[\overline{\left(X^0_{^\cdot}\right)^2}\right]\leq C\left(\Ee\left[\overline{\left(X^0_{^\cdot}\right)^2}-\overline{X^0_{^\cdot}}^2\right] \right)$ with some $C>1$ we find that
\be\label{typical}
\lim_{n\to\infty}\frac{n}{\|~\alpha_n\|}\Ee\left[\overline{X^0_{^\cdot}}^2\right]
= c
<\frac{1}{\lambda_u^2}=\Ee\left[(x^k_{j,r})^2\right]=\lim_{n\to\infty}\frac{n}{\|~\alpha_n\|}\Ee\left[\overline{\left(X^0_{^\cdot}\right)^2}\right],
\ee
that is,
\be
\Ee\left[\frac{n}{\|~\alpha_n\|}\left[\overline{\left(X^0_{^\cdot}\right)^2}-\overline{X^0_{^\cdot}}^2\right]\right]\to\lambda_u^{-2}-c>0.
\ee

To see the dominant contribution to $\tilde{G}^N_n(0)$ one must estimate the weight of all the $~\alpha^0_n$ and $~\alpha^+_n$ with a given norm. This weight is composed of their number multiplied by $\prod (\beta\|\hat{u}\|_1)^{\alpha^k_j}/\alpha^k_j!$, and can compensate the loss due to a decaying
$\exp\left\{-\pi n \lambda_\beta^2\left[\overline{\left(X^0_{^\cdot}\right)^2}-\overline{X^0_{^\cdot}}^2\right]\right\}$.
Consider first $~\alpha^0_n$. As a typical example, let $\alpha^k_{k-j}=a\geq 1$ for $j=1,\dots,j_{~\alpha^0_n}$ and for $i_{~\alpha^0_n}$ different values of $k$ (each larger than $j_{~\alpha^0_n}$), and $\alpha^k_{k-j}=0$ otherwise. Then
\[
\|~\alpha^0_n\|=\frac{a}{2} i_{~\alpha^0_n}j_{~\alpha^0_n}(j_{~\alpha^0_n}+1)
\]
and the associated weight is
\[
\left[\frac{(\beta\|\hat{u}\|_1)^a}{a!}\right]^{i_{~\alpha^0_n}j_{~\alpha^0_n}}{n-j_{~\alpha^0_n}\choose i_{~\alpha^0_n}}
\]
whose logarithm
\bea\label{weight}
\ln \left\{ \left[\frac{(\beta\|\hat{u}\|_1)^a}{a!}\right]^{i_{~\alpha^0_n}j_{~\alpha^0_n}}{n-j_{~\alpha^0_n}\choose i_{~\alpha^0_n}} \right\}
&\approx&
a\, i_{~\alpha^0_n}j_{~\alpha^0_n}(\ln \beta\|\hat{u}\|_1-\ln a+1)-\frac{1}{2}i_{~\alpha^0_n}j_{~\alpha^0_n}\ln 2\pi a+\ln {n-j_{~\alpha^0_n}\choose i_{~\alpha^0_n}}
\nonumber\\
&=&
\|~\alpha^0_n\| \frac{2}{j_{~\alpha^0_n}+1}\left(\ln\frac{\beta\|\hat{u}\|_1}{a}+1-\frac{\ln 2\pi a}{2a}\right)
+\ln {n-j_{~\alpha^0_n}\choose i_{~\alpha^0_n}}
\eea
is to be added to $-\pi \lambda_\beta^2 n \left[\overline{\left(X^0_{^\cdot}\right)^2}-\overline{X^0_{^\cdot}}^2\right]$.

\vspace{3pt}
\noindent
(vii) We start by proving that if $\Ee\left[\overline{\left(X^0_{^\cdot}\right)^2}-\overline{X^0_{^\cdot}}^2\right]\propto \|~\alpha_n\|/n$
goes to infinity, the contribution to the partition function is negligible.
Assume that $\|~\alpha_n\|\sim \|~\alpha^0_n\|$; we shall see that this is indeed the case, the contribution of $\|~\alpha^+_n\|$ to $\|~\alpha_n\|$ is finite. For a reference we note that adding
$
-\pi n\lambda_\beta^2 \Ee\left[\overline{\left(X^0_{^\cdot}\right)^2}-\overline{X^0_{^\cdot}}^2\right]
$
to (\ref{weight})
yields the averaged exponent
\be\label{E-av}
E_{\rm av}=-\|~\alpha^0_n\|\left[\pi\lambda_\beta^2 \left(\lambda_u^{-2}-c\right)
-\frac{2}{j_{~\alpha^0_n}+1}\left(\ln\frac{\beta\|\hat{u}\|_1}{a}+1-\frac{\ln 2\pi a}{2a}\right)\right]
+\ln {n-j_{~\alpha^0_n}\choose i_{~\alpha^0_n}}.
\ee
Because $i_{~\alpha^0_n}\leq n$, $\|~\alpha^0_n\|/n=a\, i_{~\alpha^0_n}j_{~\alpha^0_n}(j_{~\alpha^0_n}+1)/2n \to\infty$ can occur only if
$j_{~\alpha^0_n}$ increases with $n$ ($a$ should not increase beyond $\beta\|\hat{u}\|_1$ because $(\beta\|\hat{u}\|_1)^a/a!$ is maximal roughly at this value.)
Then the quantity in the square bracket tends to $\pi\lambda_\beta^2 \left(\lambda_u^{-2}-c\right)$; therefore, if $n$ is large enough, $E_{\rm av}$ is a negative multiple of $\|~\alpha^0_n\|$ with a positive
$O(n)=o(\|~\alpha^0_n\|)$ correction coming from the entropy.

Consider now the random exponent
\bea\label{random-E}
E=
-\|~\alpha^0_n\|\left[\pi\lambda_\beta^2 \left(\frac{n}{\|~\alpha^0_n\|}\left[\overline{\left(X^0_{^\cdot}\right)^2}-\overline{X^0_{^\cdot}}^2\right]\right)
-\frac{2}{j_{~\alpha^0_n}+1}\left(\ln\frac{\beta\|\hat{u}\|_1}{a}+1-\frac{\ln 2\pi a}{2a}\right)\right]
+\ln {n-j_{~\alpha^0_n}\choose i_{~\alpha^0_n}}
\nonumber\\
=
-\pi\lambda_\beta^2 n\left[\overline{\left(X^0_{^\cdot}\right)^2}-\overline{X^0_{^\cdot}}^2\right]
+a\, i_{~\alpha^0_n} j_{~\alpha^0_n}\left(\ln\frac{\beta\|\hat{u}\|_1}{a}+1-\frac{\ln 2\pi a}{2a}\right)
+\ln {n-j_{~\alpha^0_n}\choose i_{~\alpha^0_n}}.
\eea
Assume first that there is no concentration of probability at zero, i.e.
\be\label{no-concentration}
\Pp\left(\lim_{n\to\infty}\frac{n}{\|~\alpha_n\|}\left[\overline{\left(X^0_{^\cdot}\right)^2}-\overline{X^0_{^\cdot}}^2\right]>0  \right)=1.
\ee
Then, due to $j_{~\alpha^0_n}\to\infty$, the
quantity
in the square bracket in the middle member of (\ref{random-E}) becomes positive and the random exponent also tends to minus infinity as $n$ increases.
Would (\ref{no-concentration}) fail, $~\alpha$ with $\|~\alpha_n\|/n\to\infty$ would still be
asymptotically
irrelevant. If in Eq.~(\ref{random-E}) $E<0$ or $E=o(n)$ positive, the contribution is negligible compared to that of other $~\alpha$'s yielding $E\propto n$, see below. $E$ cannot be positive and increase faster than $n$ because that would be incompatible with a stable pair potential.
It remains the possibility that $i_{~\alpha^0_n} j_{~\alpha^0_n}\sim n$, but then $i_{~\alpha^0_n}=o(n)$ and therefore the entropy is also $o(n)$. If $\overline{\left(X^0_{^\cdot}\right)^2}-\overline{X^0_{^\cdot}}^2$ could be of order one with a non-vanishing probability then for $\beta$ large enough $E$ would tend to minus infinity. At last, suppose that the only term of order $n$ is the middle one. This is independent of the variables $\{x^k_{j,r},\,x^{+k}_{j,r}\}$ and, hence, of the pair interactions. With the choice $a=\beta\|\hat{u}\|_1$ and supposing $n\propto N$, the only relevant case (or counting with a similar contribution from the other cycles), a multiple of $\|\hat{u}\|_1=u(0)$ would be subtracted from the free energy per particle and, in effect, from the ground state energy per particle. This, however, is in contradiction with some earlier results [Li7], [Su5] saying that for positive type pair potentials at asymptotically high densities the ground state energy per particle is $\rho\hat{u}(0)/2$. This is exactly the quantity in the exponential factor that appears in $G[n, \{n_l\}_1^p](0)$ and in the partition function. $\rho\hat{u}(0)/2$ is the mean-field result, at $\beta<\infty$ it overshoots the real value, and the negative correction to it should depend on the temperature. We conclude that {\em the contribution of all the $~\alpha_n$ whose norm diverges faster than $n$ becomes asymptotically negligible compared to that coming from $\|~\alpha_n\|=O(n)$}.

\vspace{3pt}
\noindent
(viii) On physical grounds we expect that the leading-order $e^{O(n)}$ factor by which a cycle of length $n$ contributes to the partition function is actually of order $e^{cn}$ with some $c=c(\beta)>0$. Before proving this let us recall that previously we have already met terms of order $e^{o(n)}$.
A rational $~\alpha$ leads to $n\left[\overline{\left(X^0_{^\cdot}\right)^2}-\overline{X^0_{^\cdot}}^2\right]=O(1)$ and at the same time the entropy is of order $\ln n$, so altogether we have a positive exponent of order $\ln n$ for such terms. Terms with $~\alpha$ irrational but $\|~\alpha_n\|=o(n)$ also produce an entropy-dominated factor $e^{o(n)}>1$. Let $g\geq 1$ be any monotone increasing function that tends to infinity and let $K\subset\Nn_+$ be a lacunary sequence satisfying
\[
\lim_{x\to \infty}\frac{\#\{k\in K|k\leq x\}}{x/g(x)}=1.
\]
Define
\[
\alpha^k_{k-j}=\left\{\begin{array}{ll}
a & {\rm if}\ j\leq j_0, k\in K\\
0 & {\rm otherwise}
\end{array}\right.
\]
Then
\be\label{relevant}
\|~\alpha^0_n\|\propto i_{~\alpha^0_n}\sim \frac{n}{g(n)},\quad \ln {n-j_{~\alpha^0_n}\choose i_{~\alpha^0_n}}\sim \frac{n}{g(n)}(\ln g(n)+1),
\ee
i.e. the entropy wins. All these terms become asymptotically independent of $\beta$ and will not appear in the limiting free energy density.

To show that the leading-order contribution is due to $\|~\alpha_n\|\propto n$ we apply the Markov inequality to $\frac{n}{\|~\alpha_n\|}\left[\overline{\left(X^0_{^\cdot}\right)^2}-\overline{X^0_{^\cdot}}^2\right]$. Choose some $A>1$, then
\[
\lim_{n\to\infty}\Pp\left(\frac{n}{\|~\alpha_n\|}\left[\overline{\left(X^0_{^\cdot}\right)^2}-\overline{X^0_{^\cdot}}^2\right]
< A (\lambda_u^{-2}-c) \right)\geq 1-\frac{1}{A}.
\]
Thus, for large enough systems
\be\label{exponent}
E
\gtrsim
\|~\alpha^0_n\|
\left[
-A \pi\lambda_\beta^2(\lambda_u^{-2}-c)
+ \frac{2}{j_{~\alpha^0_n}+1}\left(\ln\frac{\beta\|\hat{u}\|_1}{a}+1-\frac{\ln 2\pi a}{2a}\right) \right]
+\ln {n-j_{~\alpha^0_n}\choose i_{~\alpha^0_n}}
\ee
with a probability not smaller than $1-1/A$. We want $E$ to be positive.
A typical example for $\|~\alpha^0_n\|\propto n$ is (\ref{intracycle}); in this case $c=0$ and $Y^0_0$ tends to zero with probability one. However, for $\beta$ large enough the first term on the right-hand side of (\ref{exponent}) is negative, to keep it smaller in modulus than the entropy the prefactor of $n$ in $\|~\alpha^0_n\|$ should decrease as $\beta$ increases. Let $i_{~\alpha^0_n}=\epsilon n$, then the entropy, computed with $j_{~\alpha^0_n}=O(1)$ is
\be
\ln{n-j_{~\alpha^0_n}\choose \epsilon n}\approx-n[\epsilon\ln\epsilon+(1-\epsilon)\ln(1-\epsilon)]
=
\epsilon n\ln \left[\epsilon^{-1}\left(1-\epsilon\right)^{-\frac{1-\epsilon}{\epsilon}} \right].
\ee
The best chance for the exponent (\ref{exponent}) to be positive is if $\|~\alpha^0_n\|$ is minimal under the condition that $i_{~\alpha^0_n}=\epsilon n$. The minimum is attained with $\alpha^k_{k-1}=1$ if $k\in K$ where $K\subset\{2,\dots,n\}$, $|K|=\epsilon n$, and $\alpha^k_{k-j}=0$ otherwise, resulting $\|~\alpha^0_n\|=\epsilon n$. The exponent is then
\be\label{energy+entropy}
\epsilon\left[ \ln \left(\epsilon^{-1}\left(1-\epsilon\right)^{-\frac{1-\epsilon}{\epsilon}} \right)
-A\pi(\lambda_\beta/\lambda_u)^2 +\ln\frac{e\beta\|\hat{u}\|_1}{\sqrt{2\pi}}  \right]n,
\ee
which is positive if
\be\label{bound-epsilon}
\epsilon(1-\epsilon)^{\frac{1-\epsilon}{\epsilon}}<e^{-A\pi(\lambda_\beta/\lambda_u)^2}.
\ee
Using $\frac{1-\epsilon}{\epsilon}\ln(1-\epsilon)=\sum_{k=1}^\infty\frac{\epsilon^k}{k(k+1)}-1$ it is seen that the left side of this inequality is a monotone increasing function of $\epsilon$ with the bounds
\[
\epsilon/e\leq \epsilon(1-\epsilon)^{\frac{1-\epsilon}{\epsilon}}\leq \epsilon.
\]
It follows that (\ref{bound-epsilon}) holds if $\epsilon<e^{-A\pi(\lambda_\beta/\lambda_u)^2}$ and fails if $\epsilon>e^{-A\pi(\lambda_\beta/\lambda_u)^2+1}$.

We mentioned already an exceptional occurrence of $\|~\alpha^0_n\|\sim n$ when  Eq.~(\ref{typical}) fails, the difference (\ref{variance}) is $O(1/n)$ while the separate terms are of order one. We return to it with the simplest example, $\alpha^n_1=1$ and $\alpha^k_{j}=0$ otherwise. Then $\|~\alpha^0_n\|=n-1$, $\Ee[(Y^0_0)^2]=\lambda_u^{-2} (n-1)^2/n^2$. Moreover, (keeping $~\alpha^+=0$) $\Ee\left[\,\overline{\left(X^0_{^\cdot}\right)^2}\,\right]=\lambda_u^{-2}(n-1)/n$ and thus $\Ee\left[\overline{\left(X^0_{^\cdot}\right)^2}-\overline{X^0_{^\cdot}}^2\right]=\lambda_u^{-2}(n-1)/n^2$. This is exactly the same value as the one we get if $\alpha^k_{k-1}=1$ for a single $k\geq 2$ and $\alpha^{k'}_{k'-j'}=0$ otherwise.  A similar agreement can be found between any rational sequence and a properly chosen $~\alpha^0_n$ with a finite number of nonzero $\alpha^k_{k-j}$ where all the $k$ increase with $n$ (so $~\alpha^0_n\to~0$) but all the $j$ are of order 1. This is because $\|~\alpha^0_n\|$ does not depend on the value of $k$, only on the number of those for which $\alpha^k_{k-j}>0$.
The associated entropy is also the same as that of rational sequences, of the order of $\ln n$. Therefore either the exponent is negative or it is positive and of order $\ln n$ with the conclusion that the overall contribution is asymptotically negligible.

\vspace{4pt}
\noindent
(ix) Concerning $Y^+_0$ and $~\alpha^+_n$, a non-negligible contribution may come from here. As a typical example,
assume that $j_{~\alpha^+_{n}}$ particles of cycle 0 are coupled with $i_{~\alpha^+_{n}}$ particles of cycle 1, so the number of nonzero $\alpha^{+k}_j$ is $i_{~\alpha^+_{n}}j_{~\alpha^+_{n}}$.
Suppose also that $n_1$ tends to infinity. The same estimations can be made as for $~\alpha^0$. Keeping $j_{~\alpha^+_{n}}$ bounded, the largest contribution is for $i_{~\alpha^+_{n}}=\epsilon n_1$ with $\epsilon<e^{-(\pi/\gamma)(\lambda_\beta/\lambda_u)^2}$.
However, computing the analogue of the exponent (\ref{random-E}) for cycle 1 we find it tending to $-\infty$. The reason is that
viewed from cycle 1 the roles of $j$ and $k$ are interchanged, $\alpha^{+k}_j>0$ for $\epsilon n_1$ different $k$ has the same effect on cycle 1
as $\alpha^{+k}_j>0$ for $\epsilon n$ different $j$ has on cycle 0: the sum to consider,
$\sum_{j=1}^n\sum_{k=1}^{n_1}k\,\alpha^{+k}_j$ diverges as $n_1^2$ while the maximum entropy associated with cycle 1 can increase only as $n_1$.
The conclusion is that $\|~\alpha^+_{n}\|$ must remain bounded.
{\em Thus, the leading contribution of cycle 0 to $\tilde{G}^N_n(0)$ is a factor
$\sim \exp\left\{c e^{-(\pi/\gamma) (\lambda_\beta/\lambda_u)^2}n \right\}$ with some $0<\gamma<1$,
coming from $~\alpha=\{~\alpha^0, ~\alpha^+\}$, where $~\alpha^0$ is irrational with $\|~\alpha^0_n\| \propto e^{-(\pi/\gamma)(\lambda_\beta/\lambda_u)^2}n$ and $~\alpha^+$ is rational, i.e. $\|~\alpha^+\|<\infty$}. For the exponential increase in $n$ the non-decay of $f_n(0;L\overline{X^0_{^\cdot}})$ is also necessary. If $n\lambda_\beta^2/L^2$ is bounded, this holds true irrespective of the value of $\overline{X^0_{^\cdot}}$, cf. (\ref{f_n(0)-asymp}); if $n\lambda_\beta^2/L^2$ tends to infinity, this is a consequence of $|\overline{X^0_{^\cdot}}|=O(1/\sqrt{n})$ to be shown below.

Although the role of the coupling of any individual cycle of a diverging length to the other cycles appears to be negligible, for the whole system the coupling among cycles results in a decrease of the free energy density, see the proof of Theorem~\ref{second-step}. This is because for $\beta<\infty$ the partitions of $N$ to $p\propto N$ elements dominate the partition function: single particles and particles forming finite cycles carry the uncondensed density which is positive at positive temperatures.
It will be seen that for any $\beta$ and $\rho$ the dominant contribution to the partition function comes from $p\propto N$ and $~\alpha$ chosen so that  in the graph ${\cal G}_{~\alpha}$ a macroscopic number of cycles occur in couplings.

\vspace{5pt}
\noindent
(x) In all the cases of $\|~\alpha_n\|/n=O(1)$ when $~\alpha^0_n$ is the restriction of an infinite $~\alpha^0$ to $\{\alpha^k_{k-j}|1\leq j<k\leq n\}$ and $~\alpha^+$ is rational $\overline{X^0_{^\cdot}}\to 0$ almost surely. To establish the rate of decay
the simplest result is obtained by using Chebyshev's inequality: for any $A>1$
\be\label{P-X0}
\Pp\left(|\overline{X^0_{^\cdot}}|< A \sqrt{\Ee\left[\overline{X^0_{^\cdot}}^2\right]}  \right) >1-1/A^2.
\ee
A bound on the standard deviation can be inferred from a comparison of Eqs.~(\ref{alpha-norms}) and (\ref{EY0-EY+}). Setting $\alpha^k_{k-j}=0$ for $j>j_{~\alpha^0_n}$ and $\alpha^{+k}_j=0$ for $j>j_{~\alpha^+_n}$ with $j_{~\alpha^0_n},\, j_{~\alpha^+_n}=O(1)$ justified by the previous discussion, we find
\be\label{E-X0-square}
\sqrt{\Ee\left[\overline{X^0_{^\cdot}}^2\right]}
\leq
\frac{1}{n\lambda_u}\sqrt{j_{~\alpha^0_n}\|~\alpha^0_n\|+j_{~\alpha^+_n}\|~\alpha^+_n\|}
\leq \frac{c_1 e^{-c_2(\lambda_\beta/\lambda_u)^2}}{\lambda_u\sqrt{n}}.
\ee

\vspace{5pt}
\noindent
(xi) The sign oscillation seen in $\tilde{G}\left[n,\{n_l\}_1^p\right]$ is to a large extent due to the stability of the interaction. In the Fourier-representation we use here instability shows up in $\int \hat{u}(x)\d x=u(0)<0$. In the extreme case when $\hat{u}\leq 0$ all the terms of $\tilde{G}\left[n,\{n_l\}_1^p\right]$ are positive. The positive-type $u$ considered in this paper represents the opposite extremity: $\hat{u}\geq 0$ defines a stable interaction, and the signs alternate according to the parity of the sum  $\sum_{j<k}\alpha^k_j$. In the finite-volume canonical ensemble the difference between stable and unstable interactions is not striking. Stability is in the background, it guaranties the existence of the thermodynamic limit of the free energy density and, more generally, of the Gibbs measure, missing in unstable systems. $\tilde{G}\left[n,\{n_l\}_1^p\right]>0$, so the negative terms cannot flip the overall sign.
Because the variables occurring in terms of different signs are partly independent, exact cancellation between positive and negative terms is an event of zero probability, and therefore the order-of-magnitude estimates are not altered by them.

To summarize, $\nu^{N,L}_n(y)$ is asymptotically concentrated to an $O(1/\sqrt{n})$ neighborhood of the origin, confirming thereby the lower bound (\ref{lower-bound-to-sigma1}).
This outcome is physically satisfying: if and only if $\|~\alpha_n\|\propto n$ (and then $|\overline{X^0_{^\cdot}}|\propto 1/\sqrt{n}$)  the contributions of cycle 0 to the energy and the entropy are of the same order, and this is the only case when the temperature has an influence on what the cycle adds to the free energy density.

\subsection{Condition for BEC}

Here we do not use the asymptotic form of $f(x;L\overline{X^0_{^\cdot}})$, because the exact expressions are directly obtained:
\bea\label{L-dfn(0)}
f_n\left(0;L\overline{X^0_{^\cdot}}\right)
&=&
\sum_{z\in\Zz^d}\exp\left\{-\frac{\pi n \lambda_\beta^2}{L^2}\left(z+ L\overline{X^0_{^\cdot}}\right)^2\right\}
\nonumber\\
&=&
\exp\left\{-\pi n \lambda_\beta^2 \overline{X^0_{^\cdot}}^2\right\}
\sum_{z\in\Zz^d}\exp\left\{-\frac{\pi n \lambda_\beta^2}{L^2}\left(z^2+2z\cdot L\overline{X^0_{^\cdot}}\right)\right\}
\eea
and
\be\label{int-f}
\int_\Lambda f_n\left(x;L\overline{X^0_{^\cdot}}\right)\d x
=L^d \exp\left\{-\pi n \lambda_\beta^2 \overline{X^0_{^\cdot}}^2\right\}
=\frac{L^d f_n\left(0;L\overline{X^0_{^\cdot}}\right)}{\sum_{z\in\Zz^d}\exp\left\{-\frac{\pi n \lambda_\beta^2}{L^2}\left(z^2+2z\cdot L\overline{X^0_{^\cdot}}\right)\right\}}.
\ee
Therefore
\bea\label{intFn(x)-asymp}
\lefteqn{
\int_\Lambda \tilde{G}\left[n_0,\{n_l\}_1^p\right](x)\d x  =        }
\nonumber\\
&&L^d \sum_{\{\alpha^k_j\in\Nn_0|1\leq j<k\leq N\}}\ \Delta_{\{\alpha^k_j\},\{n_l\}_0^p} \left(L^{-d}\right)^{K_{\{\alpha^k_j\}}}
\prod_{1\leq j<k\leq N}
\frac{\left(-\beta\right)^{\alpha^k_j}}{\alpha^k_j !} \prod_{r=1}^{\alpha^k_j}
\int\d x^k_{j,r}\ \hat{u}\left(x^k_{j,r}\right) \int_0^1\d t^k_{j,r}
\nonumber\\
&&\hspace{1cm}\left[
\delta(X^0_1,\dots,X^p_1) \prod_{l=0}^p
\exp\left\{-\pi n_l \lambda_\beta^2\left[\overline{\left(X^l_{^\cdot}\right)^2}-\overline{X^l_{^\cdot}}^2\right]\right\}
f_{n_l}\left(0;L\overline{X^l_{^\cdot}}\right)
\right.
\nonumber\\
&&\hspace{3cm}\left.
\times\frac{1}{\sum_{z\in\Zz^d}\exp\left\{-\frac{\pi n_0 \lambda_\beta^2}{L^2}\left(z^2+2z\cdot L\overline{X^0_{^\cdot}}\right)\right\}}
\right]\,.
\nonumber\\
\eea
The right-hand side without the last fraction is $L^d\,\tilde{G}\left[n_0,\{n_l\}_1^p\right](0)$. Thus, it is the asymptotic behavior of this fraction that decides about BEC. If $n_0=o(N^{2/d})$, the denominator tends to infinity irrespective of the value of $\overline{X^0_{^\cdot}}$: if $L\overline{X^0_{^\cdot}}$ remains bounded or increases slower than $N^{2/d}/n_0$ then each term tends to 1; otherwise an increasing number of terms with $z\cdot L\overline{X^0_{^\cdot}}<0$ will exceed 1. As a result, such cycles do not add to the condensate. Therefore, we focus on cycles of length $n\geq cN^{2/d}$ where $c>0$.
With the definitions (\ref{HN-n}),~(\ref{Theta})
\bea
\lefteqn{
\int_\Lambda \tilde{G}^N_n(x)\d x
=
L^d\sum_{\{\alpha^k_j\in\Nn_0|1\leq j\leq n,\, j+1\leq k\leq N\}} \left(\prod_{ j=1}^n\prod_{k=j+1}^N\frac{\left(-\beta\right)^{\alpha^k_j}}{\alpha^k_j !}\right)
\int_0^1 \prod_{ j=1}^n\prod_{k=j+1}^N\prod_{r=1}^{\alpha^k_j}\d t^k_{j,r}     }
\nonumber\\
&&\int \prod_{j=1}^n\prod_{k=j+1}^N \prod_{r=1}^{\alpha^k_j}\d x^k_{j,r}\hat{u}\left(x^k_{j,r}\right)
H_{N-n}\
\frac{\exp\left\{-\pi n \lambda_\beta^2\left[\overline{\left(X^0_{^\cdot}\right)^2}-\overline{X^0_{^\cdot}}^2\right]\right\}
f_n\left(0;L\overline{X^0_{^\cdot}}\right)}{\sum_{z\in\Zz^d}\exp\left\{-\frac{\pi n \lambda_\beta^2}{L^2}\left(z^2+2z\cdot L\overline{X^0_{^\cdot}}\right)\right\}}
\nonumber\\
&&=
L^d\int_{\Rr^d}\d y \frac{\Theta^N_n(y)}{\sum_{z\in\Zz^d}\exp\left\{-\frac{\pi n \lambda_\beta^2}{L^2}\left(z^2+2z\cdot Ly\right)\right\}}.
\eea
Dividing with $L^d\tilde{G}^N_n(0)=L^d\int\Theta^N_n(y')\d y'$
\be
\rho^{N,L}_0
\geq
\sum_{n\geq cN^{2/d}}\rho^{N,L}_n\, \frac{\int_\Lambda \tilde{G}^N_n(x)\d x}{L^d\tilde{G}^N_n(0)}
=
\sum_{n\geq cN^{2/d}}\rho^{N,L}_n\,\int_{\Rr^d} \frac{\nu^{N,L}_n(y)\,\d y}{\sum_{z\in\Zz^d}\exp\left\{-\frac{\pi n \lambda_\beta^2}{L^2}z\cdot\left(z+2Ly\right)\right\}}.
\ee
The result (\ref{cond2}) for BEC follows from the fact that asymptotically $\nu^{N,L}_n$ is concentrated to an $O(1/\sqrt{n})$ neighborhood of the origin. For $n\propto N^{2/d}$ this means that $L|y|=O(1)$, and if $n/N^{2/d}\to\infty$ then $L|y|=o(1)$; thus, for any $n\geq cN^{2/d}$ the sum over $\Zz^d$ remains finite in the thermodynamic limit.

This ends the proof for pair potentials with $\hat{u}\geq 0$ of compact support. The latter condition can be removed preserving $\int\hat{u}(x)x^2\d x<\infty$, with the only effect that the sure convergence to zero is changed into almost sure one.

\newpage
\newsec{Proof of Lemma~\ref{bounds-to-free-energy}}

The lower bound is trivial except for the multipliers of $\rho$ and $\rho^2$. It follows from superstability: there are positive constants $B$, $C$ such that for $L$ sufficiently large and any $N$
\[
\sum_{1\leq j<k\leq N}u_L(x_k-x_j)\geq -BN+CN(N-1)/L^d.
\]
The largest $C$ is
\be\label{C_Lambda[u]}
C_\Lambda[u]=\liminf_{N\to\infty}\frac{L^d}{N(N-1)} \sum_{1\leq j<k\leq N}u_L(x_k-x_j)
\ee
that we called in [Su5] the best superstability constant (the Fekete constant in potential theory [Ch]). In general $C_\Lambda[u]\leq \hat{u}(0)/2$, but in Section 9 of [Su5] it was shown that for $u$ of the positive type $C_\Lambda[u]=\hat{u}(0)/2$ independently of $L$.
Even if $u\geq 0$, $B>0$ must be chosen, otherwise the inequality could fail for small $N$, e.g. if $u$ is of finite range. Also, for $N$ large the quadratic term alone with $C=C_\Lambda[u]$ can be too large. If $u$ is bounded, $C=C_\Lambda[u]$ and $B=u_L(0)/2$ provide a valid lower bound.

For any partition of $N$
\[
\sum_{l=0}^p U(\omega_l)+\sum_{0\leq l'<l\leq p}U(\omega_{l'},\omega_l)
\geq - BN+C_\Lambda[u]  N(N-1)/L^d,
\]
hence
\begin{eqnarray*}
\lefteqn{
Q_{N,L}
\leq e^{-\beta\left[- BN+C_\Lambda[u] N(N-1)/L^d\right]}
\frac{1}{N}\sum_{n=1}^N \int_\Lambda\d x \int W^{n\beta}_{xx}(\d\omega_0)
}\nonumber\\
&&\sum_{p=1}^{N-n}\frac{1}{p!}\sum_{n_1,\dots,n_p\geq 1:\sum n_l= N-n}\prod_{l=1}^p\frac{1}{n_l} \int_\Lambda\d x_l \int W^{n_l\beta}_{x_lx_l}(\d\omega_l)
= e^{-\beta\left[- BN+C_\Lambda[u]  N(N-1)/L^d\right]} Q^0_{N,L}
\end{eqnarray*}
from which the lower bound for $f(\rho,\beta)$ follows with $u_L\to u$ as $L\to\infty$ and
\be\label{C[u]}
C[u]=\liminf_{L\to\infty}C_\Lambda[u].
\ee

The upper bound in (\ref{FNL-upper-lower}) and (\ref{f-upper-lower}) is based on Jensen's inequality. First, on the torus $Q_{N-n,L}(\omega_0+y)=Q_{N-n,L}(\omega_0)$, because any shift of $\omega_0$ can be defused by the same shift of the integration variables $x_1,\dots,x_p$ and hence of $\omega_1,\dots,\omega_p$; therefore
\[
Q_{N-n,L}(\omega_0)=\frac{1}{L^d}\int_\Lambda Q_{N-n,L}(\omega_0+y) \d y.
\]
The dependence on $\omega_0$ is only in the last exponential factor of $Q_{N-n,L}(\omega_0)$. Applying Jensen's inequality and
\[
\int_\Lambda u_L(y)\d y=\int u(y)\d y \leq \|u\|_1,
\]
\[
\frac{1}{L^d}\int_\Lambda e^{-\beta \sum_{l=1}^p U(\omega_0+y,\omega_l)} \d y
\geq e^{-\beta \frac{1}{L^d}\int_\Lambda \sum_{l=1}^p U(\omega_0+y,\omega_l) \d y}
= e^{-\beta \frac{n(N-n)}{L^d}\int_\Lambda u_L(y)\d y}
\geq e^{-\beta \frac{n(N-n)}{L^d} \|u\|_1}.
\]
Thus, for any $\omega_0$
\be
Q_{N-n,L}(\omega_0)\geq e^{-\beta \frac{n(N-n)}{L^d} \|u\|_1} Q_{N-n,L}
\ee
and consequently
\be\label{QNL-lower}
Q_{N,L}\geq \frac{1}{N}\sum_{n=1}^N  \exp\left\{-\beta \|u\|_1\frac{n(N-n)}{L^d} \right\}  Q_{N-n,L}\,
\int_\Lambda\d x \int W^{n\beta}_{xx}(\d\omega_0)e^{-\beta U(\omega_0)}.
\ee
Recall from Eqs.~(\ref{norms-P-W})-(\ref{open-trajectory}) the elementary properties of  the conditional Wiener measures $P^\beta_{xy}(\d\omega)$ and $W^\beta_{xy}(\d\omega)$. We will supplement them with two more.
Let
$
\psi_t(x)=\lambda_t^{-d}e^{-\pi x^2/\lambda_t^2},
$
then
\bea\label{norms}
\int P^\beta_{xy}(\d\omega)
=\psi_\beta(y-x).
\eea
Throughout the proof we use the following identity.
If $f(\omega)=f(\omega(t_0))$ with $0<t_0<\beta$, then
\be\label{one-point}
\int P^\beta_{xy}(\d\omega)f(\omega)=\int\d x'\psi_{t_0}(x-x')f(x')\psi_{\beta-t_0}(x'-y).
\ee
By Jensen's inequality,
\bea\label{Jensen}
\int_\Lambda \d x \int W^{n\beta}_{xx}(\d\omega) e^{-\beta U(\omega)} = L^d\int W^{n\beta}_{00}(\d\omega) e^{-\beta U(\omega)}
=L^d \sum_{z\in\Zz^d}\int P^{n\beta}_{0,Lz}(\d\omega) e^{-\beta U(\omega)}
\nonumber\\
\geq
L^d\sum_{z\in\Zz^d}\left(\int P^{n\beta}_{0,Lz}(\d\omega)\right)\exp\left\{-\beta
\frac{\int P^{n\beta}_{0,Lz}(\d\omega)U(\omega)}{ \int P^{n\beta}_{0,Lz}(\d\omega)}\right\}
\nonumber\\
=
\frac{L^d}{n^{d/2}\lambda_{\beta}^d}\sum_{z\in\Zz^d}\exp\left\{-\frac{\pi L^2 z^2}{n\lambda_\beta^2}\right\}
\exp\left\{-\beta\langle U\rangle_{P^{n\beta}_{0,Lz}}\right\}.
\eea
We turn to $\langle U\rangle_{P^{n\beta}_{0,Lz}}$.
With a slight extension of (\ref{one-point})
one can show that equal-time increments have the same distribution. Let $0<t_1<t_2<\beta$, $\omega(0)=x$, $\omega(\beta)=y$, and consider any $f$ depending only on $\omega(t_2)-\omega(t_1)$. Then
\begin{eqnarray}
\int P^\beta_{xy}(\d\omega)f(\omega(t_2)-\omega(t_1))
&=&\int\d x_1\d x_2\ \psi_{t_1}(x_1-x)\psi_{t_2-t_1}(x_2-x_1)\psi_{\beta-t_2}(y-x_2)f(x_2-x_1)
\nonumber\\
&=&\int\d z\ \psi_{t_2-t_1}(z)f(z)\int\d x_1\ \psi_{t_1}(x_1-x)\psi_{\beta-t_2}(y-z-x_1)
\nonumber\\
&=&\int\d z\ \psi_{t_2-t_1}(z)f(z)\psi_{\beta-(t_2-t_1)}(y-x-z)
=\int P^\beta_{0,y-x}(\d\omega)f(\omega(t_2-t_1)).
\nonumber\\
\end{eqnarray}
Thus,
\be
\int P^{n\beta}_{0,Lz}(\d\omega) u_L(\omega(k\beta+t)-\omega(j\beta+t))
=\int P^{n\beta}_{0,Lz}(\d\omega) u_L(\omega((k-j)\beta))
\ee
and
\be\label{U-average-two-forms}
\langle U\rangle_{P^{n\beta}_{0,Lz}}
=\sum_{0\leq j<k\leq n-1}\langle u_L(\omega((k-j)\beta))\rangle_{P^{n\beta}_{0,Lz}}
=\sum_{k=1}^{n-1}(n-k)\langle u_L(\omega(k\beta))\rangle_{P^{n\beta}_{0,Lz}}
=\sum_{k=1}^{n-1}k\langle u_L(\omega((n-k)\beta))\rangle_{P^{n\beta}_{0,Lz}}.
\ee
For any $z\in\Zz^d$,
$
u_L(x)=u_L(Lz-x),
$
from which with (\ref{one-point}) one obtains
\be
 \int P^{n\beta}_{0,Lz}(\d\omega) u_L(\omega((n-k)\beta))
=\int P^{n\beta}_{0,Lz}(\d\omega)  u_L(\omega(k\beta)).
\ee
Summing the two forms (\ref{U-average-two-forms}) of $\langle U\rangle_{P^{n\beta}_{0,Lz}}$ and using the above equality,
\be\label{sum-of-two-forms}
\langle U\rangle_{P^{n\beta}_{0,Lz}}=\frac{n}{2}\sum_{k=1}^{n-1}\langle u_L(\omega(k\beta))\rangle_{P^{n\beta}_{0,Lz}}.
\ee
Next,
\begin{eqnarray}
\langle u_L(\omega(k\beta))\rangle_{P^{n\beta}_{0,Lz}}
&=&\left(\int P^{n\beta}_{0,Lz}(\d\omega) \right)^{-1}\int P^{n\beta}_{0,Lz}(\d\omega) u_L(\omega(k\beta))
\nonumber\\
&=&\psi_{n\beta}(Lz)^{-1}\int  \psi_{k\beta}(x)\psi_{(n-k)\beta}(Lz-x) u_L(x) \d x.
\end{eqnarray}
A straightforward calculation yields
\be
\psi_{n\beta}(Lz)^{-1} \psi_{k\beta}(x)\psi_{(n-k)\beta}(Lz-x)
=\alpha_{n,k}^{d/2}\exp\{-\pi\alpha_{n,k}(x-Lkz/n)^2\}
\ee
and therefore
\be\label{av-u_L}
\langle u_L(\omega(k\beta))\rangle_{P^{n\beta}_{0,Lz}}
=\alpha_{n,k}^{d/2}\int u_L(x)\exp\left\{-\pi\alpha_{n,k}(x-Lkz/n)^2\right\}\d x
\ee
where
\be
\alpha_{n,k}=\left(\frac{1}{k}+\frac{1}{n-k}\right)\frac{1}{\lambda_\beta^2}.
\ee
Substituting $u_L(x)=\sum_{v\in\Zz^d}u(x-Lv)$ into (\ref{av-u_L}),
\begin{eqnarray}
\langle u_L(\omega(k\beta))\rangle_{P^{n\beta}_{0,Lz}}
&=&
\alpha_{n,k}^{d/2}\sum_{v\in\Zz^d}\int u(x-Lv)\exp\left\{-\pi\alpha_{n,k}(x-Lkz/n)^2\right\}\d x
\nonumber\\
&=&
\int u(y)\ \alpha_{n,k}^{d/2}\sum_{v\in\Zz^d}\exp\left\{-\pi\alpha_{n,k}(y-L\{kz/n\}+Lv)^2\right\}\d y
\end{eqnarray}
where $\{kz/n\}$ is the fractional part of $kz/n$ for $z\in\Zz^d$, each component of which is bounded in modulus by $1/2$.
Applying twice the Poisson summation formula
\begin{eqnarray}
\alpha_{n,k}^{d/2}\sum_{v\in\Zz^d}\exp\left\{-\pi\alpha_{n,k}(y-L\{kz/n\}+Lv)^2\right\}
=L^{-d}\sum_{v\in\Zz^d}\exp\left\{-\frac{\pi v^2}{\alpha_{n,k}L^2}+\i \frac{2\pi}{L}(y-L\{kz/n\})\cdot v\right\}
\nonumber\\
\leq L^{-d}\sum_{v\in\Zz^d}\exp\left\{-\frac{\pi v^2}{\alpha_{n,k}L^2}\right\}
=\alpha_{n,k}^{d/2}\sum_{v\in\Zz^d}\exp\left\{-\pi\alpha_{n,k}L^2v^2\right\}.
\end{eqnarray}
So
\be
\langle u_L(\omega(k\beta))\rangle_{P^{n\beta}_{0,Lz}}
\leq \|u\|_1\alpha_{n,k}^{d/2}\sum_{v\in\Zz^d}\exp\left\{-\pi\alpha_{n,k}L^2v^2\right\},
\ee
the $z$-dependence dropped from the upper bound.
Now
\begin{eqnarray}
\sum_{v\in\Zz^d}e^{-\pi\alpha_{n,k}L^2v^2}=\left(1+2\sum_{v=1}^\infty e^{-\pi\alpha_{n,k}L^2v^2}\right)^d
\leq \left(1+2\int_0^\infty e^{-\pi\alpha_{n,k}L^2v^2}\d v  \right)^d
=\left(1+\frac{1}{L\sqrt{\alpha_{n,k}}}\right)^d.
\end{eqnarray}
Inserting this into the expression for $\langle U\rangle_{P^{n\beta}_{0,Lz}}$,
\begin{eqnarray}
\langle U\rangle_{P^{n\beta}_{0,Lz}}
&\leq&\frac{n}{2} \|u\|_1\sum_{k=1}^{n-1}\alpha_{n,k}^{d/2}\sum_{v\in\Zz^d}\exp\left\{-\pi\alpha_{n,k}L^2v^2\right\}
\leq \frac{n}{2} \|u\|_1\sum_{k=1}^{n-1}\alpha_{n,k}^{d/2}\left(1+\frac{1}{L\sqrt{\alpha_{n,k}}}\right)^d
\nonumber\\
&=&\frac{n}{2} \|u\|_1\sum_{l=0}^d {d\choose l}\frac{1}{L^{d-l}}\sum_{k=1}^{n-1}\alpha_{n,k}^{l/2}
=\frac{n}{2} \|u\|_1 \left[\frac{n-1}{L^d}+\sum_{l=1}^{d-1}{d\choose l}\frac{1}{L^{d-l}}\sum_{k=1}^{n-1}\alpha_{n,k}^{l/2}+\sum_{k=1}^{n-1}\alpha_{n,k}^{d/2}\right].
\nonumber\\
\end{eqnarray}
For $d\geq 3$ and $1\leq l\leq d-1$
\be
\frac{1}{L^{d-l}}\sum_{k=1}^{n-1}\alpha_{n,k}^{l/2}
= \frac{1}{\lambda_\beta^l L^{d-l}}\sum_{k=1}^{n-1}\left(\frac{1}{k}+\frac{1}{n-k}\right)^{l/2}
\leq \frac{2^{l/2}}{\lambda_\beta^l L^{d-l}}\sum_{k=1}^{n/2}\frac{1}{k^{l/2}}=o(1)\quad (L\to\infty)
\ee
which finally yields
\be\label{av-U-final-upper-bound}
\langle U\rangle_{P^{n\beta}_{0,Lz}}
\leq \frac{\|u\|_1}{2} n \left[\frac{n-1}{L^d}+\sum_{k=1}^{n-1}\alpha_{n,k}^{d/2}+o(1)  \right]
\leq \frac{\|u\|_1}{2} n \left[\frac{n-1}{L^d}+2^{d/2}\zeta(d/2)/\lambda_\beta^d +o(1)\right].
\ee
Dropping the $o(1)$ term,
the inequality (\ref{QNL-lower}) can be continued as
\bea
Q_{N,L}
&\geq&
\frac{1}{N}\sum_{n=1}^N q_n  Q_{N-n,L}
\exp\left\{- \frac{\beta\|u\|_1}{L^d} \left[n(N-n)+\frac{1}{2}n(n-1)\right]\right\}
\exp\left\{-2^{d/2-1}\zeta(d/2)\frac{\beta \|u\|_1}{\lambda_{\beta}^d}n\right\}
\nonumber\\
&\equiv&
\frac{1}{N}\sum_{n=1}^N q_n  Q_{N-n,L}
\exp\left\{- C \left[n(N-n)+\frac{1}{2}n(n-1)\right]\right\}
\exp\left\{-Dn\right\}
\equiv
\frac{1}{N}\sum_{n=1}^N q_n  Q_{N-n,L}\, e^{-\beta\Psi^+_{n,N-n}}.
\nonumber\\
\eea
We define an auxiliary function $Q_{N,L}^-$ recursively by $Q_{0,L}^-=1$ and
\be\label{QNL-minus}
Q_{N,L}^-=\frac{1}{N}\sum_{n=1}^N q_nQ_{N-n,L}^- e^{-\beta\Psi^+_{n,N-n}}.
\ee
$Q_{N,L}^-$ has two useful properties.

\vspace{3pt}
\noindent
(i) $Q_{N,L}\geq Q_{N,L}^-$. To see it, write $Q_{N,L}$ in an analogous form,
\be\label{QN-bis}
Q_{N,L}=\frac{1}{N}\sum_{n=1}^N q_nQ_{N-n,L}\,e^{-\beta\Psi_{n,N-n}}.
\ee
By definition
\[
e^{-\beta\Psi_{n,N-n}}=\frac{\int_\Lambda\d x \int W^{n\beta}_{xx}(\d\omega_0)e^{-\beta U(\omega_0)} Q_{N-n,L}(\omega_0)}{q_nQ_{N-n,L}},
\]
and we just have proved that $\Psi_{n,N-n}\leq\Psi^+_{n,N-n}$. Therefore
\bea
Q_{N,L}-Q_{N,L}^-
&=&\frac{1}{N}\sum_{n=1}^N q_n\left[Q_{N-n,L}e^{-\beta\Psi_{n,N-n}}-Q_{N-n,L}^-e^{-\beta\Psi^+_{n,N-n}}\right]
\nonumber\\
&=&\frac{1}{N}\sum_{n=1}^N q_n e^{-\beta\Psi_{n,N-n}}\left[Q_{N-n,L}-Q_{N-n,L}^-e^{-\beta(\Psi^+_{n,N-n}-\Psi_{n,N-n})}\right]
\nonumber\\
&\geq&\frac{1}{N}\sum_{n=1}^N q_n e^{-\beta\Psi_{n,N-n}}\left[Q_{N-n,L}-Q_{N-n,L}^-\right].
\eea
For $N=1$ this reads
\be
Q_{1,L}-Q_{1,L}^-\geq q_1 e^{-\beta\Psi_{1,0}}[Q_{0,L}-Q_{0,L}^-]=0,
\ee
and $Q_{N,L}\geq Q_{N,L}^-$ follows by induction.

\vspace{3pt}
\noindent
(ii)
\be\label{QNL-minus-vs-QNL0}
Q_{N,L}^-= \exp\left\{-\frac{1}{2}CN(N-1)-DN\right\}Q^0_{N,L}.
\ee
Indeed, apply the identity
\begin{eqnarray*}
\frac{1}{2}CN(N-1)+DN
&=&Cn(N-n) +\frac{1}{2}Cn(n-1)+Dn+\frac{1}{2}C(N-n)(N-n-1)+D(N-n)
\nonumber\\
&=&\beta\Psi^+_{n,N-n}+ \frac{1}{2}C(N-n)(N-n-1)+D(N-n).
\end{eqnarray*}
From the definition (\ref{QNL-minus}) one can see that
$A_N=e^{\frac{1}{2}CN(N-1)+DN}Q_{N,L}^-$ satisfies the recurrence relation
\be
A_N=\frac{1}{N}\sum_{n=1}^N q_n  A_{N-n}
\ee
with the initial condition $A_0=1$.
$Q^0_{N,L}$ satisfies the same equation with the same initial condition, cf. Eq.~(\ref{Q0NL}), therefore $A_N=Q^0_{N,L}$, which proves (\ref{QNL-minus-vs-QNL0}). In sum,
\be
Q_{N,L}
\geq \exp\left\{-\frac{1}{2}CN(N-1)-DN\right\}Q^0_{N,L}
=\exp\left\{-\frac{1}{2}\beta\rho\|u\|_1(N-1)
-2^{d/2-1}\zeta(d/2)\frac{\beta\|u\|_1}{\lambda_\beta^d}N\right\} Q^0_{N,L}.
\ee
From here the upper bound for $F_{N,L}$ and $f(\rho,\beta)$ follows.

\newsec{Proof of Theorem~\ref{second-step}}

It is instructive to replace $Q_{N,L}$ with $\tilde{Q}_{N,L}=e^{\beta\hat{u}(0)N(N-1)/2L^d}Q_{N,L}$. The observables like $\rho^{N,L}_{n}$ are invariant under this change;
the difference is in the free energy densities, $f(\rho,\beta)$ and
\be\label{logQ-tilde}
\tilde{f}(\rho,\beta)=- \lim_{N,L\to\infty, N/L^d=\rho}\ \frac{1}{\beta L^d}\ln \tilde{Q}_{N,L}=f(\rho,\beta) - \rho^2 \hat{u}(0)/2.
\ee
According to Lemma~\ref{bounds-to-free-energy}, for $\tilde{f}(\rho,\beta)$ we have the bounds
\be\label{lower-upper}
-\left(\frac{\hat{u}(0)}{2}-C[u]\right) \rho^2 - B\rho + f^0(\rho,\beta)
\leq \tilde{f}(\rho,\beta)
\leq
\frac{\|u\|_1-\hat{u}(0)}{2}\rho^2+
 \frac{2^{d/2-1}\zeta(d/2)\|u\|_1}{\lambda_\beta^d}\rho + f^0(\rho,\beta)
\ee
with $C[u]\leq \hat{u}(0)/2\leq\|u\|_1/2$.
In [Su5] we studied the $\rho\to\infty$ limit of $\rho^{-2}f(\rho,\infty)$ for classical systems, and found that it was bounded above by the best superstability constant, cf. Eqs.~(\ref{C_Lambda[u]}) and (\ref{C[u]}), which in turn cannot exceed $\hat{u}(0)/2$ and reaches this value for $u$ of the positive type. If $Q_{N,L}$ is the partition function of the mean-field Bose gas with mean-field energy $(\hat{u}(0)/2)N(N-1)/L^d$
then $\tilde{Q}_{N,L}$ is the partition function of the ideal Bose gas. Its energy is purely kinetic, yet (\ref{logQ-tilde}) is negative due to the known particularity of the Bose statistics that it induces an effective attraction among the particles.
The result can be similar for a positive-type $u$ because of the uniform distribution of the particles [Su5] and also because at asymptotically high densities the ground state energy density of the Bose gas is $\rho^2\hat{u}(0)/2$, cf. Lieb [Li7].
This can also be seen on (\ref{lower-upper}) which for a positive and positive-type $u$, when $C[u]=\hat{u}(0)=\|u\|_1/2$, becomes
\be\label{bounds-on-tilde{f}}
- \frac{u(0)}{2}\rho + f^0(\rho,\beta)
\leq \tilde{f}(\rho,\beta)
\leq
 \frac{2^{d/2-1}\zeta(d/2)\hat{u}(0)}{\lambda_\beta^d}\rho + f^0(\rho,\beta),
\ee
the correction to $\rho^2\hat{u}(0)/2$ is of order $\rho$. The upper bound in (\ref{bounds-on-tilde{f}}) goes to zero with the temperature, but not fast enough to decide whether $\tilde{f}(\rho,\beta)$ could be negative (as $\beta$ increases, $\rho$ exceeds the critical density $\zeta(d/2)/\lambda_\beta^d$ of the ideal gas, above which $f^0(\rho,\beta)=-\zeta(1+d/2)/(\beta\lambda_\beta^d)$ tends to zero faster than the positive first term).
The linear upper and lower bounds in $\rho$ when both $u$ and $\hat{u}$ are nonnegative show that in analogy with the relation between the mean-field and the ideal Bose gas, the change from $Q_{N,L}$ to $\tilde{Q}_{N,L}$ deprives the model from its superstability but preserves stability.

The inequalities (\ref{bounds-on-tilde{f}}) give already a hint to the existence of BEC. They imply $\lim_{\rho\to\infty}f(\rho,\beta)/\rho^2=\hat{u}(0)/2$, extending the ground-state result to positive temperatures. This shows that at high densities a positive and positive-type pair potential acts as a perturbed mean-field interaction. $f(\rho,\beta)$ is a convex function of $\rho$ and $\partial f(\rho,\beta)/\partial \rho$ is the chemical potential of the canonical ensemble.
The pair potential $\tilde{u}_L(x)=u_L(x)-\hat{u}(0)/L^d$ is still stable, therefore it defines a normal thermodynamic system, but $\tilde{f}(\rho,\beta)=f(\rho,\beta) - \rho^2 \hat{u}(0)/2$ may not be convex. If it is, the chemical potential $\partial \tilde{f}/\partial\rho$ is an increasing function of $\rho$. However, due to (\ref{bounds-on-tilde{f}}) it is bounded from above, and if it attains its supremum at a finite $\rho$ then for higher densities the surplus particles must go into the condensate. Below we have to circumvent the question of convexity of $\tilde{f}(\rho,\beta)$.

The analogue of Eq.~(\ref{QNL-with-G}) for the de-superstabilized system reads
\be\label{QNL-with-tildeG}
\tilde{Q}_{N,L}=
\frac{1}{N}\sum_{n=1}^{N} \tilde{G}^N_{n}
\ee
where $\tilde{G}^N_{n}=\tilde{G}^N_{n}(0)$, cf. Eq.~(\ref{tildeGn(x)}). Thus,
for $n=N$
\bea\label{PhiNN}
\Tilde{G}^N_N=\sum_{\{\alpha^k_j\in \Nn_0|1\leq j<k\leq N\}}
\ \prod_{1\leq j<k\leq N}
\frac{ \left(-\beta\right)^{\alpha^k_j}}{\alpha^k_j !}
\prod_{r=1}^{\alpha^k_j}\int_0^1\d t^k_{j,r} \int \d x^k_{j,r}\ \hat{u}\left(x^k_{j,r}\right)
\nonumber\\
\exp\left\{-\pi N \lambda_\beta^2\left[\overline{\left(X^0_{^\cdot}\right)^2}-\overline{X^0_{^\cdot}}^2\right]\right\}f_{N}(0;L\overline{X^0_{^\cdot}})
\eea
and for $n<N$
\be\label{Phi^N_{n_0}}
\Tilde{G}^N_{n}=\sum_{p=1}^{N-n}\frac{1}{p!}\sum_{n_1,\dots,n_p\geq 1:\sum_1^p n_l=N-n}\ \frac{1}{\prod_{l=1}^p n_l}\ \tilde{G}[n,\{n_l\}_1^p]
\ee
where ($n_0=n$)
\bea\label{Phi^N_{n,{n_l}_1^p}}
\tilde{G}[n,\{n_l\}_1^p]
=
\sum_{\{\alpha^k_j\in \Nn_0|1\leq j<k\leq N\}}
\Delta_{\{\alpha^k_j\},\{n_l\}_0^p}\, L^{-dK_{\{\alpha^k_j\}}}
 \prod_{1\leq j<k\leq N}
\frac{ \left(-\beta\right)^{\alpha^k_j}}{\alpha^k_j !}
\prod_{r=1}^{\alpha^k_j}\int_0^1\d t^k_{j,r} \int \d x^k_{j,r}\ \hat{u}\left(x^k_{j,r}\right)    
\nonumber\\
\left[
\delta(X^0_1,\dots,X^p_1)
\prod_{l=0}^p
\exp\left\{-\pi n_l \lambda_\beta^2\left[\overline{\left(X^l_{^\cdot}\right)^2}-\overline{X^l_{^\cdot}}^2\right]\right\}
f_n\left(0;L\overline{X^l_{^\cdot}}\right)\right].
\nonumber\\
\eea
The quantities denoted by $X$ are derived from $X_q(t)$, cf. Eq.~(\ref{exp-asymp-expanded}). In $X_q(t)$ we have to separate intra- and inter-cycle  terms.
The two contributions are labelled by $l$ and $\neg l$, respectively. The partition $X_q(t)=X_q(t)|_l+X_q(t)|_{\neg l}$ can be read off from Eq.~(\ref{Zqt}).
Note that the quantities occurring in $\delta(X^0_1,\dots,X^p_1)$,
\be\label{X^l_1}
X^l_1 := X_{N_{l-1}+1}(0)
=-\sum_{j=1}^{N_{l-1}}\sum_{k\in C_l}\sum_{r=1}^{\alpha^{k}_{j}}x^{k}_{j,r}+\sum_{j\in C_l} \sum_{k=N_l+1}^N \sum_{r=1}^{\alpha^{k}_{j}}x^{k}_{j,r},
\quad l=0,\dots,p
\ee
are purely inter-cycle.
Furthermore,
\bea\label{Xav-decomposition}
\overline{X^l_{^\cdot}}
&=&
\frac{1}{n_l}\sum_{q\in C_l}\int_0^1 X_q(t)|_l\,\d t+\frac{1}{n_l}\sum_{q\in C_l}\int_0^1 X_q(t)|_{\neg l}\,\d t
= \overline{X^l_{^\cdot}|_l} + \overline{X^l_{^\cdot}|_{\neg l}}
\nonumber\\
\overline{\left(X^l_{^\cdot}\right)^2}
&=&
\overline{\left(X^l_{^\cdot}\right)^2|_l }+ \overline{\left(X^l_{^\cdot}\right)^2|_{\neg l}} + 2\,\overline{X^l_{^\cdot}|_l\cdot X^l_{^\cdot}|_{\neg l}}\,.
\eea

\subsection{Cycle-decoupling model}\label{dcp}

We first prove the occurrence of cycles whose length increases proportionally to $N$ in a simplified model, retaining from the multiple sum that constitutes $\tilde{G}[n,\{n_l\}_1^p]$ the single term $\alpha^k_j=0$ if $j$ and $k$ are in different cycles. We shall refer to it as the cycle-decoupling model and denote its partition function by $Q_{N,L}^{\rm dcp}$.
For a while we still continue with a general integrable $u$.
Compared to $Q_{N,L}$,
\be
Q_{N,L}^{\rm dcp}=\frac{\exp\left\{-\frac{\beta\hat{u}(0)N(N-1)}{2L^d}\right\}}{N}
\left(\Tilde{G}^N_N+\sum_{n=1}^{N-1}\Tilde{G}^n_n
\sum_{p=1}^{N-n}\frac{1}{p!}\sum_{n_1,\dots,n_p\geq 1:\sum_1^p n_l=N-n}\prod_{l=1}^p\frac{1}{n_l}\Tilde{G}^{n_l}_{n_l}\right).
\ee
Here $\Tilde{G}^{n_l}_{n_l}$ is the precise analogue of $\Tilde{G}^N_N$, cf.  Eq.~(\ref{PhiNN}), all the particles (to be considered) are in a single cycle.
That is,
\be\label{Phi^{n_l}_{n_l}}
\Tilde{G}^{n_l}_{n_l}
=\sum_{\{\alpha^k_j \in\Nn_0| \{j<k\}\subset C_l\}}
\,\prod_{\{j<k\}\subset C_l}
\frac{ \left(-\beta\right)^{\alpha^k_j}}{\alpha^k_j !}
\prod_{r=1}^{\alpha^k_j}\int_0^1\d t^k_{j,r} \int \d x^k_{j,r}\ \hat{u}\left(x^k_{j,r}\right)
e^{-\pi n_l \lambda_\beta^2\left[\overline{\left(X^l_{^\cdot}\right)^2|_l}-\left(\overline{X^l_{^\cdot}|_l}\right)^2\right]}
f_n\left(0;L\overline{X^l_{^\cdot}|_l}\right).
\ee
Note that the only $L$-dependence of $\Tilde{G}^{n_l}_{n_l}$ is in $f_n\left(0;L\overline{X^l_{^\cdot}|_l}\right)$.
By comparison with the case of $n_l=N$, for L so large that $L^{-d}\sum_{z^k_{j,r}\in\Zz^d\setminus\{0\}}\hat{u}(z^k_{j,r}/L)$ can be replaced with $\int \d x^k_{j,r}\ \hat{u}\left(x^k_{j,r}\right)$,
\bea
\lefteqn{
\int_\Lambda \d x_l \int W_{x_l x_l}^{n_l\beta}(\d\omega_l) e^{-\beta U(\omega_l)}
=
e^{-\beta\hat{u}(0)n_l(n_l-1)/2L^d}     }
\nonumber\\
&&\times
\sum_{\{\alpha^k_j \in\Nn_0| \{j<k\}\subset C_l\}}
\,\prod_{\{j<k\}\subset C_l}
\frac{ \left(-\beta\right)^{\alpha^k_j}}{\alpha^k_j !}
\prod_{r=1}^{\alpha^k_j}\int_0^1\d t^k_{j,r} \int \d x^k_{j,r}\ \hat{u}\left(x^k_{j,r}\right)
e^{-\pi n_l \lambda_\beta^2\left[\overline{\left(X^l_{^\cdot}\right)^2|_l}-\left(\overline{X^l_{^\cdot}|_l}\right)^2\right]}
f_n\left(0;L\overline{X^l_{^\cdot}|_l}\right),
\nonumber\\
\eea
which yields the Feynman-Kac form of $\Tilde{G}^{n_l}_{n_l}$,
\be\label{Phi^{n_l}_{n_l}-FK}
\Tilde{G}^{n_l}_{n_l}
=e^{\beta\hat{u}(0)n_l(n_l-1)/2L^d}\int_\Lambda \d x_l \int W_{x_l x_l}^{n_l\beta}(\d\omega_l) e^{-\beta U(\omega_l)}.
\ee
Let us remark that because $N(N-1)-\sum_{l=0}^p n_l(n_l-1)=\sum_{0\leq l'< l\leq p} n_ln_{l'}$, in the cycle-decoupling model there remains a repulsive mean-field type pair interaction between the cycles. Since typically $p\propto N$, the total energy of this interaction is
\[
(\hat{u}(0)/4L^d)\sum_{0\leq l'<l\leq p}n_l n_{l'}\geq p(p+1)\hat{u}(0)/8L^d\propto \rho\hat{u}(0) N.
\]

We can incorporate the global mean-field factor into the partition function by defining
\[
\tilde{Q}_{N,L}^{\rm dcp}:=e^{\frac{\beta\hat{u}(0) N(N-1)}{2L^d}}Q_{N,L}^{\rm dcp}.
\]
Then
\be\label{Qdcp-tilde}
\tilde{Q}_{N,L}^{\rm dcp}
=\frac{1}{N}
\left(\Tilde{G}^N_N+\sum_{n=1}^{N-1}\Tilde{G}^n_n
\sum_{p=1}^{N-n}\frac{1}{p!}\sum_{n_1,\dots,n_p\geq 1:\sum_1^p n_l=N-n}\prod_{l=1}^p\frac{1}{n_l}\Tilde{G}^{n_l}_{n_l}\right)
\ee
and in the average over the partitions of $N-n$ one can recognize $\tilde{Q}_{N-n,L}^{\rm dcp}$. Defining $\tilde{Q}_{0,L}^{\rm dcp}=1$, we have therefore
\be\label{Q-decoupling}
\tilde{Q}_{N,L}^{\rm dcp}=\frac{1}{N}\sum_{n=1}^{N}\Tilde{G}^n_{n}\tilde{Q}_{N-n,L}^{\rm dcp}.
\ee
The estimates done in Lemma~\ref{bounds-to-free-energy} for $\int_\Lambda \d x \int W_{x x}^{n\beta}(\d\omega) e^{-\beta U(\omega)}$ provide us with bounds on $\Tilde{G}^{n}_{n}$.
First, the superstability bound
\[
U(\omega)
\geq - Bn+C_\Lambda[u] n(n-1)/L^d
\]
implies
\bea\label{upperbound-on-phi^l_nl}
\Tilde{G}^n_{n}
\leq q_n  e^{\beta Bn}\,
e^{\beta\left[\hat{u}(0)/2-C_\Lambda[u]\right]n(n-1)/L^d}.
\eea
From Eqs.~(\ref{Jensen}) and (\ref{av-U-final-upper-bound}) we can infer the lower bound
\bea\label{lowerbound-on-phi^l_nl}
\Tilde{G}^n_{n}
\geq  q_n e^{-2^{d/2-1}\zeta(d/2)\beta \|u\|_1 n/\lambda_\beta^d}\,
e^{-\beta\left[\|u\|_1-\hat{u}(0)\right] n(n-1)/2L^d}.
\eea
It is at this point that we abandon the general potential and continue with $u\geq 0$ and $\hat{u}\geq 0$, when $\hat{u}(0)=\|u\|_1=2C_\Lambda[u]$, $B=u_L(0)/2$. The upper and lower bounds for $\Tilde{G}^n_{n}$ become
\be\label{upperlower-on-phi^l_nl}
q_n e^{-2^{d/2-1}\zeta(d/2)\beta \hat{u}(0) n/\lambda_\beta^d}
\leq\Tilde{G}^n_{n}
\leq
q_n  e^{\beta u_L(0)n/2}.
\ee
Upper and lower bounds for $\tilde{Q}_{N,L}^{\rm dcp}$ can be obtained by substituting (\ref{upperlower-on-phi^l_nl}) into Eq.~(\ref{Qdcp-tilde}):
\be\label{Qdcp-tilde-upper}
e^{-2^{d/2-1}\zeta(d/2)\beta \hat{u}(0) N/\lambda_\beta^d}Q^0_{N,L}
\leq
\tilde{Q}_{N,L}^{\rm dcp}
\leq e^{\beta u_L(0)N/2} Q^0_{N,L}\qquad (u\geq 0,\ \hat{u}\geq 0).
\ee
These are the same as those for $\tilde{Q}_{N,L}=e^{\frac{\beta\hat{u}(0) N(N-1)}{2L^d}}Q_{N,L}$, so any modification of $\tilde{Q}_{N,L}^{\rm dcp}$ due to coupling must fit within these bounds.

Equation~(\ref{Q-decoupling}) has the same form as (\ref{Q0NL}) for the partition function of the ideal Bose gas and, given $\Tilde{G}^n_{n}$, it defines $\tilde{Q}_{N,L}^{\rm dcp}$ recursively.
We can proceed as we did for the ideal gas.
Writing $\tilde{Q}_{N,L}^{\rm dcp}=\exp\{-\beta L^d \tilde{f}_{N,L}^{\rm dcp}\}$, if $N$ and $L$ are large and $n=o(N)$ then
\be
\frac{\tilde{Q}_{N-n,L}^{\rm dcp}}{\tilde{Q}_{N,L}^{\rm dcp}}=\exp\left\{\beta n\frac{\tilde{f}_{N,L}^{\rm dcp}-\tilde{f}_{N-n,L}^{\rm dcp}}{(n/L^d)}\right\}
= \exp\left\{\beta n[\partial \tilde{f}^{\rm dcp}/\partial \rho+o(1)]\right\}.
\ee
Here we anticipate that $\partial \tilde{f}^{\rm dcp}/\partial \rho$ exists. It is the chemical potential of the cycle-decoupling model, and below it will be obtained as the solution of Eq.~(\ref{mu=tilde-f}).
Let $g_N=o(N^{2/d})$ be a sequence of positive integers that tends to infinity. Repeating the argument given in the discussion of the ideal gas,
\bea\label{lim1-dcp}
\lim_{N,L\to\infty, N/L^d=\rho}\frac{1}{N}\sum_{n=1}^{g_N}\Tilde{G}^n_{n} \frac{\tilde{Q}_{N-n,L}^{\rm dcp}}{\tilde{Q}_{N,L}^{\rm dcp}}
=\frac{1}{\rho\lambda_\beta^d}\sum_{n=1}^\infty \frac{\varphi^0_{n}\exp\left\{\beta n\partial \tilde{f}^{\rm dcp}/\partial \rho\right\}}{n^{d/2}}\leq 1
\eea
where $\varphi^0_1=1$ and for $n\geq 2$
\bea
\varphi^0_{n}
&=&
\sum_{\{\alpha^k_j \in\Nn_0| 1\leq j<k\leq n\}}
\frac{ \left(-\beta\right)^{\alpha^k_j}}{\alpha^k_j !}
\prod_{r=1}^{\alpha^k_j}\int_0^1\d t^k_{j,r} \int \d x^k_{j,r}\ \hat{u}\left(x^k_{j,r}\right)
e^{-\pi n \lambda_\beta^2\left[\overline{\left(X^0_{^\cdot}\right)^2|_0}-\left(\overline{X^0_{^\cdot}|_0}\right)^2\right]}.
\eea
In Eq.~(\ref{lim1-dcp}) it was possible to replace $\Tilde{G}^n_n/N$ with $\varphi^0_{n}/(\rho\lambda_\beta^d n^{d/2})$ because $n\lambda_\beta^2/L^2\to 0$ and therefore
$f_n(0;L\overline{X^0_{^\cdot}|_0})\sim L^d/(\lambda_\beta^d n^{d/2})$ as $L$ increases, cf. Eq.~(\ref{f_n(0)-asymp}).
As in Eq.~(\ref{lim1}), the limit of the sum up to $g_N$ contains all the contribution of finite cycles and nothing else, the sum from $M$ to $g_N$ goes to zero if first $N$ and $L$ and then $M$ tends to infinity; the consequence is that the asymptotic weight of cycles whose length diverges with $N$ but the divergence is slower than $N^{2/d}$ is zero.
The bounds (\ref{upperlower-on-phi^l_nl}) show that $\varphi^0_{n}$ cannot increase or decrease faster than exponentially in $n$.
Thus, $e^{-b_ln}\leq\varphi^0_n\leq e^{b_un}$ with positive numbers $b_l$, $b_u$.
Because $\varphi^0_{n}$ is independent of the density, the infinite sum in (\ref{lim1-dcp}) can increase with $\rho$ only due to an increase of $\partial \tilde{f}^{\rm dcp}/\partial \rho$. However, $-\limsup_{n\to\infty}n^{-1}\ln\varphi^0_n\leq b_l$ is an upper bound to
$\beta\partial \tilde{f}^{\rm dcp}/\partial \rho$, otherwise the infinite sum could not be convergent for $\rho$ large enough. It follows that
\be\label{sup}
\sup_\rho\sum_{n=1}^\infty \frac{\varphi^0_{n}\exp\left\{\beta n\partial \tilde{f}^{\rm dcp}/\partial \rho\right\}}{n^{d/2}}
=\zeta^{\rm dcp}(\beta)<\infty.
\ee
If $\rho\leq\zeta^{\rm dcp}(\beta)/\lambda_\beta^d$, the equation
\be\label{mu=tilde-f}
\frac{1}{\rho\lambda_\beta^d}\sum_{n=1}^\infty \frac{\varphi^0_{n}\exp\{\beta n\mu\}}{n^{d/2}}=1
\ee
has a unique solution for $\mu$ that can be identified with $\partial \tilde{f}^{\rm dcp}/\partial \rho$. It is an increasing function of $\rho$ which reaches its maximum $\bar{\mu}=\bar{\mu}(\beta)$ at $\rho=\zeta^{\rm dcp}(\beta)/\lambda_\beta^d$, where $\bar{\mu}$ satisfies the equation
\be
\frac{1}{\zeta^{\rm dcp}(\beta)}\sum_{n=1}^\infty \frac{\varphi^0_{n}\exp\{\beta n\bar{\mu}\}}{n^{d/2}}=1.
\ee
So $\partial \tilde{f}^{\rm dcp}(\rho,\beta)/\partial \rho\equiv\bar{\mu}(\beta)$ if $\rho\geq\zeta^{\rm dcp}(\beta)/\lambda_\beta^d$.
From Theorem~\ref{first-step}
\[
\lim_{n\to\infty}n^{-1}\ln\varphi^0_n= c  e^{-\epsilon\beta}
\]
with some positive constants $c$ and $\epsilon$. Hence, $\bar{\mu}(\beta)=-c\, e^{-\epsilon\beta}/\beta$.
If $\rho>\zeta^{\rm dcp}(\beta)/\lambda_\beta^d$, in Eq.~(\ref{lim1-dcp}) the inequality is strict, and the completion to 1 comes from cycle lengths diverging at least as fast as $N^{2/d}$.
More is true, however. Choose any $\varepsilon>0$, then
\[
\lim_{N\to\infty}\frac{1}{N}\sum_{n=\lfloor\varepsilon N^{2/d}\rfloor}^N 1=1,\quad
\lim_{N,L\to\infty, N/L^d=\rho}\frac{1}{N}\sum_{n=\lfloor\varepsilon N^{2/d}\rfloor}^N \Tilde{G}^n_{n} \frac{\tilde{Q}_{N-n,L}^{\rm dcp}}{\tilde{Q}_{N,L}^{\rm dcp}}<1.
\]
This tells us that on average $\Tilde{G}^n_{n}\,\tilde{Q}_{N-n,L}^{\rm dcp}/\tilde{Q}_{N,L}^{\rm dcp}<1$ if $n\geq\varepsilon N^{2/d}$, which suggests that
for any $h_N=o(N)$, $h_N>\varepsilon N^{2/d}$
\be\label{no-submacroscopic}
\lim_{N,L\to\infty, N/L^d=\rho}\frac{1}{N}\sum_{n=\lfloor\varepsilon N^{2/d}\rfloor}^{h_N} \Tilde{G}^n_{n} \frac{\tilde{Q}_{N-n,L}^{\rm dcp}}{\tilde{Q}_{N,L}^{\rm dcp}}=0.
\ee
Now $f_n(0;L\overline{X^0_{^\cdot}|_0})=O(1)$ for $n\geq \varepsilon N^{2/d}$, cf. Eq.~(\ref{f_n(0)-asymp}), so $\Tilde{G}^n_{n}\leq C\varphi^0_n$ with some constant $C$, and $\varphi^0_n$ is independent of $N$. At the same time
\[
\tilde{Q}_{N-n,L}^{\rm dcp}/\tilde{Q}_{N,L}^{\rm dcp}\sim e^{n\beta\bar{\mu}(\beta)},
\]
also independent of $N$. Thus, we are dealing with a sequence $a_n>0$ such that
\bea\label{a_n}
\lim_{N\to\infty}\frac{1}{N}\sum_{n=\lfloor\varepsilon N^{2/d}\rfloor}^N a_n
&=&\lim_{N\to\infty}\frac{N-\varepsilon N^{2/d}}{N}\frac{1}{N-\varepsilon N^{2/d}}\sum_{n=\lfloor\varepsilon N^{2/d}\rfloor}^N a_n
\nonumber\\
&=&\lim_{N\to\infty}\frac{1}{N-\varepsilon N^{2/d}}\sum_{n=\lfloor\varepsilon N^{2/d}\rfloor}^N a_n \leq 1.
\eea
Supposing the opposite of (\ref{no-submacroscopic}),
\[
\lim_{N\to\infty}\frac{1}{N}\sum_{n=\lfloor\varepsilon N^{2/d}\rfloor}^{h_N}a_n
= \lim_{N\to\infty}\frac{h_N-\varepsilon N^{2/d}}{N}\frac{1}{h_N-\varepsilon N^{2/d}}
\sum_{n=\lfloor\varepsilon N^{2/d}\rfloor}^{h_N}a_n>0.
\]
This implies ($m<M$)
\[
\lim_{N\to\infty}\frac{1}{h_N-\varepsilon N^{2/d}}
\sum_{n=\lfloor\varepsilon N^{2/d}\rfloor}^{h_N}a_n
=\lim_{m, M\to\infty}\frac{1}{M-m}\sum_{n=m}^{M}a_n=\infty,
\]
in contradiction with (\ref{a_n}).
Therefore (\ref{no-submacroscopic}) holds true, and
\be\label{part-1-decoupled}
\frac{\zeta^{\rm dcp}(\beta)}{\rho\lambda_\beta^d}
+
\lim_{\varepsilon\downarrow 0}\lim_{N,L\to\infty, N/L^d=\rho}\frac{1}{N}
\sum_{n=\lfloor \varepsilon N\rfloor}^N\Tilde{G}^n_n \frac{\tilde{Q}_{N-n,L}^{\rm dcp}}{\tilde{Q}_{N,L}^{\rm dcp}}=1   \qquad (\rho>\zeta^{\rm dcp}(\beta)/\lambda_\beta^d).
\ee
As in the noninteracting gas, in infinite volume the cycles are either finite or macroscopic.

\subsection{Coupling of the cycles}\label{cp}

Consider the full model. We write the partition functions in their form
\be
\tilde{Q}_{N,L}=\sum_{p=0}^N\frac{1}{p!}\sum_{n_0,\dots,n_p\geq 1:\sum_0^p n_l=N}\frac{1}{\prod_{l=0}^p n_l}\Tilde{G}\left[\{n_l\}_0^p\right],
\qquad
\tilde{Q}_{N,L}^{\rm dcp}=\sum_{p=0}^N\frac{1}{p!}\sum_{n_0,\dots,n_p\geq 1:\sum_0^p n_l=N}\frac{1}{\prod_{l=0}^p n_l}\prod_{l=0}^p \Tilde{G}^{n_l}_{n_l}.
\ee
Although we preserve cycle 0 for a better comparison with the cycle-decoupling model, now it is not distinguished.
Let $\{n_l\}_{l=0}^p$ be given.
$\Delta_{\{\alpha^k_j\},\{n_l\}_0^p}$ and $K_{\{\alpha^k_j\}}$ depend only on
\[
~\alpha^{\rm cp}=\{\alpha^k_j\in\Nn_0| \mbox{$j$ and $k$ are in different cycles}\}.
\]
For the set of $~\alpha^{\rm cp}$ that satisfy $\Delta_{\{\alpha^k_j\},\{n_l\}_0^p}=1$ we use the notation
$
A^{\rm cp}_{\{n_l\}_0^p},
$
and the $d\left(\sum_{\alpha^k_j\in~\alpha^{\rm cp}}\alpha^k_j-K_{\{\alpha^k_j\}}\right)$ dimensional manifold in $\Rr^{d\sum_{\alpha^k_j\in~\alpha^{\rm cp}}\alpha^k_j}$ on which $X^0_1=X^1_1=\cdots=X^p_1=0$ will be denoted by
$X_{~\alpha^{\rm cp}}$. There is no constraint for
\[
~\alpha_l=\{\alpha^k_j\in\Nn_0| \{j<k\}\subset C_l\}.
\]
Also, let
\be
\|~\alpha_l\|_1=\sum_{\alpha^k_j\in~\alpha_l}\alpha^k_j,\qquad
\|~\alpha^{\rm cp}\|_1=\sum_{\alpha^k_j\in~\alpha^{\rm cp}}\alpha^k_j,\qquad
\|\hat{u}^{~\alpha^{\rm cp}}\|= \int_{X_{~\alpha^{\rm cp}}}\prod_{j,k,r}\hat{u}(x^k_{j,r})\prod_{j,k,r}\d x^k_{j,r}.
\ee
The coupling brings in $\|~\alpha^{\rm cp}\|_1$ factors $\beta$ and $\hat{u}$, and $\beta^{\|~\alpha^{\rm cp}\|_1}\,\|\hat{u}^{~\alpha^{\rm cp}}\|$ is a volume raised to the power $K_{~\alpha^{\rm cp}}$. Denoting this volume by $v_{\beta,~\alpha^{\rm cp}}$,
\be
\beta^{\|~\alpha^{\rm cp}\|_1}\,\|\hat{u}^{~\alpha^{\rm cp}}\|=(v_{\beta,~\alpha^{\rm cp}})^{K_{~\alpha^{\rm cp}}}.
\ee
With these notations
\bea\label{Phi^N_n-new-notation}
\Tilde{G}\left[\{n_l\}_0^p\right]
=
\prod_{l=0}^p \sum_{~\alpha_l}
\frac{[-\beta u(0)]^{\|~\alpha_l\|_1}}{\prod_{\alpha^k_j\in~\alpha_l}\alpha^k_j !}
\prod_{\alpha^k_j\in~\alpha_l}
\prod_{r=1}^{\alpha^k_j}\int_0^1\d t^k_{j,r}
\int\d x^k_{j,r}\frac{\hat{u}(x^k_{j,r})}{u(0)}
e^{-\pi n_l \lambda_\beta^2\left[\overline{\left(X^l_{^\cdot}\right)^2|_l}-\left(\overline{X^l_{^\cdot}|_l}\right)^2\right]}
f_n\left(0;L\overline{X^l_{^\cdot}|_l}\right)
\nonumber\\
\sum_{~\alpha^{\rm cp}\in A^{\rm cp}_{\{n_l\}_0^p}}
\frac{(-1)^{\|~\alpha^{\rm cp}\|_1}}{\prod_{\alpha^k_j\in ~\alpha^{\rm cp}}\alpha^k_j !}
\left[\frac{\rho v_{\beta,~\alpha^{\rm cp}}}{N}\right]^{K_{~\alpha^{\rm cp}}}
\int_{X_{~\alpha^{\rm cp}}}\frac{\prod_{j,k,r}\d x^k_{j,r}\hat{u}(x^k_{j,r})}{\|\hat{u}^{~\alpha^{\rm cp}}\|}
\prod_{\alpha^k_j\in ~\alpha^{\rm cp}}\prod_{r=1}^{\alpha^k_j}\int_0^1\d t^k_{j,r}
\nonumber\\
\exp\left\{-\pi \lambda_\beta^2\sum_{l=0}^p n_l\left[
\overline{\left(X^l_{^\cdot}\right)^2}-\overline{X^l_{^\cdot}}^2-\left(\overline{\left(X^l_{^\cdot}|_{l}\right)^2}-\left(\overline{X^l_{^\cdot}|_{l}}\right)^2  \right) \right]\right\}
\prod_{l=0}^p\frac{f_{n_l}\left(0;L\overline{X^l_{^\cdot}}\right)}{f_{n_l}\left(0;L\overline{X^l_{^\cdot}|_l }\right)}.
\nonumber\\
\eea
Observe that
\be\label{averages}
\int_{X_{~\alpha^{\rm cp}}}\prod_{j,k,r}\d x^k_{j,r} \frac{\prod_{j,k,r}\hat{u}(x^k_{j,r})}{\|\hat{u}^{~\alpha^{\rm cp}}\|}
\prod_{\alpha^k_j\in ~\alpha^{\rm cp}}\prod_{r=1}^{\alpha^k_j}\int_0^1\d t^k_{j,r} =1,\qquad
\prod_{\alpha^k_j\in~\alpha_l}
\prod_{r=1}^{\alpha^k_j}\int_0^1\d t^k_{j,r}\int_{\Rr^d}\d x^k_{j,r}\frac{\hat{u}(x^k_{j,r})}{u(0)}=1
\ee
($u(0)=\|\hat{u}\|_1$.) In Eq.~(\ref{Phi^N_n-new-notation}) the first line alone is $\prod_{l=0}^p\Tilde{G}^{n_l}_{n_l}$.
The vectors $x^k_{j,r}$ for $j$ and $k$ in the same cycle are independent identically distributed random variables; those with $j$ and $k$ in different cycles follow a coupled distribution. The probability measures can be read off from (\ref{averages}).
The sum over $A^{\rm cp}_{\{n_l\}_0^p}$ and the coupled distribution still factorize according to the maximal connected components of coupled cycles.
The sign of a term depends only on the parity of $\|~\alpha^{\rm cp}\|_1$ and, because $\Tilde{G}\left[\{n_l\}_0^p\right]>0$, the positive terms dominate.
Otherwise, the inclusion of $~\alpha^{\rm cp}$ has a threefold effect:
(i) the fluctuations of the momenta $\overline{\left(X^l_{^\cdot}\right)^2}-\overline{X^l_{^\cdot}}^2$ increase on average, (ii) For $K_{~\alpha^{\rm cp}}>0$ there appears a dimensionless factor $(\rho v_{\beta,~\alpha^{\rm cp}}/N)^{K_{~\alpha^{\rm cp}}}$, and (iii) the large number of different ways to couple the cycles contributes to the entropy.
The analysis in Theorem~\ref{first-step} showed that in the infinite-volume limit of ${\cal G}_{~\alpha^{\rm cp}}$ each vertex is of finite degree, i.e. the exponentially dominant contribution to the partition function comes from terms in which $\sum_{j\in C_l}\left[\sum_{l'>l}\sum_{k\in C_{l'}}\alpha^k_j+\sum_{l'<l}\sum_{k\in C_{l'}}\alpha^j_k\right]$ remains finite for every $l$ as $N$ tends to infinity.
A trivial reason is the rapid decrease of $1/\alpha^k_j!$ for any given pair $(j, k)$; a nontrivial reason is that
keeping the number of edges incident on every vertex finite the increase of fluctuations can be controlled. For $~\alpha^{\rm cp}$ thus chosen the entropy wins the competition between (ii) and (iii) for all $p$ of order $N$ and all values of $\rho$ and $\beta$, but the increase of $\overline{\left(X^l_{^\cdot}\right)^2}-\overline{X^l_{^\cdot}}^2$ influences the dependence on them. In a first time we disregard this latter and focus on the interplay between (ii) and (iii).

Let us recall from Eq.~(\ref{K-sum-V}) that a connected merger graph of $V$ vertices contributes $V-1$ to $K_{~\alpha^{\rm cp}}$. So if the coupling regroups the $p+1$ vertices (each representing a cycle) into $m_{~\alpha^{\rm cp}}$ connected components then $K_{~\alpha^{\rm cp}}=p+1-m_{~\alpha^{\rm cp}}$. The connected components are isolated vertices and clusters (circles composed of two or more vertices and their mergers).
Let $N_{\rm isl}$ be the number of isolated vertices, $N_{\rm cls}$ the number of clusters and $N_{\rm nisl}$ the total number of non-isolated vertices (i.e. those in clusters).
Then $p+1=N_{\rm isl}+N_{\rm nisl}$, $m_{~\alpha^{\rm cp}}=N_{\rm isl}+N_{\rm cls}$ and
\[
K_{~\alpha^{\rm cp}}=N_{\rm nisl}-N_{\rm cls}=\sum_{i=1}^{N_{\rm cls}}(V_i-1)
\]
where $V_i\geq 2$ is the number of vertices in the $i$th cluster. We must analyze partial sums of $A^{\rm cp}_{\{n_l\}_0^p}$ running over $~\alpha^{\rm cp}$ such that $\|~\alpha^{\rm cp}\|_1$ is even and increases linearly with $N_{\rm nisl}$.
Our aim is to show that the coupling among the cycles produces a global factor $e^{CN}$ ($C>0$) that appears in $\tilde{Q}_{N,L}$ compared to $\tilde{Q}_{N,L}^{\rm dcp}$.
We demonstrate the presence of such a factor on an example as follows.

Let $p+1=cN$ where $c<1$.
Let $N_{\rm isl}=aN$ ($0<a<c$) and suppose that the remaining $N_{\rm nisl}=(c-a)N$ vertices are coupled in $N_{\rm cls}=\frac{1}{2}N_{\rm nisl}$ two-circles. Then $m_{~\alpha^{\rm cp}}=(a+(c-a)/2)N=(c+a)N/2$ and $K_{~\alpha^{\rm cp}}=(c-a)N/2$.
For $\prod_{\alpha^k_j\in ~\alpha^{\rm cp}}(1/\alpha^k_j !)$ we can substitute $\epsilon_0^{N_{\rm nisl}}$ with some $\epsilon_0<1$.
The number of $~\alpha^{\rm cp}$ that differ only in the permutations of the (labelled) vertices is
\[
{N_{\rm isl}+N_{\rm nisl}\choose N_{\rm isl}}\frac{N_{\rm nisl}!}{N_{\rm cls}!\, 2^{N_{\rm cls}}}
=
{cN\choose aN} \frac{[(c-a)N]!}{\left[\frac{(c-a)N}{2}\right]!\,2^{\frac{(c-a)N}{2}}}
\]
so altogether we get
\be\label{loss-times-gain}
\epsilon_0^{N_{\rm nisl}}\left[\frac{\rho v_{\beta,~\alpha^{\rm cp}}}{N}\right]^{N_{\rm nisl}-N_{\rm cls}}\frac{(N_{\rm isl}+N_{\rm nisl})!}{N_{\rm isl}!\, N_{\rm cls}!\,2^{N_{\rm cls}} }
=
\left[\frac{\epsilon\rho v_{\beta,~\alpha^{\rm cp}}}{N}\right]^{\frac{(c-a)N}{2}}\frac{(cN)!}{(aN)!\,\left[\frac{(c-a)N}{2}\right]!\,2^{\frac{(c-a)N}{2}}}
\sim
 \left[\left(\frac{\epsilon\rho v_{\beta,~\alpha^{\rm cp}}}{e(c-a)}\right)^{\frac{c-a}{2}}\frac{c^c}{a^a}\right]^N.
\ee
Here $\epsilon=\epsilon_0^2$.
For any nonzero $\rho v_{\beta,~\alpha^{\rm cp}}$, choosing $a$ so that $0<c-a<\epsilon\rho v_{\beta,~\alpha^{\rm cp}}/e$, the expression above increases exponentially with $N$.
Note also that it equals 1 if $a=c$ (no coupling) and tends to $[c\epsilon\rho v_{\beta,~\alpha^{\rm cp}}/e]^{cN/2}$ if $a$ goes to zero (full coupling). This shows that in order to obtain an exponentially large contribution for arbitrarily small $\rho v_{\beta,~\alpha^{\rm cp}}$ both $N_{\rm isl}$ and $N_{\rm cls}$ must be proportional to $N$.
Our choice of $(c-a)N/2$ two-circles as clusters exemplifies the general case.

We still must count with the increase of $\overline{\left(X^l_{^\cdot}\right)^2}-\overline{X^l_{^\cdot}}^2$ appearing in Eq.~(\ref{Phi^N_n-new-notation}) in the mean value of
\bea\label{origin-corr-factor}
\lefteqn{
\exp\left\{
-\pi\lambda_\beta^2\sum_{l=0}^p n_l\left[
\overline{\left(X^l_{^\cdot}\right)^2}-\overline{X^l_{^\cdot}}^2-\left(\overline{\left(X^l_{^\cdot}\right)^2|_{l}}-\overline{X^l_{^\cdot}|_{l}}^2  \right) \right]\right\}
\prod_{l=0}^p\frac{f_{n_l}\left(0;L\overline{X^l_{^\cdot}}\right)}{f_{n_l}\left(0;L\overline{X^l_{^\cdot}|_l }\right)}          }
\nonumber\\
&&=
\exp\left\{
\sum_{l=0}^p\left[-\pi\lambda_\beta^2 n_l
\left(\overline{\left(X^l_{^\cdot}\right)^2|_{\neg l}} - \overline{X^l_{^\cdot}|_{\neg l}}^2 +2\,\left[\overline{X^l_{^\cdot}|_l \cdot X^l_{^\cdot}|_{\neg l}}- \overline{X^l_{^\cdot}|_l} \cdot \overline{X^l_{^\cdot}|_{\neg l}}\right]\right) \right] \right\}
\prod_{l=0}^p\frac{f_{n_l}\left(0;L\overline{X^l_{^\cdot}}\right)}{f_{n_l}\left(0;L\overline{X^l_{^\cdot}|_l }\right)}.
\nonumber\\
\eea
The mean value of the above expression can be represented by
\be\label{corr-factor}
e^{-c_1 N_{\rm nisl}\lambda_\beta^2\rho^{2/d}}
=\left[ e^{-c_1\lambda_\beta^2\rho^{2/d}} \right]^{(c-a)N}
\ee
with a suitably chosen positive constant $c_1$.
This is explained as follows.\\
(i) In the exponent $n_l$ multiplies averages that involve division with $n_l$, see Eq.~(\ref{Xav-decomposition}), so it only neutralizes this division in $\overline{\left(X^l_{^\cdot}\right)^2}$, $\overline{\left(X^l_{^\cdot}\right)^2|_{l}}$, $\overline{\left(X^l_{^\cdot}\right)^2|_{\neg l}}$ and $\overline{X^l_{^\cdot}|_l \cdot X^l_{^\cdot}|_{\neg l}}$, and leaves behind a factor $1/n_l$ in the other terms. \\
(ii) $f_{n_l}\left(0;L\overline{X^l_{^\cdot}}\right)/f_{n_l}(0;L\overline{X^l_{^\cdot}|_l })$ tends to 1 if $n_l$ remains finite or increases slower than $L^2$ as $L\to\infty$. Also, $\ln f_{n_l}\left(0;L\overline{X^l_{^\cdot}}\right)-\ln f_{n_l}(0;L\overline{X^l_{^\cdot}|_l })\approx -\pi n_l\lambda_\beta^2\left(\overline{X^l_{^\cdot}}^2-\overline{X^l_{^\cdot}|_l }^2\right)$ is of order 1 if $n_l$ increases faster than $L^2$, cf. Eqs.~(\ref{f_n(0)-asymp}), (\ref{P-X0}) and (\ref{E-X0-square}). 
\\
(iii) $\overline{\left(X^l_{^\cdot}\right)^2}-\overline{X^l_{^\cdot}}^2-\left(\overline{\left(X^l_{^\cdot}\right)^2|_{l}}-\overline{X^l_{^\cdot}|_{l}}^2  \right)=0$ if cycle $l$ is uncoupled, therefore the number of terms contributing to the sum is proportional to $N_{\rm nisl}$.\\
(iv)
The $\rho$-dependence can be found by a physical argument.
In the exponent $\lambda_\beta^2$ must be divided with a squared length, and the only relevant length here is the mean distance between neighboring cycles which scales as $\rho^{-1/d}$.
In $~\alpha^{\rm cp}$ there is no information about the spatial position of the pairs that appear with $\alpha^k_j>0$, but such an information is present in $x^k_{j,r}$, the dual-space vector associated with the difference of the position vectors of particles $j$ and $k$. The only physically meaningful interpretation of $\|~\alpha^{\rm cp}\|_1\propto N_{\rm nisl}$ is that the clusters are formed by neighboring cycles, and the larger the density, the stronger is the interaction of a cycle with its neighbors:
The exponent originates from the kinetic energy, the repulsive interaction has the tendency to confine the particles, and due to the uncertainty principle the kinetic energy increases with $\rho$ as $\rho^{2/d}$. This adds to the mean-field repulsion already found among the uncoupled cycles.

To fully account for the effect of coupling (\ref{corr-factor}) must still multiply (\ref{loss-times-gain}). The result is
\bea\label{corr-1}
\mbox{(\ref{loss-times-gain})$\times$(\ref{corr-factor})}
 &=&
 \left[\left(\frac{\epsilon\rho v_{\beta,~\alpha^{\rm cp}}}{e(c-a)}\right)^{\frac{c-a}{2}}\frac{c^c}{a^a}
 \left[e^{-c_1\lambda_\beta^2\rho^{2/d}} \right]^{c-a}\right]^N.
\eea
This is exponentially increasing if
\[
\sqrt{\frac{\epsilon\rho v_{\beta,~\alpha^{\rm cp}}}{c-a}} \left(\frac{c^c}{a^a} \right)^{\frac{1}{c-a}}e^{-c_1\lambda_\beta^2\rho^{2/d}-\frac{1}{2}} >1,
\]
which can be attained for any $\beta$ and $\rho$ by choosing $a<c$ close enough to $c$. Because $c^c/a^c>1$, it holds for
\[
c-a\leq  \epsilon\rho v_{\beta,~\alpha^{\rm cp}}e^{-2c_1\lambda_\beta^2\rho^{2/d}-1}.
\]
The expression (\ref{corr-1}) has a maximum as a function of $a$ somewhere between 0 and $c$. Neglecting $c^c/a^a$, (\ref{corr-1}) is maximal at
\[
c-a=\epsilon\rho v_{\beta,~\alpha^{\rm cp}}e^{-2(c_1\lambda_\beta^2\rho^{2/d}+1)}.
\]
With the maximizing $c-a$ one obtains the exponential factor by which $\tilde{Q}_{N,L}$ exceeds $\tilde{Q}_{N,L}^{\rm dcp}\,$:
\be\label{QNL-increased-1}
\tilde{Q}_{N,L}
=e^{CN} \tilde{Q}_{N,L}^{\rm dcp},
\qquad\qquad
C=\frac{1}{2}\epsilon \rho v_{\beta,~\alpha^{\rm cp}}e^{-2(c_1\lambda_\beta^2\rho^{2/d}+1)}.
\ee

Let us examine another example which is in some sense the opposite of the one we discussed above:
the total number of cycles is still $p+1=cN$ but all are coupled into a single circle. Then, we have $m_{~\alpha^{\rm cp}}=1$, $K_{~\alpha^{\rm cp}}=cN-1$ and the number of $~\alpha^{\rm cp}$ that differ only in the permutations of the vertices is $[cN-1]!$. So the loss times the gain -- still without the factor (\ref{corr-factor}) -- is
\be\label{loss-times-gain-2}
\epsilon_0\left[\frac{\epsilon_0\rho v_{\beta,~\alpha^{\rm cp}}}{N}\right]^{cN-1}[cN-1]!   \sim \epsilon_0\left[\frac{c\epsilon_0\rho v_{\beta,~\alpha^{\rm cp}}}{e} \right]^{cN-1}.
\ee
If $c\epsilon_0\rho v_{\beta,~\alpha^{\rm cp}}<e$, this is vanishingly small. On the other hand, for $\rho v_{\beta,~\alpha^{\rm cp}}$ large enough (\ref{loss-times-gain-2}) is much larger than (\ref{loss-times-gain}), because $\rho v_{\beta,~\alpha^{\rm cp}}/e$ is raised to a power more than twice as large as in (\ref{loss-times-gain}). (Actually, this is the largest possible power: once $p$ is given, the largest exponent is obtained for the smallest $m_{~\alpha^{\rm cp}}$ which is 1, so $\max K_{~\alpha^{\rm cp}}=p$. To compensate $N^{-p}$ we then need the largest multiplicity, $p!$, that we can get by coupling the $p+1$ vertices into a single circle.)
Together with (\ref{corr-factor}),
\bea\label{corr-2}
\mbox{(\ref{loss-times-gain-2})$\times$(\ref{corr-factor})}
=
\frac{e}{c\rho v_{\beta,~\alpha^{\rm cp}}}
\left[c\epsilon_0\rho v_{\beta,~\alpha^{\rm cp}}
e^{-c_1\lambda_\beta^2\rho^{2/d}-1} \right]^{cN}.
\eea
The base of the exponential is smaller than 1 if both $\beta$ and $\rho$ are small and also if any of them is large. One may have an idea of what should be substituted for $c$, i.e. for $p$:
it can be the average number of cycles,
\be\label{expected-p}
\langle p\rangle_{N,L}=\frac{N}{\rho}\sum_{k=1}^{N}\frac{\rho^{N,L}_k}{k}.
\ee
Indeed, $N\rho^{N,L}_k/\rho$ is the expected number of particles in cycles of length $k$, $N\rho^{N,L}_k/k\rho$ is the expected number of k-cycles formed by $N$ particles, so (\ref{expected-p}) is the expected total number of elements in the partition of $N$. If there are no infinite cycles in the infinite system, by interchanging limit and summation
\be\label{full-no-infinite-cycles}
\lim_{N, L\to\infty, N/L^d=\rho}\frac{\langle p\rangle_{N,L}}{N}=
\frac{1}{\rho}\sum_{k=1}^{\infty}\frac{\rho_k}{k}=:B(\rho,\beta)<1.
\ee
Thus, a reasonable choice is $c=B(\rho,\beta)$. However, because we do not know the interplay among all the parameters (including those of $\hat{u}$), we cannot assert with certainty that for some intermediate values of $\beta$ and $\rho$ the base in (\ref{corr-2}) cannot be larger than 1. If this was true, (\ref{corr-2}) would dominate (\ref{corr-1}) with the implication of a curious reentrant phase transition in addition to BEC. Still, there would result a factor $e^{CN}$ multiplying $\tilde{Q}_{N,L}^{\rm dcp}$, and from the point of view of BEC the precise $\rho$- and $\beta$-dependence of $C$ is unimportant.
Dividing
\be
\tilde{Q}_{N,L}
=
e^{CN}\tilde{Q}_{N,L}^{\rm dcp}=e^{CN}\frac{1}{N}\sum_{n=1}^{N}\Tilde{G}^n_{n}\tilde{Q}_{N-n,L}^{\rm dcp}
\ee
with $\tilde{Q}_{N,L}$ we are back to the equation
\be
1=\frac{1}{N}\sum_{n=1}^N\Tilde{G}^n_{n} \frac{\tilde{Q}_{N-n,L}^{\rm dcp}}{\tilde{Q}_{N,L}^{\rm dcp}}
\ee
and the result about BEC in the cycle-decoupling model. In particular, $\zeta_c(\beta)=\zeta^{\rm dcp}(\beta)$, the critical density $\zeta_c(\beta)/\lambda_\beta^d$ for the full model is the same as for the cycle-decoupling model. What changes is the free energy and the relation between the density and the chemical potential, just as with the inclusion of the mean-field term. However, contrary to the latter there can be a nonanalytic change also in the inter-cycle contribution to $f(\rho,\beta)$.
This ends the proof of the theorem.

\newsec{Physical meaning of the permutation cycles}

Here we propose a possible interpretation of the permutation cycles as physical entities.
When in the stochastic description $n$ physical particles form an effective single-particle trajectory then quantum-mechanically they occupy one and the same one-particle state. About the nature of this state we can be guided by inspecting the noninteracting gas and its partition function
\be
Q^0_{N,L}=\sum_{p=1}^N
\frac{1}{p!}\sum_{n_1,\dots,n_p\geq 1:\sum_1^p n_l=N}\ \prod_{l=1}^p\frac{1}{n_l}
\sum_{z\in\Zz^d}\exp\left\{-\frac{\pi n_l \lambda_\beta^2}{L^2}z^2\right\}.
\ee
Given $\{n_l\}_1^p$ and uniformly distributed random vectors $\{y_l\}_1^p$ in $\Lambda$, all the $n_l$ particles of the $l$th cycle are in a superposition of the plane wave states $L^{-d/2}e^{\i \frac{2\pi}{L}z\cdot (x_l-y_l)}$ with weights $\exp\{-\pi n_l\lambda_\beta^2 z^2/(2L^2)\}$,
\be\label{cycle-state-ideal}
\psi^{L}_{n_l,y_l}(x_l)
=\frac{
\sum_{z\in\Zz^d}\exp\left\{-\frac{\pi n_l\lambda_\beta^2}{2L^2} z^2\right\}L^{-d/2}e^{\i \frac{2\pi}{L}z\cdot (x_l-y_l)}}
{\left[\sum_{z\in\Zz^d}\exp\left\{-\frac{\pi n_l\lambda_\beta^2}{L^2} z^2\right\}  \right]^{1/2}}
=
\left(\frac{2}{\sqrt{n_l}\lambda_\beta}\right)^{d/2}
\frac{\sum_{z\in\Zz^d}\exp\left\{-\frac{2\pi(x_l-y_l+Lz)^2}{n_l\lambda_\beta^2}\right\}}
{\left[\sum_{z\in\Zz^d}\exp\left\{-\frac{\pi L^2z^2}{n_l\lambda_\beta^2}\right\}\right]^{1/2}}.
\ee
The thermal equilibrium state (density matrix) of the noninteracting Bose gas is a mixed state of the form
\be
{\cal D}^0_{N,L}=\sum_{p=1}^N \frac{1}{p!}\sum_{n_1,\dots,n_p\geq 1:\sum_1^p n_l=N}\ \frac{1}{\prod_{l=1}^pn_l}
\bigotimes_{l=1}^p L^{-d}\int_\Lambda |\psi^{L}_{n_l,y_l}\rangle\langle \psi^{L}_{n_l,y_l}|\, \d y_l
\ee
If $\rho<\zeta(d/2)/\lambda_\beta^d$, the cycle lengths remain finite in the infinite system, and
\be\label{psi^{infty}_{n_l,y_l}(x_l)}
\lim_{L\to\infty}\psi^{L}_{n_l,y_l}(x_l)=\left(\frac{2}{\sqrt{n_l}\lambda_\beta}\right)^{d/2}
\exp\left\{-\frac{2\pi(x_l-y_l)^2}{n_l\lambda_\beta^2}\right\},
\ee
the $l$th cycle represents a Gaussian localized state of width $\propto\lambda_{n_l\beta}$ centered at $y_l$. If $\rho>\zeta(d/2)/\lambda_\beta^d$, some $n_l$ will diverge proportionally to $N$. As seen on its first form, $\psi^{L}_{n_l,y_l}(x_l)$ tends to the zero momentum plane wave,
$
\psi^{L}_{n_l,y_l}(x_l)\sim L^{-d/2}
$
as $L$ increases.

In the interacting Bose gas $\prod_{l=1}^p q_{n_l}$ is replaced with $\Tilde{G}\left[{\{n_l\}_1^p}\right]$ and $Q^0_{N,L}$ with
\[
\tilde{Q}_{N,L}
=\sum_{p=1}^N \frac{1}{p!}\sum_{n_1,\dots,n_p\geq 1:\sum_1^p n_l=N} \frac{1}{\prod_{l=1}^p n_l}\
\Tilde{G}\left[{\{n_l\}_1^p}\right],
\]
see Eq.~(\ref{QNL-average-1-to-p}).
However, once $\{\alpha^k_j, x^k_{j,r}, t^k_{j,r}\}$ are given, for every $l$ it suffices to consider
\[
\sum_{z\in\Zz^d} \exp\left\{-\frac{\pi \lambda_\beta^2}{L^2}\sum_{q\in C_l}\int_0^1\left[z+Z_q(t)\right]^2\d t\right\}
=
\exp\left\{-\pi n_l \lambda_\beta^2\left[\overline{\left(X^l_{^\cdot}\right)^2}-\overline{X^l_{^\cdot}}^2\right]\right\}
\sum_{z\in\Zz^d}\exp\left\{-\pi n_l \lambda_\beta^2\left(\frac{z}{L}+\overline{X^l_{^\cdot}}\right)^2\right\}.
\]
In Section 3 we already have interpreted $\hbar(2\pi/L)Z_q(t)=hX_q(t)$ as the shift due to interactions of the momentum of the $q$th particle at "time" $t$ compared to its value in the ideal gas, and $\overline{X^l_{^\cdot}}$ is the average over time and particles in the $l$th cycle of $X_q(t)$. Accordingly, the common wave function of the $n_l$ particles in cycle $l$ is
\bea
\psi^{L,\overline{X^l_{^\cdot}}}_{n_l,y_l}(x_l)
&=&\frac{\sum_{z\in\Zz^d}\exp\left\{-\frac{\pi n_l\lambda_\beta^2}{2L^2} (z+L\overline{X^l_{^\cdot}})^2\right\}L^{-d/2}e^{\i \frac{2\pi}{L}(z+L\overline{X^l_{^\cdot}})\cdot (x_l-y_l)}}
{\left[\sum_{z\in\Zz^d}\exp\left\{-\frac{\pi n_l\lambda_\beta^2}{L^2} (z+L\overline{X^l_{^\cdot}})^2\right\}  \right]^{1/2}}
\nonumber\\
&=&\left(\frac{2}{\sqrt{n_l}\lambda_\beta}\right)^{d/2}\
\frac{\sum_{z\in\Zz^d}\exp\left\{-\frac{2\pi(x_l-y_l+Lz)^2}{n_l\lambda_\beta^2}\right\}\exp\left\{-\i 2\pi z\cdot L\overline{X^l_{^\cdot}}\right\}}
{\left[\sum_{z\in\Zz^d}\exp\left\{-\frac{\pi L^2z^2}{n_l\lambda_\beta^2}\right\}\cos2\pi z\cdot L\overline{X^l_{^\cdot}} \right]^{1/2}}.
\eea
Here (as in (\ref{cycle-state-ideal})) we applied the identity
\be
\frac{1}{L^d}\sum_{z\in\Zz^d}e^{-\frac{\pi\lambda^2}{L^2}(z+a)^2}e^{\i\frac{2\pi}{L}(z+a)\cdot x}
=\frac{1}{\lambda^d}\sum_{z\in\Zz^d}e^{-\frac{\pi}{\lambda^2}(x+Lz)^2}e^{-\i 2\pi z\cdot a}.
\ee
With
\be
\Tilde{G}\left[{\{n_l\}_1^p}\right]=\sum_{\{\alpha^k_j\in\Nn_0|1\leq j< k\leq N\}}\int_0^1\prod_{j,k,r}\d t^k_{j,r}\int\prod_{j,k,r}\d x^k_{j,r} H\left(\{\alpha^k_j, t^k_{j,r},x^k_{j,r}\}\right),
\ee
where $H$ can be read off from Eq.~(\ref{Phi^N_{n,{n_l}_1^p}}), the density matrix is
\bea
\lefteqn{
{\cal D}_{N,L}
=\sum_{p=1}^N \frac{1}{p!}\sum_{n_1,\dots,n_p\geq 1:\sum_1^p n_l=N}\ \frac{1}{\prod_{l=1}^pn_l}
}\nonumber\\
&&\frac{1}{\Tilde{G}\left[{\{n_l\}_1^p}\right]} \sum_{\{\alpha^k_j\in\Nn_0|1\leq j< k\leq N\}}\int_0^1\prod_{j,k,r}\d t^k_{j,r}\int\prod_{j,k,r}\d x^k_{j,r} H\left(\{\alpha^k_j, t^k_{j,r},x^k_{j,r}\}\right)
\bigotimes_{l=1}^p L^{-d}\int_\Lambda |\psi^{L,\overline{X^l_{^\cdot}}}_{n_l,y_l}\rangle\langle \psi^{L,\overline{X^l_{^\cdot}}}_{n_l,y_l}|\, \d y_l.
\nonumber\\
\eea
If $\rho<\zeta_c(\beta)/\lambda_\beta^d$, the cycle lengths remain finite in the infinite system, and the wave function in infinite volume reduces to the $z=0$ term (\ref{psi^{infty}_{n_l,y_l}(x_l)}) independent of the interaction. It is only when $n_l\lambda_\beta^2/L^2\to\infty$ and $\overline{X^l_{^\cdot}}=O(1/\sqrt{n_l})$, and thus $L\overline{X^l_{^\cdot}}\to 0$, that the wave function tends to the pure zero-momentum state. These are precisely the conditions that we found for BEC and proved to be met for positive and positive-type pair potentials at densities $\rho> \zeta_c(\beta)/\lambda_\beta^d$.
Although asymptotically the individual cycles represent the same states as in the noninteracting gas, their distribution is different because of their coupling via wave vectors.

\newsec{Historical notes}

In 1924 Bose gave a deceptively simple statistical physical derivation of Planck's radiation formula [Bos]. To obtain the good result all he had to do was to count the cells of volume $h^3$ of the phase space in which the light quanta are distributed somewhat differently than usual. In an endnote the German translator Einstein praised the work and promised to apply its method to an ideal atomic gas, that he indeed did in three papers [E1-3].
The surprising result, today known as Bose-Einstein condensation,
was not received with much enthusiasm. Distinguished colleagues as Halpern, Schr\"odinger or Smekal had difficulty to understand the new "cell counting" that Bose, and Einstein in [E1], used instead of Boltzmann's. Einstein answered the objections in papers [E2,3] and also in letters, see e.g. [Schr] and the very clear response [E4].
Somewhat later another blow came from Uhlenbeck. To quote London [Lon2],

\noindent
\emph{"This very interesting discovery, however, has not appeared in the textbooks, probably because Uhlenbeck in his thesis {\rm [Uh]} questioned the correctness of Einstein's argument. Since, from the very first, the mechanism appeared to be devoid of any practical significance, all real gases being condensed at the temperature in question, the matter has never been examined in detail; and it has been generally supposed that there is no such condensation phenomenon."}

The regard onto Einstein's work changed in 1938 with the discovery of superfluidity [All, Kap]. Fritz London promptly reacted [Lon1], and in a follow-up paper [Lon2] he detailed his view, that superfluidity must have to do with Bose-Einstein condensation -- the name was coined by him. To support his idea, he computed the critical temperature of the ideal gas with the mass of the He4 atom and the density of liquid helium, and found it not very far off, 1K above the $\lambda$ point separating the He I and He II phases. Prior to that he had to reexamine the controversy between Einstein and Uhlenbeck, and take Einstein's side. In his derivation Einstein arrived at an equation connecting the number of particles $N$ to the chemical potential $\mu$, that one must solve for the latter. $\mu$ appears in an infinite sum over the allowed discrete values of the single-particle momentum. Einstein approximated the sum with an integral and observed the convergence of the integral in three dimensions at $\mu=0$, the largest possible value of the chemical potential: as if $N$ could not go beyond a maximum. At the beginning of his second paper he resolved this paradox by assigning the surplus particles to the zero momentum mode. Uhlenbeck, certainly unaware of Ehrenfest's earlier and identical criticism, argued that without the approximation by an integral the paradox disappears, the original equation can be solved with a $\mu<0$ for arbitrarily large $N$. At that time two crucial mathematical elements were missing from the weaponry of theoretical physicists: a clear notion of the thermodynamic limit and its importance to see a sharp phase transition in the framework of statistical physics, and the appearance of the Dirac delta in probability theory, the fact that in the limit of a sequence of discrete probability distributions an atomic measure can emerge on a continuous background. Einstein's intuition worked correctly, he tacitly performed the thermodynamic limit. The separate treatment of the zero momentum state bothered physicists for a long time, including Feynman, who proposed an alternative derivation based on the statistics of permutation cycles [Fe2]. Simultaneously with London's publications, in a paper written with Kahn, Uhlenbeck also admitted that Einstein was right [Kah]. Tisza published his two-fluid theory about the same time [T1,2], and attributed the specific transport properties of helium II to BEC. However, the argument against describing a strongly interacting dense system of atoms with an ideal gas persisted and set the task: prove BEC in the presence of interaction.

The first consistent theory of superfluidity was given by Landau in 1941 [Lan1]. This extremely influential work denied all connection with BEC (clearly, a position taken against London and even more Tisza, who was earlier in his group in Kharkiv; see also Kadanoff [Kad]). The research on BEC for interacting bosons started after World War II and produced a huge number of papers that we can review only in great lines. Monographs about it and its connection with superfluidity extend over decades, some of them are [Lon4, No, Gr, Sew, Pet, Pi1, Li5, Leg, Ued, Ver, Kag]; those written after 2000 usually cover also the theory of trapped dilute ultra-cold gases of alkaline atoms, that we will not discuss.

Maybe the first and certainly one of the most important contributions was that of Bogoliubov, who described superfluidity on the basis of BEC of weakly interacting bosons [Bog1]; for a review see [Z]. His theory had an immense impact on the forthcoming research in quantum statistical physics. Bogoliubov initiated the algebraic approach, the use of second quantization. Writing the Hamiltonian in terms of creation and annihilation operators proved to be very fruitful, because it opened the way for diverse approximations. The interaction, which now must be integrable, appears in a quartic expression. Applying $c$-number substitution and truncation of the quartic sum made it possible to understand many features of the physics of the interacting Bose gas.
These models, including those expressible with number operators and known under the names of mean-field, imperfect, perturbed mean-field, full diagonal Bose gas model, lead naturally to BEC [Hu4, Dav, Fa1, Buf, Ber1-Ber4, Lew1, Do1, Do2]. (For a discussion of some problems arising with the truncation of the Hamiltonian see the Introduction of [Su10].) The justification of the $c$-number substitution was the subject of later papers [G5, Su8, Su9, Li6]. Another contribution of Bogoliubov, his inequality and $1/q^2$-theorem [Bog2] became the main tool to prove the absence of BEC -- and the breakdown of a continuous symmetry in general -- at positive temperatures in one- and two-dimensional quantum-mechanical models.

Analytical methods permit to get closer to the problem of liquid helium. A popular one of the fifties was the use of pseudo-potentials [Hu3, Hu4]. However, the {\em par excellence} analytical method is functional integration -- the one we employed in this paper. Feynman devised it to solve the time-dependent Schr\"odinger equation and to compute $\left\langle x\left|e^{-it H/\hbar}\right| x\right\rangle$ [Fe1]. The mathematical justification for imaginary time $it=\hbar\beta$, where $\beta$ is real positive, as a functional integral with the Wiener measure came from Kac [Kac1,2]. The acknowledgement in [Fe2] reveals that it must have been Kac who convinced Feynman to apply what we call today the Feynman-Kac formula to the $\lambda$-transition of liquid helium. The result was three seminal papers [Fe2-4], the first of which was devoted to a first-principle study of the $\lambda$-transition. Although Feynman was not interested in mathematical rigor, he seized a point which gained importance with time: the appearance of long permutation cycles during the transition.

In 1954 London published his macroscopic theory of superfluidity [Lon3] whose emblematic element is a macroscopic wave function associated with the condensate. This and Landau's theory [Lan1,2] constitute the basis for the phenomenological description of superfluidity. Another cardinal contribution from the fifties was due to Oliver Penrose and Onsager [Pen]; we discussed it in the Introduction.
Their characterization of BEC with the non-decay of the off-diagonal element of the integral kernel of $\sigma_1=\lim_{N,L\to\infty}\sigma^{N,L}_1$ was later extended into the notion of off-diagonal long-range order [Ya]. They gave the first and surprisingly precise estimate of the condensate fraction in the ground state of liquid helium, $\sim 8\%$, only slightly modified later by a sophisticated numerical calculation [Ce] and deduction from experiment [Sn], [So]. They extended their study to positive temperatures and found a connection between BEC and the fraction of particles in "large" permutation cycles.

In 1960 the start of the Journal of Mathematical Physics marked the adulthood of a new discipline. While the theory of BEC seemed more or less settled for most physicists, their more math-minded colleagues saw there a field to explore. The mathematical results of this decade are mostly negative, they prove the absence of BEC in different situations: at low fugacity, in one and two dimensions at positive temperatures, and in one dimension in the ground state. In three thorough and difficult papers Ginibre adapted the method of combining Banach space technics with the Kirkwood-Salzburg equation [Bog3, R2, R4, R5, Bog4] to quantum statistics, and proved the existence, analyticity and exponential clustering of the reduced density matrices at low fugacity in the thermodynamic limit [G2-4], see also [G1]. This remains until today the most elaborate application of the path integral method in statistical physics. For the quantal version of Tonks' hard rod model [Ton] Girardeau established the ground state, easily deducible from that of the free Fermi gas [Gi]. However, even the explicit knowledge of the ground state did not permit to decide quickly about BEC; finally, it was shown not to exist [Schu, Len]. The analogous model with soft-delta interaction is more difficult, it was solved with Bethe Ansatz by Lieb and Liniger for the ground state [Li1] and by Lieb for the excited states [Li2]. There is probably no BEC in this model either, approximate methods predict an algebraic decay of the off-diagonal correlation [Hal, Cr, Ko]. The general one-dimensional case was studied much later, with the conclusion that neither diagonal nor off-diagonal long-range order can exist in the ground state provided that the compressibility is finite [Pi2]. The question of the breakdown of a continuous symmetry at positive temperatures in one and two dimensions was discussed in generality by Wagner [W]. BEC can be considered as an instance of such a breakdown: that of gauge invariance [Fa2, Su8, Su9, Li6]. Its absence was shown by Hohenberg [Ho], the completion of the proof with the extension of Bogoliubov's inequality to unbounded operators was done later [Bou].

The most important result of the seventies was the long-awaited first proof of BEC in a system of interacting bosons in three dimensions. This was done on the cubic lattice for hard-core bosons at half-filling by Dyson, Lieb and Simon [Dy]. The extension to the ground state on the square lattice took another decade [Ken, Kub].
The system can also be considered as that of half spins with XY- or antiferromagnetic interaction, and the work was preceded by the proof of ordering in the classical Heisenberg model [Fr], applying partly the same methods.
In the seventies two schools, one in Leuven directed by Andr\'e Verbeure and another in Dublin under John Lewis' leadership started a research program in quantum statistical physics, with special emphasis on different aspects of BEC. The main tool of the Leuven School was operator algebraic. The Dublin School dominantly worked on BEC in the free Bose gas with different domain shapes, boundary conditions, imperfection, for bosons with spins, etc. Later this group pioneered the application of the Large Deviation Principle to problems of the Bose gas [Lew1, Ber2, Ber3, Ber4]. The intertwined activity of these two schools is reviewed in Verbeure's book [Ver].

In the mathematical literature of the interacting Bose gas most often either the pair potential or its Fourier transform or both are chosen to be nonnegative, see e.g. [Li7, Car, MP1, MP2, Su10]  -- in Theorem~\ref{second-step} we needed both. Beside the argument we gave in the Introduction another reason is specific to dilute systems: the effect of a nonnegative spherical pair potential in the dilute limit reduces to $s$-wave scattering, so the interaction can be characterized by a single parameter, the $s$-wave scattering length $a$. This made it possible to obtain rigorously the ground state energy [Li3, Li4] and the free energy [Se1, Yi] in the dilute gas limit. Yet another question in which the positivity of the interaction played a role is the sign of the shift $\Delta T_c=T_c-T^0_c$ of the critical temperature in the dilute limit. Here $T_c$ and $T^0_c$ are the critical temperatures of the interacting and the ideal gas, respectively. There is a well-known argument saying that $r$-space repulsion implies $k$-space attraction [Hu1-Hu4]; so BEC should be easier for repulsive bosons, $T_c>T^0_c$. In a long debate a consensus has formed that the shift was positive, and $\Delta T_c/T^0_c\approx 1.3\sqrt{a\rho^{1/3}}$ in three dimensions. Although the question is pertinent, it cannot be decided without proving $T_c>0$, which was done only for the free Bose gas in a nonnegative external field, where $\Delta T_c>0$ follows from the min-max principle [Kac3]. A rigorous upper bound on $\Delta T_c$ and a review of the related literature can be found in [Se2]. The only case offering the possibility of a comparison with experiment is the $\lambda$-transition of liquid helium. Here the shift is negative [Lon2], but the system is dense, and the pair potential acting in it has an attractive tail.

Explicit summation over the symmetric group is part of the first-quantized treatment of quantum many-body systems, in particular when the Feynman-Kac formula is applied. The more elegant algebraic method has its limits, and from the early nineties there has been a substantial reappearance of the path integral technic in quantum statistical mechanics.
A paper by Aizenman and Lieb [Ai1] used it to prove the partial survival of Nagaoka ferromagnetism in the Hubbard model at positive temperatures. T\'oth [Tth] proved BEC of hard-core bosons on the complete graph. Aizenman and Nachtergaele [Ai2] studied ordering in the ground state of quantum spin chains. Ceperly [Ce] applied the path-integral Monte Carlo method to a thorough numerical analysis of the superfluidity of liquid helium. The present author picked up the thread left by Feynman and discussed BEC of particles in continuous space in connection with the probability distribution of permutation cycles [Su1, Su2].
The revival of interest in the relation between BEC and infinite cycles gave rise to many other papers, e.g. [Bun], [Scha], [Uel1], [Uel2], [Ben], [Do3], [Ad1]. There appeared also a new field of research on Hamiltonian models of random permutations, apparently more amenable to study by functional integration and large deviations analysis; see e.g. [Bet1], [Bet2], [Bet3], [El], [Ad2]. It is quite possible that with time this interesting new subject will acquire importance in pure probability theory.

\vspace{10pt}\noindent
{\bf Acknowledgement}

\vspace{5pt}
\noindent
This work was supported by the Hungarian Scientific Research Fund (OTKA) through Grant No. K128989.

\newpage
\noindent{\Large\bf References}
\begin{enumerate}
\item[{[Ad1]}] Adams S., Collevecchio A., and K\"onig W.: {\em A variational formula for the free energy of an interacting many-particle system.} Ann. Prob. {\bf 39}, 683-728 (2011).
\item[{[Ad2]}] Adams S. and Dickson M.: {\em An explicit large deviations analysis of the spatial cycle Huang-Yang-Luttinger model.} Ann. Henri Poincar\'e {\bf 22}, 1535-1560 (2021).
\item[{[Ai1]}] Aizenman M. and Lieb E. H.: {\em Magnetic properties of some itinerant-electron systems at $T>0$.} Phys. Rev. Lett. {\bf 65}, 1470-1473 (1990).
\item[{[Ai2]}] Aizenman M. and Nachtergaele B.: {\em Geometric aspects of quantum spin states.} Commun. Math. Phys. {\bf 164}, 17-63 (1994).
\item[{[All]}] Allen J. F. and Misener D.: {\em Flow of liquid helium II.} Nature {\bf 141},  75 (1938), and {\bf 142}, 643-644 (1938).
\item[{[Ben]}] Benfatto G., Cassandro M., Merola I. and Presutti E.: {\em Limit theorems for statistics of combinatorial partitions with applications to mean field Bose gas.} J. Math. Phys. {\bf 46}, 033303 (2005).
\item[{[Ber1]}] van den Berg M., Lewis J. T and de Smedt Ph: {\em Condensation in the imperfect boson gas.} J. Stat. Phys. {\bf 37}, 697-707 (1984).
\item[{[Ber2]}] van den Berg M., Lewis J. T. and Pul\'e J. V.: {\em The Large Deviation Principle and some models of an interacting Boson gas.} Commun. Math. Phys. {\bf 118}, 61-85 (1988).
\item[{[Ber3]}] van den Berg M., Dorlas T. C., Lewis J. T. and Pul\'e J. V.: {\em A perturbed mean field model of an interacting boson gas and the Large Deviation Principle.} Commun. Math. Phys. {\bf 127}, 41-69 (1990).
 \item[{[Ber4]}] van den Berg M., Dorlas T. C., Lewis J. T. and Pul\'e J. V.: {\em The pressure in the Huang-Yang-Luttinger model of an interacting Boson gas.} Commun. Math. Phys. {\bf 128}, 231-245 (1990).
\item[{[Bet1]}] Betz V. and Ueltschi D.: {\em Spatial random permutations and infinite cycles.} Commun. Math. Phys. {\bf 285}, 469-501 (2009).
\item[{[Bet2]}] Betz V. and Ueltschi D.: {\em Spatial random permutations and Poisson-Dirichlet law of cycle lengths.} Electr. J. Probab. {\bf 16}, 1173-1192 (2011).
\item[{[Bet3]}] Betz V., Ueltschi D. and Velenik Y.: {\em Random permutations with cycle weights.} Ann. Appl. Probab. {\bf 21}, 312-331 (2011).
\item[{[Bog1]}] Bogoliubov N. N.: {\em On the theory of superfluidity.} J. Phys. USSR {\bf 11}, 23-32 (1947).
\item[{[Bog2]}] Bogoliubov N. N.: {\em Quasi-averages in problems of statistical mechanics.} Dubna Report No. D-781, (1961), Ch. II (in Russian); Phys. Abhandl. Sowijetunion, 1962, {\bf 6}, 1-110; ibid., 1962, {\bf 6}, 113-229 (in German); {\em Selected Works, vol.II: Quantum Statistical Mechanics.} Gordon and Breach (N.Y. 1991); Collection of Scientific Papers in 12 vols.: Statistical Mechanics, vol.6, Part II. Nauka (Moscow 2006).
\item[{[Bog3]}] Bogolyubov N. N. and Khatset B. I.: {\em On some mathematical problems of the theory of statistical equilibrium.} Dokl. Akad. Nauk SSSR {\bf 66}, 321 (1949).
\item[{[Bog4]}] Bogolyubov N. N., Petrina D. Ya. and Khatset B. I.: {\em Mathematical description of the equilibrium state of classical systems on the basis of the canonical ensemble formalism.} Teor. Mat. Fiz. {\bf 1}, 251-274 (1969) and Ukr. J. Phys. {\bf 53}, 168-184 (2008).
\item[{[Bos]}] Bose S. N.: {\em Plancks Gesetz und Lichtquantenhypothese.} Z. Phys. \textbf{26}, 178-181 (1924).
\item[{[Bou]}] Bouziane M. and Martin Ph. A.: {\em Bogoliubov inequality for unbounded operators and the Bose gas.} J. Math. Phys. {\bf 17}, 1848-1851 (1976).
\item[{[Buf]}] Buffet E. and Pul\'e J. V.: {\em Fluctuation properties of the imperfect Bose gas.} J. Math. Phys. {\bf 24}, 1608-1616 (1983).
\item[{[Bun]}] Bund S. and Schakel M. J.: {\em String picture of Bose-Einstein condensation.} Mod. Phys. Lett. B {\bf 13}, 349 (1999).
\item[{[Car]}] Carlen E. A., Holzmann M., Jauslin I., Lieb E. H.: {\em A fresh look at a simplified approach to the Bose gas.} Physical Review A {\bf 103}, 053309 (2021).
\item[{[Ch]}] Choquet G.:  \emph{Diam\`etre transfini et comparaison de diverses capacit\'es.} S\'eminaire Brelot-Choquet-Deny. Th\'eorie du potentiel \textbf{3}, No. 4, 1-7 (1958-1959).
\item[{[Ce]}] Ceperley D. M.: {\em Path integrals in the theory of condensed helium.} Rev. Mod. Phys. {\bf 67}, 279-355 (1995).
\item[{[Cr]}] Creamer D. B., Thacker H. B. and Wilkinson D.: {\em A study of correlation functions for the delta-function Bose gas.} Physica {\bf 20D}, 155-186 (1986).
\item[{[Dav]}] Davies E. B.: {\em The thermodynamic limit for an imperfect boson gas.} Commun. Math. Phys. {\bf 28}, 69-86 (1972).
\item[{[Do1]}] Dorlas T. C., Lewis J. T. and Pul\'e J. V.: {\em Condensation in some perturbed meanfield models of a Bose gas.} Helv. Phys. Acta {\bf 64}, 1200-1224 (1991).
\item[{[Do2]}] Dorlas T. C., Lewis J. T. and Pul\'e J. V.: {\em The full diagonal model of a Bose gas.} Commun. Math. Phys. {\bf 156}, 37-65 (1993).
\item[{[Do3]}] Dorlas T. C., Martin Ph. A. and Pul\'e J. V.: {\em Long cycles in a perturbed mean field model of a boson gas.} J. Stat. Phys. {\bf 121}, 433-461 (2005).
\item[{[Dy]}] Dyson F. J., Lieb E. H. and Simon B.: {\em Phase transitions in quantum spin systems with isotropic and nonisotropic interactions.} J. Stat. Phys. {\bf 18}, 335-383 (1978).
\item[{[E1]}] Einstein A.: {\em Quantentheorie des einatomigen idealen Gases.} Sitz.ber. Preuss. Akad. Wiss. {\bf 1924}, 261-267.
\item[{[E2]}] Einstein A.: {\em Quantentheorie des einatomigen idealen Gases. II.} Sitz.ber. Preuss. Akad. Wiss. {\bf 1925}, 3-14.
\item[{[E3]}] Einstein A.: {\em Quantentheorie des idealen Gases.} Sitz.ber. Preuss. Akad. Wiss. {\bf 1925}, 18-25.
\item[{[E4]}] Einstein A.: {\em Answer to Schr\"odinger on 28 February 1925.} The collected papers of Albert Einstein, Vol. 14, Document 446, p. 438.
\item[{[El]}] Elboim D. and Peled R.: {\em Limit distributions for Euclidean random permutations.} Commun. Math. Phys. {\bf 369}, 457-522 (2019).
\item[{[Fa1]}] Fannes M. and Verbeure A.: {\em The condensed phase of the imperfect Bose gas.} J. Math. Phys. {\bf 21}, 1809-1818 (1980).
\item[{[Fa2]}] Fannes M., Pul\'e J. V. and Verbeure A.: {\em On Bose condensation.} Helv. Phys. Acta {\bf 55}, 391-399 (1982).
\item[{[Fe1]}] Feynman R. P.: {\em Space-time approach to non-relativistic quantum mechanics.} Rev. Mod. Phys. {\bf 20}, 367-387 (1948).
\item[{[Fe2]}] Feynman R. P.: {\em Atomic theory of the $\lambda$ transition in helium.} Phys. Rev. {\bf 91}, 1291-1301 (1953).
\item[{[Fe3]}] Feynman R. P.: {\em Atomic theory of liquid helium near absolute zero.} Phys. Rev. {\bf 91}, 1301-1308 (1953).
\item[{[Fe4]}] Feynman R. P.: {\em Atomic theory of the two-fluid model of liquid helium.} Phys. Rev. {\bf 94}, 262-277 (1954).
\item[{[Fr]}] Fr\"ohlich J., Simon B. and Spencer T.: {\em Infrared bounds, phase transitions and continuous symmetry breaking.} Commun. Math. Phys. {\bf 50}, 79-85 (1976).
\item[{[G1]}] Ginibre J.: {\em Some applications of functional integration in Statistical Mechanics.} In: {\em Statistical Mechanics and Quantum Field Theory}, eds. C. De Witt and R. Stora, Gordon and Breach (New York 1971).
\item[{[G2]}] Ginibre J.: {\em Reduced density matrices of quantum gases. I. Limit of infinite volume.} J. Math. Phys. {\bf 6}, 238-251 (1965).
\item[{[G3]}] Ginibre J.: {\em Reduced density matrices of quantum gases. II. Cluster property.} J. Math. Phys. {\bf 6}, 252-262 (1965).
\item[{[G4]}] Ginibre J.: {\em Reduced density matrices of quantum gases. III. Hard-core potentials.} J. Math. Phys. {\bf 6}, 1432-1446 (1965).
\item[{[G5]}] Ginibre J.: {\em On the asymptotic exactness of the Bogolyubov approximation for many boson systems.} Commun. Math. Phys. {\bf 8}, 26-51 (1968).
\item[{[Gi]}] Girardeau M.: {\em Relationship between systems of impenetrable bosons and fermions in one dimension.} J. Math. Phys. {\bf 1}, 516-523 (1960).
\item[{[Gr]}] Griffin A.: {\em Excitations in a Bose-condensed liquid} (Cambridge University Press, 1993).
\item[{[Hal]}] Haldane F. D. M.: {\em Effective harmonic-fluid approach to low-energy properties of one-dimensional quantum fluids.} Phys. Rev. Lett. {\bf 47}, 1840-1843 (1981).
\item[{[Ho]}] Hohenberg P. C.: {\em Existence of long-range order in one and two dimensions.} Phys. Rev. {\bf 158}, 383-386 (1967).
\item[{[Hu1]}] Huang K.: {\em Imperfect Bose gas.} In: {\em Studies in Statistical Mechanics, Vol II.} eds. J. de Boer and G. E. Uhlenbeck, North-Holland (Amsterdam 1964), pp. 1-106.
\item[{[Hu2]}] Huang K.: {\em Statistical Mechanics.} 2nd ed. Wiley (New York 1987), p. 303, Prob. 12.7.
\item[{[Hu3]}] Huang K. and Yang C. N.: {\em Quantum-mechanical many-body problem with hard-sphere interaction.} Phys. Rev. {\bf 105}, 767-775 (1957).
\item[{[Hu4]}] Huang K., Yang C. N. and Luttinger J. M.: {\em Imperfect Bose with hard-sphere interactions.} Phys. Rev. {\bf 105}, 776-784 (1957).
\item[{[Kac1]}] Kac M.: {\em On distributions of certain Wiener functionals.} Trans. Amer. Math. Soc. {\bf 65}, 1-13 (1949).
\item[{[Kac2]}] Kac M.: {\em On some connections between probability theory and differential and integral equations.} In: Proceedings of the Second Berkeley Symposium on Probability and Statistics, J. Neyman ed., Berkeley, University of California Press (1951).
\item[{[Kac3]}] Kac M. and Luttinger J. M.: {\em Bose-Einstein condensation in the presence of imputities.} J. Math. Phys. {\bf 14}, 1626-1628 (1973).
\item[{[Kad]}] Kadanoff L. P.: {\em Slippery wave functions.} J. Stat. Phys. {\bf 152}, 805-823 (2013).
\item[{[Kag]}] Kagan M. Yu.. {\em Modern trends in superconductivity and superfluidity} (Springer-Verlag, 2013).
\item[{[Kah]}] Kahn B. and Uhlenbeck G. E.: {\em On the theory of condensation.} Physica {\bf 5}, 399-416 (1938).
\item[{[Kap]}] Kapitza P.: {\em Viscosity of liquid helium below the $\lambda$-point.} Nature {\bf 141}, 74 (1938).
\item[{[Ken]}] Kennedy T., Lieb E. H. and Shastry B. S.: {\em The XY model has long-range order for all spins and all dimensions greater than one.} Phys. Rev. Lett. {\bf 61}, 2582-2584 (1988).
\item[{[Ko]}] Korepin V. E., Bogoliubov N. M. and Izergin A. G.: {\em Quantum inverse scattering method and correlation functions.} Cambridge University Press (1993) Ch. XVIII.2.
\item[{[Kub]}] Kubo K. and Kishi T.: {\em Existence of long-range order in the XXZ model.} Phys. Rev. Lett. {\bf 61}, 2585-2587 (1988).
\item[{[Lan1]}] Landau L. D.: {\em The theory of superfluidity of helium II.} J. Phys. USSR {\bf 5}, 71-90 (1941).
\item[{[Lan2]}] Landau, L.D.: {\em On the theory of supefluidity of helium II.} J. Phys. USSR {\bf 11}, 91-92 (1947).
\item[{[Leg]}] Leggett A. J.: {\em Quantum liquids} (Oxford University Press, 2006).
\item[{[Len]}] Lenard A.: {\em Momentum distribution in the ground state of the one-dimensional system of impenetrable bosons.} J. Math. Phys. {\bf 5}, 930-943 (1964).
\item[{[Lew1]}] Lewis J. T., Zagrebnov V. A. and Pul\'e J. V.: {\em The large deviation principle for the Kac distribution.} Helv. Phys. Acta {\bf 61} 1063-1078 (1988).
\item[{[Li1]}] Lieb E. H. and Liniger W.: {\em Exact analysis of an interacting Bose Gas. I. The general solution and the ground state.} Phys. Rev. {\bf 130}, 1605-1616 (1963).
\item[{[Li2]}] Lieb E. H.: {\em Exact analysis of an interacting Bose gas. II. The excitation spectrum.} Phys. Rev. {\bf 130}, 1616-1624 (1963).
\item[{[Li3]}] Lieb E. H. and Yngvason J.: {\em Ground state energy of the low density Bose gas.} Phys. Rev. Lett. {\bf 80}, 2504-2507 (1998).
\item[{[Li4]}] Lieb E. H. and Yngvason J.: {\em The ground state energy of a dilute two-dimensional Bose gas.} J. Stat. Phys. {\bf 103}, 509-526 (2001).
\item[{[Li5]}] Lieb E. H., Seiringer R., Solovey J. P. and Yngvason J.: {\em The mathematics of the Bose gas and its condensation.}  Birkh\"auser Verlag (Basel-Boston-Berlin 2005), and arXiv:cond-mat/0610117.
\item[{[Li6]}] Lieb E. H., Seiringer R. and Yngvason J.: {\em Justification of {\rm c}-number substitution in bosonic Hamiltonians.} Phys. Rev. Lett. {\bf 94}, 080401 (2005).
\item[{[Li7]}]  Lieb E. H.: {\em  Simplified approach to the ground-state energy of an imperfect Bose gas.} Phys. Rev. {\bf 130}, 2518-2528 (1963).
\item[{[Lon1]}] London F.: {\em The $\lambda$-phenomenon of liquid helium and the Bose-Einstein degeneracy.} Nature {\bf 141}, 643-644 (1938).
\item[{[Lon2]}] London F.: {\em On the Bose-Einstein condensation.} Phys. Rev. {\bf 54}, 947-954 (1938).
\item[{[Lon3]}] London F.: {\em Superfluids, Vol. 2: Macroscopic theory of superfluid helium.} (Wiley, New York, 1954).
\item[{[MP1]}] Martin Ph. A. and Piasecki J.: {\em Self-consistent equation for an interacting Bose gas.} Phys. Rev. E {\bf 68}, 016113 (2003).
\item[{[MP2]}] Martin Ph. A. and Piasecki J.: {\em Bose gas beyond mean field .} Phys. Rev. E {\bf 71}, 016109 (2005).
\item[{[No]}] Nozi\`eres P. and Pines D.: {\em Theory of quantum liquids II: Superfluid Bose liquids.} Addison-Wesley (Redwood City, 1990).
\item[{[Pen]}] Penrose O. and Onsager L.: {\em Bose-Einstein condensation and liquid He.} Phys. Rev. {\bf 104}, 576-584 (1956).
\item[{[Pet]}] Pethick C. J. and Smith H.: {\em Bose-Einstein condensation in dilute gases.} Cambridge University Press (2002).
\item[{[Pi1]}] Pitaevskii L. and Stringari S.: {\em Bose-Einstein condensation.} Clarendon Press (Oxford 2003).
\item[{[Pi2]}] Pitaevskii L. and Stringari S.: {\em Uncertainty principle, quantum fluctuations, and broken symmetries.} J. Low Temp. Phys. {\bf 85}, 377-388 (1991).
\item[{[R1]}] Ruelle D.: {\em Classical statistical mechanics of a system of particles.} Helv. Phys. Acta \textbf{36}, 183-197 (1963).
\item[{[R2]}] Ruelle D.: \emph{Statistical Mechanics.} W. A. Benjamin (New York-Amsterdam 1969).
\item[{[R3]}] Ruelle D.: {\em Superstable interactions in classical statistical mechanics.} Commun. Math. Phys. {\bf 18}, 127-159 (1970).
\item[{[R4]}] Ruelle D.: {\em Correlation functions of classical gases.} Ann. Phys. {\bf 25}, 109-120 (1963).
\item[{[R5]}] Ruelle D.: {\em Cluster property of the correlation functions of classical gases.} Rev. Mod. Phys. {\bf 36}, 580-584 (1964).
\item[{[Scha]}] Schakel A. M. J.: {\em Percolation, Bose-Einstein condensation, and string proliferation.} Phys. Rev. E {\bf 63}, 026115 (2001).
\item[{[Schr]}] Schr\"odinger E.: {\em Letter to Einstein on 5 February 1925.} The collected papers of Albert Einstein, Vol. 14, Document 433, p. 429.
\item[{[Schu]}] Schultz T. D.: {\em Note on the one-dimensional gas of impenetrable point-particle bosons.} J. Math. Phys. {\bf 4}, 666-671 (1963).
\item[{[Se1]}] Seiringer R.: {\em Free energy of a dilute Bose gas: Lower bound.} Commun. Math. Phys. {\bf 279}, 595-636 (2008).
\item[{[Se2]}] Seiringer R. and Ueltschi D.: {\em Rigorous upper bound on the critical temperature of dilute Bose gases.} Phys. Rev. B {\bf 80}, 014502 (2009).
\item[{[Sew]}] Sewell G. L.: {\em Quantum mechanics and its emergent macrophysics.} Princeton University Press (2002).
\item[{[Sn]}] Snow W. M. and Sokol P. E.: {\em Density and temperature dependence of the momentum distribution in liquid Helium 4.} J. Low. Temp. Phys. {\bf 101}, 881-928 (1995).
\item[{[So]}] Sosnick T. R., Snow W. M. and Sokol P. E.: {\em Deep-inelastic neutron scattering from liquid He4.} Phys. Rev. B{\bf 41}, 11185-11202 (1990).
\item[{[Su1]}] S\"ut\H o A.: {\em Percolation transition in the Bose gas.} J. Phys. A: Math. Gen. {\bf 26}, 4689-4710 (1993).
\item[{[Su2]}] S\"ut\H o A.: {\em Percolation transition in the Bose gas: II.} J. Phys. A: Math. Gen. {\bf 35}, 6995-7002 (2002). See also arXiv:cond-mat/0204430v4 with addenda after Eqs.~(34) and (44).
\item[{[Su5]}] S\"ut\H o A.: {\em Ground state at high density.} Commun. Math. Phys. {\bf 305}, 657-710 (2011).
\item[{[Su6]}] S\"ut\H o A.: {\em The total momentum of quantum fluids.} J. Math. Phys. \textbf{56}, 081901 (2015), Section IV.
\item[{[Su8]}] S\"ut\H o A.: {\em Equivalence of Bose-Einstein condensation and symmetry breaking.} Phys. Rev. Lett. {\bf 94}, 080402 (2005).
\item[{[Su9]}] S\"ut\H o A.: {\em Bose-Einstein condensation and symmetry breaking.} Phys. Rev. A {\bf 71}, 023602 (2005).
\item[{[Su10]}] S\"ut\H o A. and Sz\'epfalusy P.: {\em Variational wave functions for homogenous Bose systems.} Phys. Rev. {\bf A77}, 023606 (2008).
\item[{[Su11]}] S\"ut\H o A.: {\em Fourier formula for quantum partition functions.} arXiv:2106.10032 [math-ph] (2021).
\item[{[Su13]}] S\"ut\H o A.: {\em Crystalline ground states for classical particles.} Phys. Rev. Lett. {\bf 95}, 265501 (2005).
\item[{[T1]}] Tisza L.: {\em Transport phenomena in helium II.} Nature {\bf 141}, 913 (1938).
\item[{[T2]}] Tisza L.: {\em La viscosit\'e de h\'elium liquide et la statistique de Bose-Einstein.} C. R. Paris {\bf 207}, 1035-1186 (1938).
\item[{[Ton]}] Tonks L.: {\em The complete equation of state of one, two and three-dimensional gases of hard elastic spheres.} Phys. Rev. {\bf 50}, 955-963 (1936).
\item[{[Tth]}] T\'oth B.: {\em Phase transition in an interacting Bose system. An application of the theory of Ventsel’ and Freidlin.} J. Stat. Phys {\bf 61}, 749–764 (1990).
\item[{[Ued]}] Ueda M.: {\em Fundamentals and new frontiers of Bose-Einstein condensation.} World Scientific (2010).
\item[{[Uel1]}] Ueltschi D.: {\em Relation between Feynman cycles and off-diagonal long-range order.} Phys. Rev. Lett. {\bf 97}, 170601 (2006).
\item[{[Uel2]}] Ueltschi D.: {\em Feynman cycles in the Bose gas.} J. Math. Phys. {\bf 47}, 123303 (2006).
\item[{[Uh]}] Uhlenbeck G. E.: {\em Dissertation Leiden} 1927, p. 69.
\item[{[Ver]}] Verbeure A.: {\em Many body boson systems: half a century later.} Springer (2011).
\item[{[Vers]}] Vershik A. M.: {\em Statistical mechanics of combinatorial partitions, and their limit shapes.} Funk. Anal. Appl. {\bf 30}, 90-105 (1996).
\item[{[W]}] Wagner H.: {\em Long-wavelength excitations and the Goldstone theorem in many-particle systems with "broken symmetries".} Z. Phys. {\bf 195}, 273-299 (1966).
\item[{[Ya]}] Yang C. N.: {\em Concept of off-diagonal long-range order and the quantum phases of liquid He and of superconductors.} Rev. Mod. Phys. {\bf 34}, 694-704 (1962).
\item[{[Yi]}] Yin J.: {\em Free energies of dilute Bose gases: Upper bound.} J. Stat. Phys. {\bf 141}, 683-726 (2010).
\item[{[Z]}] Zagrebnov V. A. and Bru J.-B.: {\em The Bogoliubov model of weakly imperfect Bose gas.} Phys. Rep. {\bf 350}, 291-434 (2001).

\end{enumerate}

\end{document}